\documentclass[11pt]{article}

\usepackage{newtxtext,newtxmath}

\usepackage{graphicx,xcolor}

\usepackage[letterpaper,margin=1in]{geometry}

\usepackage{bm}
\usepackage{braket}
\usepackage[normalem]{ulem}

	\renewcommand\bra[1]{{\langle{#1}|}}
	\renewcommand\ket[1]{{|{#1}\rangle}}

\usepackage{lscape}

\newcommand{\device}[1]{$\mathsf{ibm\_#1}$}
\newcommand{\nitrogen}{N$_2$ }
\newcommand{\fetwostwo}{[2Fe-2S] }
\newcommand{\fefoursfour}{[4Fe-4S] }

\newcommand{\rhf}{{\mathrm{RHF}}}
\newcommand{\vett}[1]{{\bf{#1}}}
\newcommand{\bts}{\vett{x}}
\newcommand{\nmo}{ N_{\mathrm{MO}} }
\newcommand{\nalpha}{ N_\alpha }

\newcommand{\nvir}{ \nmo-\nalpha }
\newcommand{\numberop}[1]{\hat{n}_{#1}}
\newcommand{\crt}[1]{ \hat{a}^\dagger_{#1} }
\newcommand{\dst}[1]{ \hat{a}^{\phantom{\dagger}}_{#1} }
\newcommand{\scientific}[2]{$#1 \cdot 10^{#2}$}
\newcommand{\abs}[1]{\left\vert#1\right\vert}
\newcommand{\norm}[1]{\left\Vert#1\right\Vert}
\newcommand{\innerprod}[2]{\langle#1\vert#2\rangle}
\newcommand{\groundstate}{\mathrm{G}}
\newcommand{\posn}[1]{\vett{R}_{#1}}
\newcommand{\pose}[1]{\vett{r}_{#1}}
\newcommand{\nocontentsline}[3]{}
\newcommand{\tocless}[2]{\bgroup\let\addcontentsline=\nocontentsline#1{#2}\egroup}

\renewenvironment{abstract}
	{\quotation}
	{\endquotation}

\date{}

\makeatletter
\renewcommand{\fnum@figure}{\textbf{Figure \thefigure}}
\renewcommand{\fnum@table}{\textbf{Table \thetable}}
\makeatother

\usepackage{scicite}

\usepackage{url}

\def\scititle{
Chemistry Beyond the Scale of Exact Diagonalization on a Quantum-Centric Supercomputer
}
\title{\bfseries \boldmath \scititle}

\author{
	Javier Robledo-Moreno$^{1\dagger}$, 
	Mario Motta$^{1\ast}$, 
	Holger Haas$^{1}$, 
	Ali Javadi-Abhari$^{1}$, \and
	Petar Jurcevic$^{1}$, 
	William Kirby$^{2}$, 
	Simon Martiel$^{3}$, 
	Kunal Sharma$^{1}$, 
	Sandeep Sharma$^{4}$, \and
	Tomonori Shirakawa$^{5, 6, 7}$, 
	Iskandar Sitdikov$^{1}$, 
	Rong-Yang Sun$^{5, 6, 7}$, 
	Kevin~J.~Sung$^{1}$, \and
	Maika Takita$^{1}$, 
	Minh C. Tran$^{2}$, 
	Seiji Yunoki$^{5, 6, 7, 8}$, 
	Antonio Mezzacapo$^{1\perp}$. \and
	\small$^{1}$IBM Quantum, IBM T.J. Watson Research Center, Yorktown Heights, NY 10598, USA.\and
	\small$^{2}$IBM Quantum, IBM Research Cambridge, Cambridge, MA 02142, USA.\and
	\small$^{3}$IBM Quantum, IBM France Lab, Orsay, France.\and
	\small$^{4}$Department of Chemistry, University of Colorado, Boulder, CO 80302, USA.\and
	\small$^{5}$Computational Materials Science Research Team, RIKEN Center for Computational Science (R-CCS), \and \small Kobe, Hyogo, 650-0047, Japan.\and
	\small$^{6}$Quantum Computational Science Research Team, RIKEN Center for Quantum Computing (RQC),\and \small  Wako, Saitama, 351-0198, Japan.\and
	\small$^{7}$RIKEN Interdisciplinary Theoretical and Mathematical Sciences Program (iTHEMS), \and \small  Wako, Saitama 351-0198, Japan. \and
	\small$^{8}$RIKEN Center for Emergent Matter Science (CEMS), Wako, Saitama 351-0198, Japan.\and
	\small$^\dagger$Corresponding author. Email: j.robledomoreno@ibm.com\and
	\small$^\ast$Corresponding author. Email: mario.motta@ibm.com\and
	\small$^\perp$Corresponding author. Email: mezzacapo@ibm.com\and
}

\begin{document} 

\maketitle

\begin{abstract}

A universal quantum computer can simulate diverse quantum systems, with electronic
structure for chemistry offering challenging problems for practical use cases around the hundred-qubit mark. While current quantum
processors have reached this size, deep circuits and large number of measurements lead to prohibitive
runtimes for quantum computers in isolation. Here, we demonstrate the use of classical distributed
computing to offload all but an intrinsically quantum component of a workflow for electronic structure
simulations. Using a Heron superconducting processor and the supercomputer Fugaku, we
simulate the ground-state dissociation of N$_2$ and the [2Fe-2S] and [4Fe-4S] clusters, with circuits up to 77
qubits and 10,570 gates. The proposed algorithm processes quantum samples to produce upper
bounds for the ground-state energy and sparse approximations to the ground-state
wavefunctions. Our results suggest that, for current error rates, a quantum-centric supercomputing architecture can tackle challenging chemistry problems beyond sizes amenable
to exact diagonalization.

\end{abstract}

\section*{Introduction}

\noindent
The most common task in theoretical quantum chemistry is the computation of ground-state energies by solving the Schr\"{o}dinger equation $H \ket{\Psi} = E \ket{\Psi}$ in the Born-Oppenheimer approximation. 
Exact numerical solutions in a finite basis set have a cost growing combinatorially in the number of electrons and orbitals. This limits exact diagonalization in the full configuration interaction (FCI) to system sizes close to 22 electrons in 22 orbitals (22e,22o)~\cite{vogiatzis2017pushing} and (26e,23o)~\cite{gao2024distributed}.
For system sizes beyond the reach of FCI, one must rely on approximate methods, e.g., diagrammatic techniques,
wavefunction ansatzes,  
and Monte Carlo integration
\cite{leblanc2015solutions,motta2017towards}.

Progress in quantum computing has triggered a flurry of theoretical proposals for computational chemistry over the last decade (e.g.,~\cite{cao2019quantum,mcardle2020quantum,bauer2020quantum}).
At the same time, attempts have been made at implementations on pre-fault-tolerant quantum processors~\cite{kandala2017hardware,google2020hartree,huggins2022unbiasing,motta2023quantum,zhao2023orbital,obrien2023pair, weaving2023contextualVQE_N2}, but these have so far been limited to small systems, for two main reasons.
First, despite numerous efforts to improve on the measurement problem (e.g.,~\cite{verteletskyi2020measurement,huggins2021efficient,dutt2023practical}), runtime for energy expectation value estimation on interesting systems remains out of any reasonable timescale.
Second, the depths of chemically-motivated quantum circuits for computations of chemistry are very high. For unitary coupled cluster~\cite{anand2022quantum} and a single step of time evolution, these quantities scale as $M^4$~\cite{hastings2014improving} on a system with $M$ spin-orbitals. While this scaling can be improved with various techniques~\cite{lee2021even}, on pre-fault-tolerant devices the signal emerging from circuits of such size is weakened by the accumulation of gate errors and qubit decoherence.

 In this manuscript we show that a quantum-centric supercomputing architecture and workflow -- which we call \textit{Sample-Based Quantum Diagonalization} (SQD) -- allow to tackle realistic electronic structure problems on system sizes beyond the reach of exact diagonalization on pre-fault-tolerant quantum processors. We conduct quantum experiments to study the ground-state properties of the \nitrogen molecule and the [2Fe-2S] and [4Fe-4S] clusters using 58, 45, and 77 qubits respectively, and a maximum number of 3.5 K two-qubit gates.

The manuscript is structured as follows. In the \textit{Results} section we provide a brief description of the problem statement, the concerted quantum-classical workflow and the configuration recovery technique, as well as of the quantum circuits run in the experiments. This section ends with the presentation of the experiment results on the ground state properties of the \nitrogen molecule in a correlation-consistent basis set and the active spaces of the [2Fe-2S] and [4Fe-4S] clusters. The \textit{Discussion} section summarizes our findings and examines some conditions for the advantage with SQD or variations thereof. The \textit{Materials and Methods} section provides detailed explanations on the subspace projection and diagonalization and approximate total spin symmetry restoration, the configuration recovery technique, and experimental details including the construction of the quantum circuits and the mapping into quantum processors.

\section*{Results}

\begin{figure*}[h!]
    \centering
    \includegraphics[width=1\linewidth]{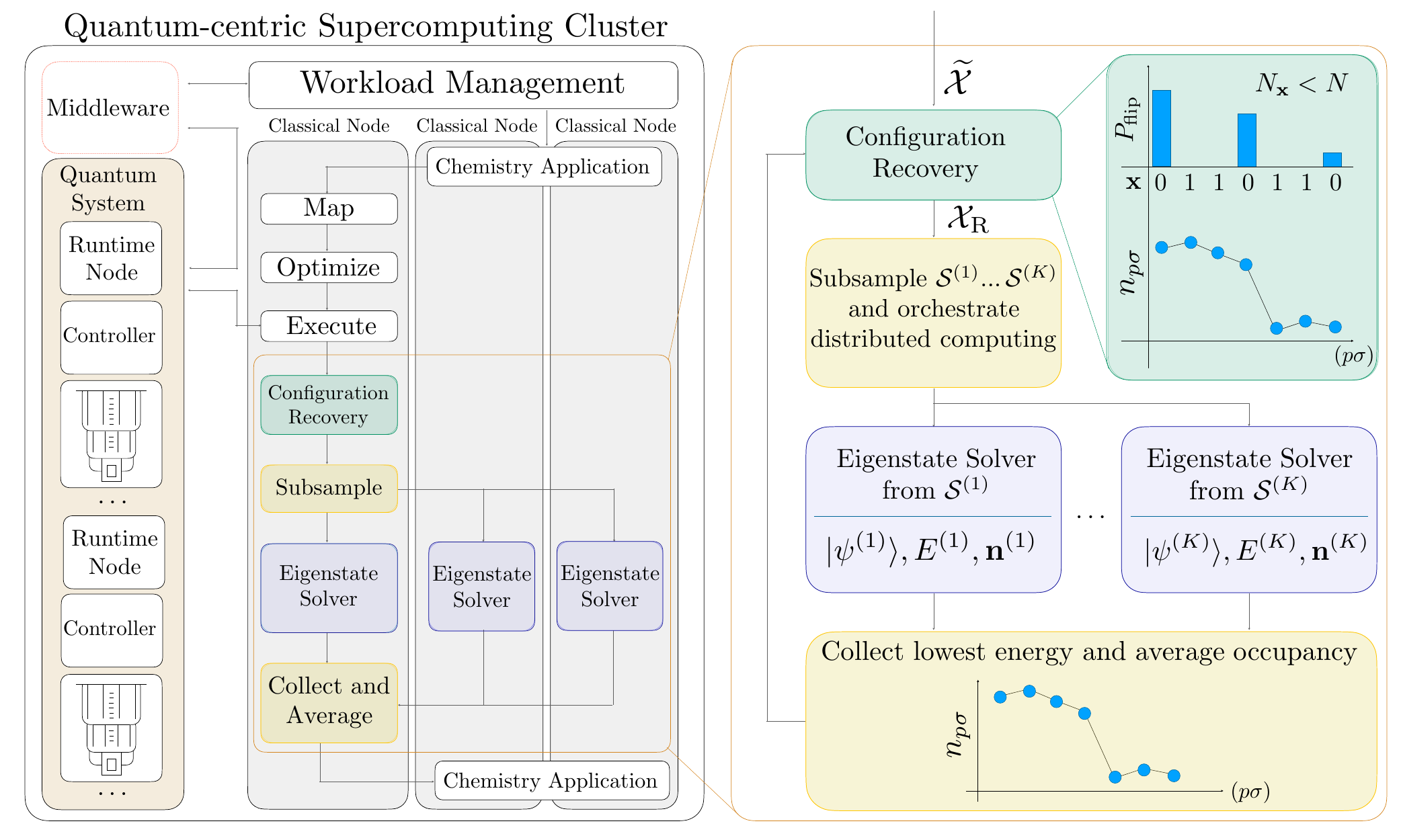}
    \caption{\textbf{Quantum-centric supercomputing architecture and sample-based quantum diagonalization workflow diagram. Left:} We illustrate a simplified architecture used to execute our workflow. The architecture has a cluster with a quantum system alongside classical runtime nodes within an isolated environment. A workload management system controls hybrid quantum-classical jobs through middleware. Our workflow is distributed on a set of classical nodes. It includes standard quantum chemistry application routines such as computing electronic integrals, mapping to qubits, and preparing circuits to be executed. \textbf{Right:} Details of the classical post-processing step. The input is a set of noisy samples $\widetilde{\mathcal{X}}$ from the quantum execution that are processed with our configuration recovery step, using information from a vector $\vett{n}$ of reference orbital occupancies. The green inset shows an example where a configuration with $N_\bts<N$ is corrected. The set of recovered configurations $\mathcal{X}_{\textrm{R}}$ is sub-sampled and distributed for projection and diagonalization on parallel classical nodes. A new average reference occupancy vector $\vett{n}$ is computed from the results, and the configuration recovery loop is repeated self-consistently until convergence.}
    \label{fig:Diagram}
\end{figure*}

We set up the discussion of our results by considering the quantum-centric supercomputing architecture~\cite{alexeev2023quantum} schematized in Fig.~\ref{fig:Diagram}. The architecture enables scaling of computational capacity by leveraging quantum processors for their natural task: executing a limited number of large quantum circuits. 
We follow the workflow in Fig.~\ref{fig:Diagram} to summarize our methods.

Our main goal is to find the ground state of chemistry Hamiltonians
\begin{equation}
\label{eq:es_ham}
\hat{H} = \sum_{ \substack{pr\\\sigma} } h_{pr} \, \crt{p\sigma} \dst{r\sigma}
+ 
\sum_{ \substack{prqs\\\sigma\tau} }
\frac{(pr|qs)}{2} \, 
\crt{p\sigma}
\crt{q\tau}
\dst{s\tau}
\dst{r\sigma}
\;,
\end{equation}
expanded over a discrete basis set.
Here we have defined the fermionic creation/annihilation operator $\crt{p\sigma}$/$\dst{p\sigma}$ associated to the $p$-th basis set element and the spin $\sigma$, while $h_{pr}$ and $(pr|qs)$ are the one- and two-body electronic integrals, obtained from standard chemistry software~\cite{sun2018pyscf}. Throughout this manuscript we use molecular orbitals as basis set elements. We map the degrees of freedom of Eq.~(\ref{eq:es_ham}) to qubits with a Jordan-Wigner (JW) transformation~\cite{JordanWigner1928}. We then construct a quantum circuit to be executed on quantum hardware, preparing a state $|\Psi\rangle$ on $M$ qubits, which represents a molecular wavefunction on $M$ molecular spin-orbitals. In the JW mapping, the single-qubit basis states $|0\rangle$/$|1\rangle$ represent empty/occupied spin-orbitals. These mapping and optimization steps are performed on classical nodes, see Fig.~\ref{fig:Diagram}. 
We execute the circuit on a quantum computer and measure $|\Psi\rangle$ in the computational basis. Repeating this produces a set of measurement outcomes
\begin{equation}
    \widetilde{\mathcal{X}} = \left\{ \bts \; | \; \bts \sim \widetilde{P}_\Psi (\bts) \right\}
\end{equation}
in the form of bitstrings $\bts \in \{0, 1\}^M$ distributed according to some $\widetilde{P}_\Psi$; the bitstrings represent electronic configurations (Slater determinants).

\subsection*{Configuration recovery}

On a pre-fault-tolerant quantum computer, the action of noise alters the distribution from its ideal form $P_\Psi = \left|\langle \bts | \Psi \rangle\right|^2$ to some other $\widetilde{P}_\Psi$, which generates the noisy set of configurations $\widetilde{\mathcal{X}}$, accessible to us via quantum measurement.
Noise in the quantum system broadens the distribution $P_\Psi$ over configurations that do not contribute to low-energy states, so-called {\it deadwood} \cite{ivanic2001identification}. As a result, only a fraction of $\widetilde{\mathcal{X}}$ contains a meaningful quantum signal.
To improve this scenario, we introduce a {\it self-consistent configuration recovery} technique, which allows a probabilistic partial recovery of noiseless configuration samples from $\widetilde{\mathcal{X}}$.

The configuration recovery scheme is inspired by the structure of chemistry problems. The Hamiltonian in Eq.~(\ref{eq:es_ham}) conserves the number of particles separately for each spin species. The recovery routine targets configurations $\bts$ that have the wrong particle number $N_\bts \neq N$ due to the accumulation of errors in the execution of the quantum circuit. 
\begin{figure}
    \centering
    \includegraphics[width=.6\linewidth]{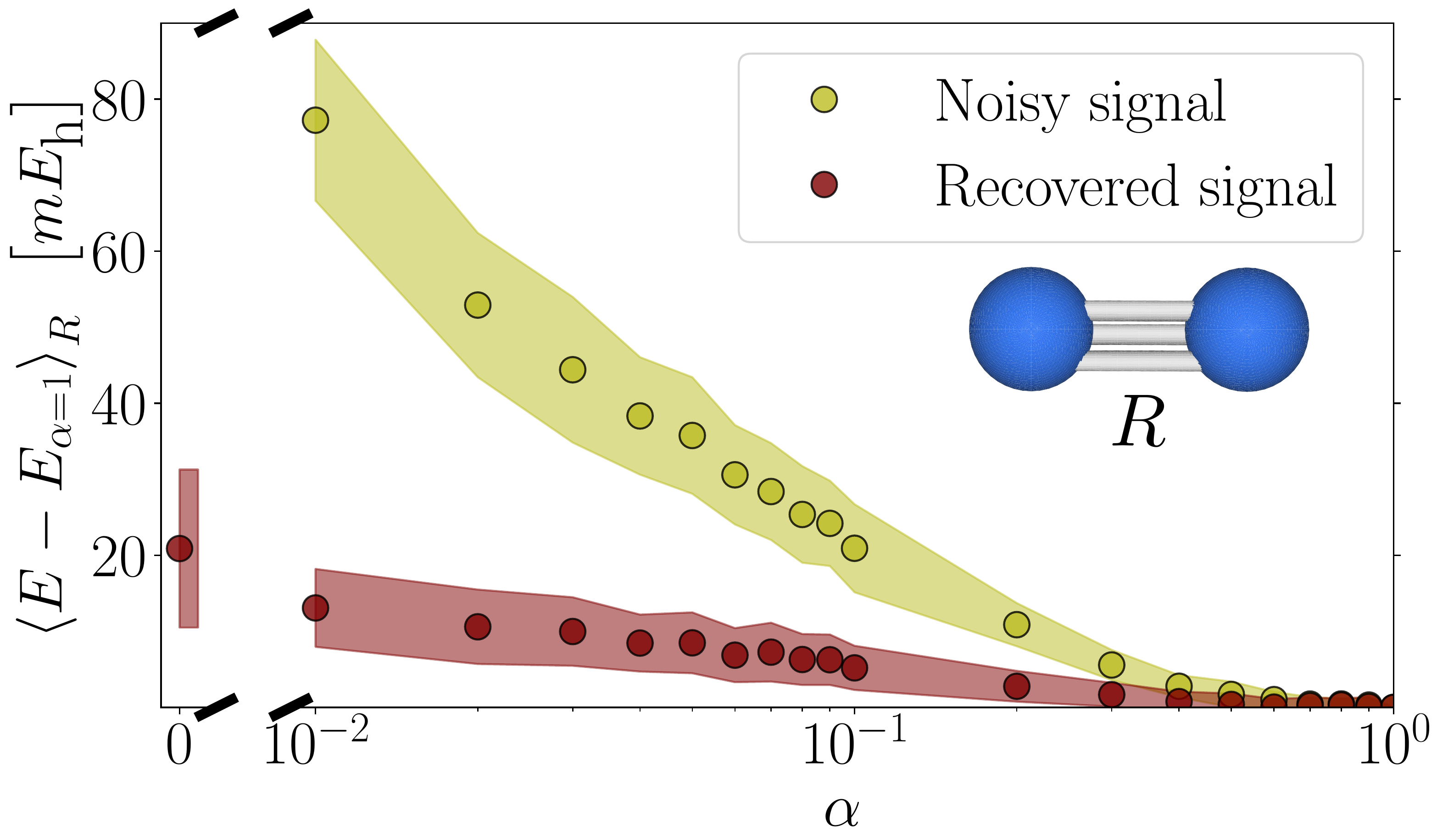}
    \caption{\textbf{Self-consistent configuration recovery.} Energy error (mean and standard deviation) in the dissociation of \nitrogen (6-31g), averaged over 24 bond lengths equally spaced between $R = 0.7 \, \textrm{\AA}$ and $R = 3.0 \, \textrm{\AA}$, as a function of quantum signal to noise ratio, parametrized by $\alpha$. The point at $\alpha = 0$ corresponds to sampling from the uniform distribution (no signal). The energies are obtained via projection and diagonalization of $d=10^6$ raw noisy samples against a subspace of the same size obtained by the configuration recovery routine. 
    \label{fig:1.5}}
\end{figure}
Repeated rounds of recovery can be carried out self-consistently. The first step of each recovery round is to iterate through the set $\widetilde{\mathcal{X}}$ and find configurations $\bts$ with $N_\bts \neq N$ particles.
If $N_\bts > N$ (or $N_\bts<N$), $|N_\bts-N|$ bits are sampled to be flipped from the set of occupied (or empty) spin-orbitals, according to a distribution proportional to a monotonically-increasing function~\cite{Supplement} of $\left|x_{p\sigma} - n_{p\sigma}\right|$, the distance from the current value of the bit to the average occupancy of the spin-orbital $p\sigma$, obtained from the previous recovery round.
This generates a new set of recovered configurations $\mathcal{X}_{\textrm{R}}$.

\begin{figure*}
    \centering
    \includegraphics[width=1\linewidth]{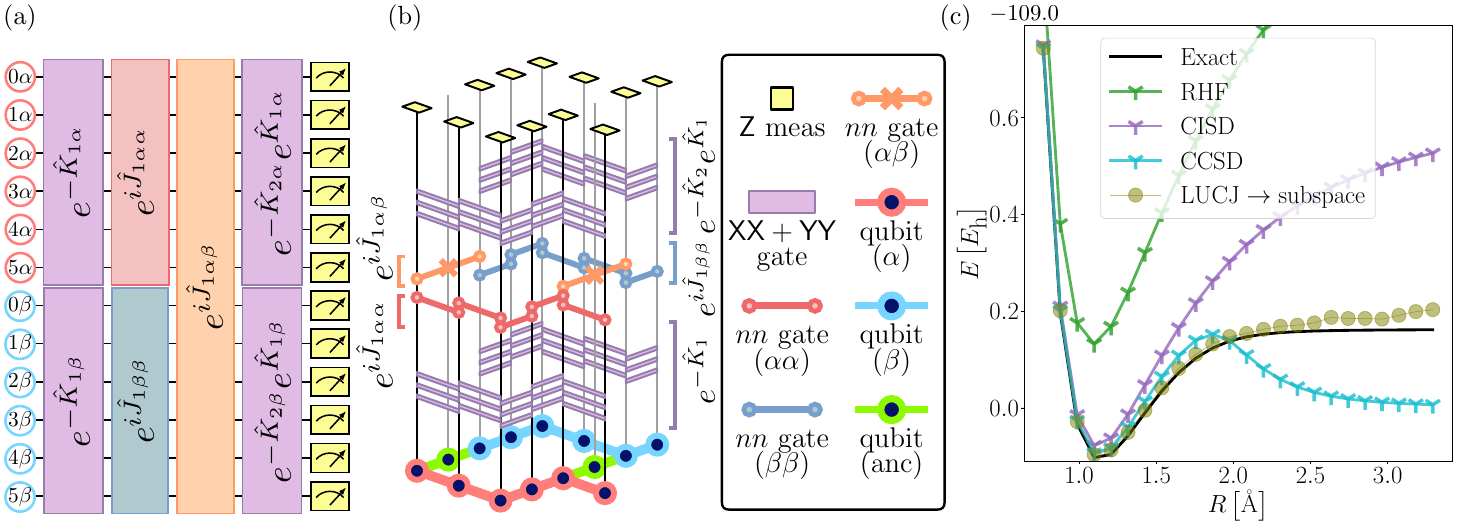}
    \caption{\textbf{Quantum circuits for chemistry: local unitary cluster Jastrow}. \textbf{(a)} Schematic representation of the truncated LUCJ circuit used to generate the set of samples. The circuit is comprised of orbital rotation unitaries $\exp ( -K_{\mu\sigma} )$, same-spin cluster operators $\exp (i\left(J_{\mu\alpha \alpha}+J_{\mu\beta \beta}\right))$, and opposite-spin cluster operators $\exp (iJ_{\mu\alpha \beta})$, with $\alpha$/$\beta$ denoting spin-up/down. \textbf{(b)} A single unit of the heavy-hex lattice with a compilation of the LUCJ circuit into single- and two-qubit gates, with the $nn$ gate defined as $U_{nn}(\varphi) = \exp(i\varphi (1-Z_0)(1-Z_1)/4)$.
    \textbf{(c)} Comparison of the potential energy surface of \nitrogen (with a 6-31G basis) obtained from a noiseless subspace simulation of the LUCJ circuit in panel (a) with $L = 1$, and optimized parameters, against restricted coupled cluster singles and doubles (CCSD). Restricted Hartree-Fock (RHF), configuration interaction singles and doubles (CISD), and exact energies are shown for reference.}
    \label{fig02}
\end{figure*}

Following the next step of Fig.~\ref{fig:Diagram}, we build $K$ batches of $d$ configurations $\mathcal{S}^{(1)}\hdots , \mathcal{S}^{(K)}$ using samples from the set $\mathcal{X}_{\textrm{R}}$, according to a distribution proportional to the empirical frequencies of each $\bts$ in $\mathcal{X}_{\textrm{R}}$.
We project and diagonalize the Hamiltonian over each $\mathcal{S}^{(k)}: k = 1, \hdots, K $, as proposed recently in the quantum selected configuration interaction method~\cite{kanno2023qQSCI, nakagawa2023adaptqsci}, which draws inspiration from the classical selected configuration interaction framework~\cite{evangelista2014adaptive,schriber2016communication,holmes2016efficient,holmes2016heat,tubman2016deterministic,schriber2017adaptive,sharma2017semistochastic}. 

Each batch of sampled configurations spans a subspace $\mathcal{S}^{(k)}$ in which the many-body Hamiltonian is projected:
\begin{equation}\label{eq:projection}
    \hat{H}_{\mathcal{S}^{(k)}} = \hat{P}_{\mathcal{S}^{(k)}} \hat{H}  \hat{P}_{\mathcal{S}^{(k)}} \textrm{, with } \hat{P}_{\mathcal{S}^{(k)}} = \sum_{\bts \in {\mathcal{S}^{(k)}}} | \bts \rangle \langle \bts | \;.
\end{equation}
The ground states and energies of $\hat{H}_{\mathcal{S}^{(k)}}$, which we label $|\psi^{(k)} \rangle$ and $E^{(k)}$, are then computed using the iterative Davidson method on multiple classical nodes.
The computational cost -- both quantum and classical -- to produce $|\psi^{(k)} \rangle$ is polynomial in $d$, the dimension of the subspace.

The ground states are then used to obtain new occupancies
\begin{equation}
\label{eq:nR_def}
    n_{p\sigma}=  \frac{1}{K}\sum_{1\leq k \leq K}  \left\langle \psi^{(k)} \right| \hat{n}_{p\sigma} \left| \psi^{(k)} \right\rangle,
\end{equation}
for each spin-orbital tuple $(p\sigma)$, averaged on the $K$ batches. These occupancies are sent back to the configuration recovery step, and this entire self-consistent iteration is repeated until convergence, realizing a sample-based quantum diagonalization of the target Hamiltonian. The initial guess for $\vett{n}$ used for the first round of recovery comes from the raw quantum samples in the correct particle sector.
The configuration recovery routine can be seen effectively as a problem-informed clustering of a noisy signal around the occupations $\vett{n}$.
In general, the convergence of the configuration recovery procedure depends on the error rates and the physical properties of the system under consideration. With current error rates, and for the systems in this study, we have observed its convergence within 3 iteration steps in all systems. We always chose a maximum number of 5 recovery iterations. We foresee that the lowering of error rates in future quantum hardware will result in faster convergence. Additional details are provided in the \textit{Materials and Methods} section.

To test the noise robustness, we perform numerical simulations that confirm the improvements of applying the configuration recovery routine to the dissociation of \nitrogen (6-31G basis set). In this test we sample from the exact ground state $P_{\Psi_\mathrm{G}}(\bts) = \left|\langle \bts | \psi_\groundstate \rangle \right|^2$ and we set a subspace dimension of $d = 10^6$. 
We use a global depolarizing noise channel to model the effect of noise, $\widetilde{P}_{\Psi_\mathrm{G}} (\bts) = \alpha  P_{\Psi_\mathrm{G}}(\bts) + (1-\alpha) \frac{1}{2^M}$, with $\alpha \in [0, 1]$ the parameter that controls the amount of noisless quantum signal. Fig.~\ref{fig:1.5} shows the error in the ground-state energy relative to the noiseless case ($\alpha = 1$), as a function of the amount of signal $\alpha$, for the estimator both with and without configuration recovery. On the \nitrogen model, errors below $10 m E_\textrm{h}$ can be obtained from $\sim 20\%$ signal using the raw noisy samples. However, by using configuration recovery we can tolerate a $\sim 2\%$ signal to reach the same error.  This numerical experiment hints that the use of configuration recovery will be crucial for large-scale experiments.

\subsection*{Quantum circuits}

Before presenting our experimental results, we discuss the circuits $|\Psi\rangle$ used to produce the candidate ground states. We employ a truncated version of the local unitary cluster Jastrow (LUCJ) ansatz~\cite{motta2023bridging}, shown in Fig.~\ref{fig02} (a),
\begin{equation}\label{equation:uCJ}
    {| \Psi \rangle = \prod_{\mu=1}^L 
     e^{\hat{K}_\mu} e^{i \hat{J}_\mu} e^{-\hat{K}_\mu} | \bts_\rhf \rangle}.
\end{equation}
Here $\hat{K}_\mu = \sum_{pr, \sigma} K_{pr}^\mu \, \crt{p \sigma} \dst{r \sigma}$ are generic one-body operators,
$\hat{J}_\mu = \sum_{pr,\sigma\tau} J_{p\sigma, r\tau}^\mu \, \numberop{p \sigma} \numberop{r \tau}$ are density-density operators restricted to spin-orbitals that are mapped onto adjacent qubits \cite{motta2023bridging}, and $\bts_\rhf$
is the bitstring representing the restricted Hartree-Fock (RHF) state in the JW mapping.
Through this local approximation, the LUCJ ansatz allows for moderate circuit depths. 
Its accuracy derives from the connection with unitary coupled cluster theory and adiabatic state preparation~\cite{matsuzawa2020jastrow,evangelista2019exact,motta2023bridging}.
The moderate depths of LUCJ are due to the use of exponentials of one-body operators, implementable in linear depth and a quadratic number of 2-qubit gates, and density-density operators, implementable in constant depth and a linear number of ZZ rotations (due to the locality approximation)~\cite{motta2023bridging}. The LUCJ circuit, compiled into one- and two-qubit gates, is shown in Fig.~\ref{fig02} (b).

In Fig.~\ref{fig02} (c) we show a numerical experiment comparing the potential energy curve of \nitrogen (6-31G basis) obtained by restricted CCSD to one obtained from the LUCJ ansatz, numerically optimized using the subspace energy as the objective function~\cite{Supplement}. The dimension of the diagonalization subspace for different bond lengths ranging from $d = 209764$ to $d = 1340964$, with a median of $d = 563250$.  Due to the presence of strong static correlation, CCSD fails in the description of the dissociation curve, while the optimized LUCJ ansatz produces a qualitatively correct dissociation curve. We simulated the LUCJ ansatz using the ffsim library \cite{ffsim}.

Throughout our experiments we use the truncated LUCJ circuit $|\Psi\rangle = e^{-\hat{K}_2} e^{\hat{K}_1} e^{i \hat{J}_1} e^{-\hat{K}_1} | \bts_\rhf \rangle$, which is the result of considering the $L=2$ circuit and removing the last orbital rotation and Jastrow operations. We parameterize the LUCJ circuits converting the CCSD wavefunction in Jastrow form and imposing a locality approximation to the resulting $J^\mu$ tensors, i.e., zeroing out the components $J^\mu_{p\sigma,r\tau}$ not corresponding to adjacent qubits~\cite{motta2023bridging}. For systems where the $J_{p\sigma,r\tau}^\mu$ parameters obtained from CCSD have small amplitude, the $e^{-\hat{K}_1}$ and $e^{\hat{K}_1}$ terms in the ansatz approximately cancel: in such situation, without $e^{\hat{K}_2}$, the resulting wavefunction is over-concentrated around the Hartree-Fock configuration (see the \textit{Materials and Methods} Section for additional information). Although we initialize the LUCJ circuit with CCSD parameters, the nature of these two methods is very different. In particular CCSD theory is non-variational while SQD is. This adds an additional difficulty in the understanding of the relative performance of CCSD and SQD with LUCJ initialized from CCSD parameters. We have performed optimization-free experiments, exploiting the connection between LUCJ and classical coupled cluster theory, yet closing a quantum-classical optimization could further improve the quality of the solutions.

\subsection*{Implementation on a quantum computer}

In the following, we present the experimental results obtained using the methods discussed so far, on Heron quantum processors and the Fugaku supercomputer.
The largest experiment is run on a subset of 77 qubits of a 133-qubit Heron quantum processor. The median fidelites for this subset are $99.77\%$ for two-qubits gates, $99.97\%$ for single-qubit, and readout fidelity of $98.37\%$, with median coherence times $T_1 = 180 \mu$s and $T_2 = 150\mu$s. 
In SQD calculations, it is particularly important to have as many measurement outcomes with correct particle number as possible. However, qubits may be initialized imperfectly, i.e., not in the $\ket{000\dots000}$ state, resulting in more measurement outcomes with incorrect particle number. To mitigate this source of error, we 
employ a reset-mitigation scheme by adding an additional measurement instruction before the circuit execution and post-selecting outcomes based on this first measurement returning the initial state $\ket{000\dots000}$. This post-selection results in $\sim1/3$ retention rate of all the executions, i.e. qubits collapse initialized in the desired state upon the additional measurement with probability $\sim1/3$. 

The classical projection and diagonalizations are obtained with the Davidson method implemented in the library PySCF~\cite{sun2018pyscf} on a single node, or DICE~\cite{sharma2017semistochastic, holmes2016heat} for distributed computing on multiple nodes. Convergence to the most accurate solution can be obtained in two ways: increasing the accuracy per diagonalization with the subspace size $d$, and increasing the number of batches $K$, which will reduce statistical errors in the analysis. For our largest experiment on the [4Fe-4S] cluster, we use up to $d=\textrm{100M}$, distributing a single projection and diagonalization to $64$ nodes of Fugaku, and $K= 100$ batches, for a total of 6400 nodes. We analyze runtime performance as a function of $d$ and $K$ versus the number of nodes used~\cite{Supplement}. At $64$ nodes per diagonalization on the largest experiments, classical runtimes are about $1.5$ hours. The largest HCI calculation that we performed on 16 nodes at $d=\textrm{2.3M}$ took about 16 minutes. 

We perform two classes of experiments: the breaking of the triple bond of \nitrogen (cc-pVDZ basis), in the top panel in Fig.~\ref{fig03} (a), 
and the ground states of [2Fe-2S] and [4Fe-4S] clusters (active spaces of the TZP-DKH basis), shown in the top panels in Fig.~\ref{fig03} (b) and (c). We study the ground-state properties of these molecular systems in the $S_z = 0$ and $S^2 = 0$ subspace, where $S_z$ is the total $\hat{z}$ component of the spin and $S^2 = S^2_x + S^2_y + S^2_z$. In this work, we used Slater determinants to define the subspaces, which in general are not eigenfunctions of $S^2$ unlike configuration state functions (CSFs). In the closed-shell systems studied here, a source of spin contamination is the fact that sampled determinants are not closed under the spin inversion operation. Therefore, we achieved an approximate restoration of the $S^2$ symmetry by extending the set of sampled determinants to ensure closure under spin inversion, as detailed in the \textit{Materials and Methods} section.

\subsubsection*{Triple bond breaking in N$_2$}

\begin{figure*}
    \centering
    \includegraphics[width=1\linewidth]{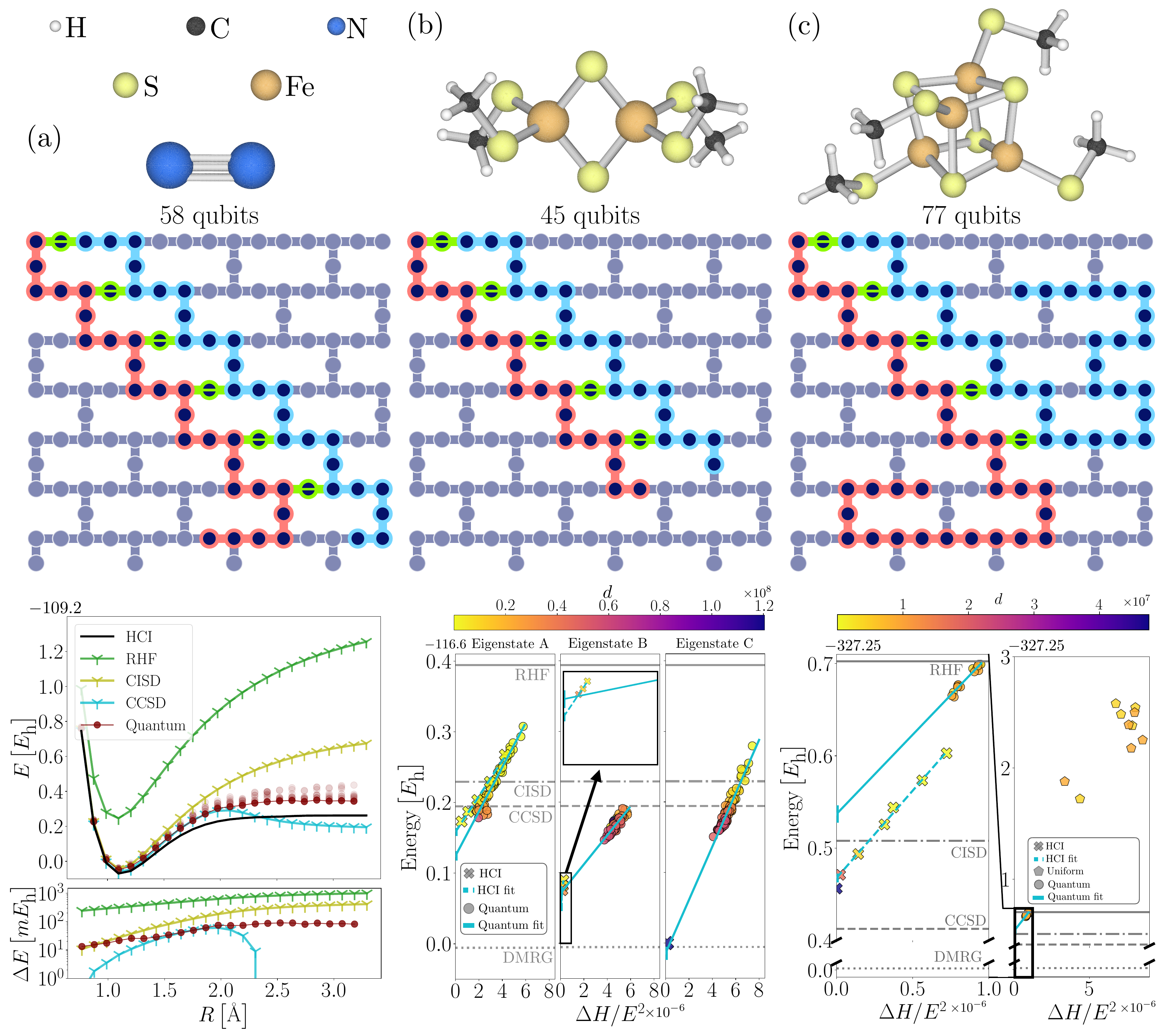}
    \caption{\footnotesize \textbf{Experiments: Chemistry on large basis sets.} \textbf{(a)} 58 qubits are used to model the \nitrogen dissociation (cc-pVDZ basis set). \textbf{(b)} 45 qubits are used for the [2Fe-2S] cluster  (TZP-DKH basis set) and \textbf{(c)} 77 qubits for the [4Fe-4S] cluster (TZP-DKH basis set). The top panels show a 3-dimensional representation of the geometry of each molecule. The middle panels show the qubits selected on a Heron quantum processor layout, following the same color convention as panel (b) in Fig.~\ref{fig02}. The \textbf{bottom panel in (a)} shows the potential energy surface comparison, as well as the energy difference $\Delta E$ between the Heat-Bath Configuration Interaction (HCI) energy and the energies obtained from different methods. The brown scatter plot shows the value of $E^{(k)}$ for all batches of configurations and the connected dots show $\min_k ( E^{(k)} )$. The \textbf{bottom panel} in \textbf{(b)} shows the energy-variance analysis for three different eigenstates that both HCI and our method find upon increasing the value of $d$, as labeled by the color bar. For each approximate eigenstate $|\psi^{(k)} \rangle$, the horizontal axis $\Delta H = \langle \psi^{(k)} | \hat{H}^2 | \psi^{(k)} \rangle - \langle \psi^{(k)} | \hat{H} | \psi^{(k)}  \rangle^2$. The \textbf{bottom panel in (c)} shows a comparison of the energy-variance analysis applied to quantum measurement outcomes and bitstrings (with the correct particle number) sampled from the uniform distribution. The DMRG energy in panels (b) and (c) is from Ref.~\cite{li2017spin}.}
    \label{fig03}
\end{figure*}

The breaking of the N$_2$ bond is a well-known test of the accuracy of electronic structure methods in the presence of static electronic correlation~\cite{fan2006usefulness,bulik2015can}.
Restricted CCSD theory, a dominant paradigm for the accurate description of weakly correlated systems in quantum chemistry, fails in the description of N$_2$ dissociation due to static correlation effects: as correlations become stronger, RHF becomes unstable toward a symmetry-broken unrestricted Hartree-Fock (UHF) state. CCSD built from RHF predicts an artificial barrier to binding and over-correlates at dissociation, whereas CCSD built from a UHF reference dissociates correctly at the cost of spin contamination, a manifestation of L\"owdin's symmetry dilemma. 
We use a correlation-consistent cc-pVDZ basis set, to place emphasis on a theory's ability to treat both dynamic and static correlation in an accurate and balanced manner. We map the \nitrogen molecule onto a Heron processor as shown in the middle panel of Fig.~\ref{fig03} (a). 
We project and diagonalize a Hamiltonian using $d = 16 \cdot 10^6$ configurations. We consider $K = 10$ batches of configurations and each point in the dissociation curve has $|\tilde{\mathcal{X}} | = 100\cdot 10^3$ measurement outcomes and 10 iterations of recovery.

The combined  quantum runtime for all points in the dissociation curve is of $\sim45$ minutes. The experimental data is reported in the bottom panel of Fig.~\ref{fig03} (a), showing the potential energy surface of \nitrogen compared to classical approximate methods. The data from our experiments are consistent with other classical methods except for CCSD, which fails in the description of the dissociation as seen for the smaller basis set considered in Fig.~\ref{fig03}(b). Amongst the classical selected configuration interaction (SCI) methods, \textit{Heat-Bath Configuration Interaction} (HCI)~\cite{holmes2016heat} obtains the best results for \nitrogen and will be our reference classical method in all the other experiments. The difference between our method and HCI energies is everywhere within tens of $mE_\textrm{h}$. We further analyze the accuracy of our experiments as a function of $d$, and the effect of orbital optimizations in the accuracy of the predictions~\cite{Supplement}. This first test demonstrates that we are capable of addressing multi-reference ground states and builds confidence for the next set of experiments, which will focus on assessing the ability of the quantum-classical architecture to do precision many-body physics. 


\subsubsection*{[2Fe-2S] cluster: precision many-body physics}

Iron–sulfur (FeS) clusters are molecular ensembles of sulfide-linked 1- to 8-iron centers in variable oxidation states. They are important cofactors in biological processes ranging from nitrogen fixation to photosynthesis and respiration~\cite{beinert1997iron}.
Their electronic structure, with multiple low-lying states of differing electronic and magnetic character, is responsible for their rich chemistry. At the same time, they pose considerable challenges for experimental studies and numerical tools.
For our experiments we consider the synthetic [Fe$_2$S$_2$(SCH$_3$)$_4$]$^{-2}$ cluster~\cite{sharma2014low},
abbreviated [2Fe-2S] and used in numerical studies to mimic the oxidized dimers prominently found in ferredoxins~\cite{venkateswara2004synthetic}. 

The qubit mapping of the LUCJ circuit on the Heron processor for [2Fe-2S] is shown in the middle panel of Fig.~\ref{fig03} (b). 
We consider $K = 10$ batches of configurations and $|\tilde{\mathcal{X}}| = 2.4576 \cdot 10^6$ measurement outcomes.  
The  quantum runtime for this system is of $\sim45$ minutes. 
For the [2Fe-2S] cluster, we perform an energy-variance analysis of the low-energy spectrum of the molecule. The energy-variance analysis is a tool routinely used in classical computational electronic structure to capture the convergence of the approximate eigenstate energy for different levels of accuracy of a computational method~\cite{Kashima2001Energy_Variance}. Here we use energy-variance analysis to assess the convergence as a function of $d$, which is directly related to quantum and classical accuracy, runtimes, and costs. If one can statistically sample from a good approximation of an eigenstate, points at finite number of samples will be distributed linearly in the energy-variance plane~\cite{Kashima2001Energy_Variance}. This gives us a tool to detect eigenstates for both quantum and classical methods. 

\subsubsection*{[4Fe-4S] cluster: a stress test for methodology and quantum processors}

The circuits considered for the \nitrogen experiments and [2Fe-2S] reached sizes of approximately 1-1.5k two-qubit gates. 
We now test the quality of the signal in noisy circuits that test the limits of Heron processors, using up to 6400 nodes of Fugaku for the classical processing. 
We consider the synthetic [Fe$_4$S$_4$(SCH$_3$)$_4$]$^{-2}$ cluster \cite{sharma2014low},
abbreviated [4Fe-4S], a representative of nature's cubanes, whose ground state deduction from experimental measurements was an early success of inorganic spectroscopy~\cite{papaefthymiou1986moessbauer}. The LUCJ circuit employed for this molecular species contains approximately 3.5k two-qubit gates. The qubit mapping on the Heron processor for [4Fe-4S] is shown in the middle panel of Fig.~\ref{fig03} (c).  As in the previous experiment, we consider $K = 10$ batches of configurations and $|\tilde{\mathcal{X}}| = 2.4576 \cdot 10^6$ measurement outcomes.
The  quantum runtime for this system is of $\sim45$ minutes. 
For $d>250 \cdot 10^3$, configuration recovery is warm started with the $\vett{n}$ obtained from the method at $d = 250 \cdot 10^3$, and only two iterations are then performed.  

This last set of experiments sheds light on the quality of the quantum signal that is passed to the configuration recovery at these large circuit sizes.  
The bottom right panel in Fig.~\ref{fig03} (c) shows a comparison of the energy-variance analysis from measurement outcomes obtained from the Heron processor and configurations sampled from the uniform distribution. 
We see that, even if the quantum solutions produced are worse than other classical methods, the energy and variance obtained from quantum data are significantly lower than those obtained from uniformly distributed configurations (i.e. pure noise), on subspaces of the same size. This confirms that there is a valuable signal at circuit sizes of approximately 3.5k two-qubit gates.

\emph{Significance for quantum computing --} Current quantum computers in isolation can perform calculations on systems sufficiently large that exact brute-force classical solutions are not available~\cite{kim2023evidence, shinjo2024unveiling}. However, these studies have targeted spin systems, leading to circuits that match the connectivity and the measurement and coherence budgets of the quantum devices.
In this work, we present electronic structure calculations on active spaces beyond the scale where full configuration interactions are available. Key to achieve this result is the use of classical and quantum computers in concert, to implement the SQD method.

SQD makes economical use of quantum computing resources by drawing samples from a single quantum circuit. Although in principle other estimators, such as the standard ones used in variational quantum eigensolvers, have bounded variance for any wavefunction, the dire scaling of number of measurements to estimate energies makes them impractical for the molecules targeted in this work~\cite{wecker2015progress}.
It is also more robust against quantum noise, because reconstructing the exact ground-state probability distribution on a quantum computer is not required to get accurate energy approximations, as long as one is sampling relevant configurations (i.e., in the ground-state support).

We have used a LUCJ class of quantum circuits that can reproduce a low-rank decomposition and sparsification of the quantum unitary CCSD~(qUCCSD), which allowed us to keep circuit depths manageable~\cite{motta2023bridging}. Lower error rates on quantum operations will allow to access deeper quantum circuits with higher connectivity, giving access to more general probability distributions. We have performed optimization-free experiments exploiting the connection between LUCJ and classical coupled cluster theory, for the purpose of assessing accuracy and scalability, yet closing a quantum-classical optimization in future work will further improve the quality of our samples.

\emph{Generalization -- }
The SQD method can be applied to simulation tasks other than quantum chemistry, if the target ground-state wavefunction can be accurately approximated by a sparse vector.
Developing quantum circuits with polynomially-sized support in the computational basis will be an important element of the generalization of SQD, as these circuits are the sources of samples processed by classical computers.

To counter wavefunction broadening on current quantum hardware, we have used a self-consistent configuration recovery method, exploiting a problem-inspired clustering that leverages the average occupation numbers of the molecular orbitals. 
We foresee generalizations of our configuration recovery technique to problems other than quantum chemistry that are not informed by the physics of the problem. Conversely, for specific applications one could use even more information about the problem.

\emph{Implications in the search for quantum advantage -- } Our study contributes to the search for quantum advantage in at least two ways: by identifying an accuracy metric, and by formulating a set of conditions to improve over SCI.

Computations can be ranked against three parameters: runtime, energy or cost, and accuracy. While the first two are often easy to measure, ranking by accuracy is in general not straightforward.
For methods that produce upper bounds to the ground-state energy, including SQD, the expectation value of the Hamiltonian defines an unconditional accuracy metric: a lower energy is ranked as a higher quality, all other conditions (e.g. total spin) being equal.
Comparing SQD energies, for example, allowed us to benchmark our results against SCI and uniform configuration sampling on a 77-qubit experiment, without access to exact solutions.
Additionally, the approximate wavefunctions produced here can be stored in classical memory, which permits a classical prover to certify them, and allows their manipulation by further classical processing. 

Using the expectation value of the Hamiltonian as an accuracy metric, one can easily and naturally rank SQD results along with those of variational classical methods, giving the search for this specific form of quantum advantage a quantitative meaning.
Since every variational classical method has a specific domain of applicability~\cite{lee2023evaluating}, identifying areas where SQD may offer an accuracy advantage is a delicate problem.
For example, variational quantum Monte Carlo methods~\cite{becca2017quantum,carleo2017solving}, of paramount importance in many-body physics, are sensitive to the structures of the probability distributions they are modeling, not just to their supports. 
Similarly, methods based on tensor networks~\cite{white1999ab,chan2011density} are successfully used to tackle strongly correlated problems in chemistry, granted the ability to converge their energies with bond dimensions, 
but convergence can be challenging in some cases~\cite{hubig2018error,li2019electronic} because of its computational cost and its sensitivity to the nature and ordering of the basis-set orbitals.
Developing an in-depth understanding of SQD through extensive numeric and methodological investigations is necessary to establish or rule out advantage in strictly variational ground-state simulations.

\begin{figure*}
    \centering
    \includegraphics[width=.75\linewidth]{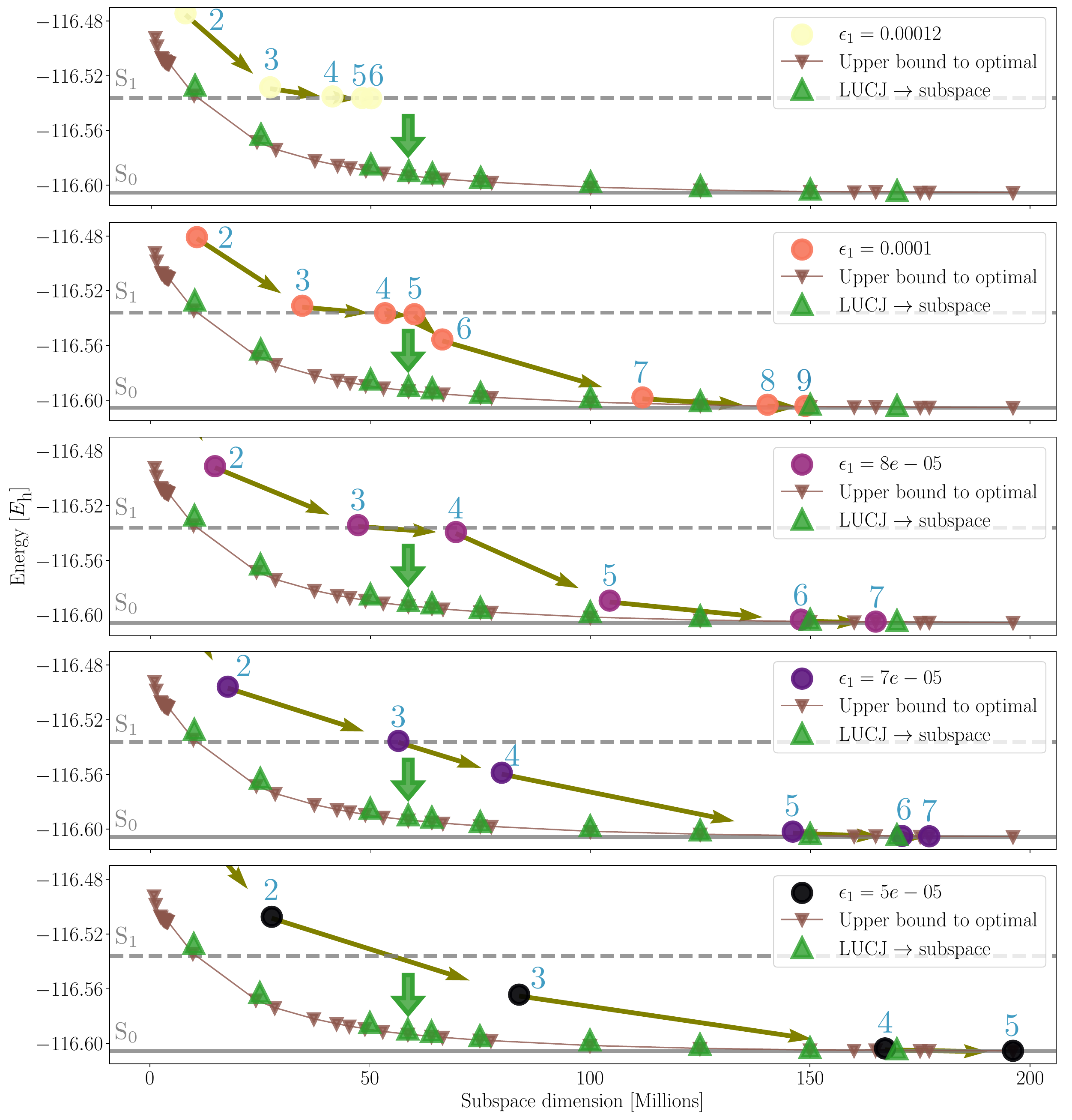}
    \caption{ Comparison of the quality of subspaces generated by HCI and an optimized LUCJ circuit in for the description of the ground state of [2Fe-2S]. Different panels show the energy as a function of subspace dimension of the corresponding diagonalization. The dots connected by arrows show the trajectory of HCI in the energy-subspace dimension plane. Each dot is labeled by the iteration it corresponds to. Different panels correspond to HCI calculations carried out with different values of the selection cutoff $\epsilon_1$, as indicated in the legend. Brown triangles show energies and subspace dimensions obtained after truncating the most accurate HCI wave function by repeatedly removing the electronic configurations corresponding to the lowest wave function amplitudes. The green triangles show the energies for different subspace dimensions  obtained from the optimized LUCJ circuit. The green arrows show the subspace dimension that the circuit was optimized for (see main text). The horizontal solid and dashed lines indicate the DMRG estimates for the S$_0$ and S$_1$ eigenstates.}
    \label{fig:SCI_vs_SQD}
\end{figure*}

SQD shares with SCI the assumption that the ground state may be approximated by a sparse linear combination of determinants, i.e. with a number of determinants much smaller than the Hilbert space dimension. Note that this assumption does not necessarily underline variational Monte Carlo or tensor networks. Therefore, it is natural to look for conditions to improve over SCI. One such condition is the existence of a quantum circuit which produces subspaces of better quality, and more efficiently, than classical heuristic selection methods. In the search for ground states, the quality of a subspace can be determined by a lower variational energy. We conduct numerical experiments that suggest that there exists LUCJ circuits whose samples produce subspaces of better quality than the HCI classical selection heuristic. The LUCJ circuit under consideration shares the same depth and connectivity as the circuits used in the experiments. In this study we consider a particular flavor of SCI, Heat bath Configuration Interaction (HCI)~\cite{holmes2016efficient}, and the [2Fe-2S] cluster (20 orbitals $\rightarrow$ 40+ qubits). 

We first perform HCI calculations with different values of the selection cutoff $\epsilon_1$ (see Ref.~\cite{smith2017cheap} for details on the definition of $\epsilon_1$), as shown in Fig.~\ref{fig:SCI_vs_SQD}. We observe that the larger and more restrictive values of $\epsilon_1$ do not allow HCI to reach the DMRG energy reference. Instead, it converges to the first excited state (S$_1$) energy. Decreasing the value of $\epsilon_1$ allows the subspace dimension to become larger and improve the energy beyond that first excited state, at the cost of a significant increase in subspace dimension, which results in a higher computational cost. From the largest HCI wave function (last point for $\epsilon_1 = 5\cdot 10^{-5}$), we repeatedly remove the electronic configurations whose wave function amplitude is below a given threshold and compute the energy in the resulting subspace, providing a collection of energy-subspace dimension pairs that we label as ``upper bound to optimal'' in Fig.~\ref{fig:SCI_vs_SQD}. The removal of tens of millions of configurations (resulting in significantly smaller subspace dimensions) does not significantly deteriorate the quality of the ground-state approximation. These results indicates that HCI does not perform an optimal search of the relevant electronic configurations. Consequently, much smaller subspaces exist that yield comparable energy values. In this particular molecule, HCI needs to explore and perform diagonalizations in subspaces larger than the optimal search. 

We optimize an LUCJ circuit to produce samples in the identified optimal subspaces with high probability. The optimal parameters are found in a two step optimization workflow. First, we optimize the Kullback–Leibler divergence between samples drawn from the LUCJ circuit and the amplitudes of 58 M-dimensional ground-state wave function estimation (green arrow in Fig.~\ref{fig:SCI_vs_SQD}). Then, the circuit parameters are further fine-tuned to minimize the SQD energy using a differential-evolution strategy.

With the samples produced by the optimized LUCJ circuit, we proceed to evaluate the energy of resulting subspaces of varying dimensionality, including the 58 M-dimensional subspace that the circuit was optimized for. We observe good agreement between the LUCJ subspaces and those that are an upper bound to the optimal ones. It is worth noting that despite being optimized for a 58 M-dimensional subspace, the resulting circuit produces subspaces of outstanding quality of dimensionalities both smaller and larger.

The active space considered for the [2Fe-2S] cluster is small enough for HCI to be able to find the relevant bitstrings by exploring and performing diagonalizations in subspaces significantly larger than the optimal ones. This shows that in some problem instances, the diagonalization itself is not the runtime bottleneck. Instead, the runtime bottleneck is a sub-optimal proposal of electronic configurations. Larger and more strongly correlated systems will pose challenges for the classical heuristics. It is for these systems that one can look for a quantum advantage in sampling.

Characterizing the domain of applicability of different classical and quantum heuristics facilitates combining them in quantum-centric supercomputing environments, shifting the search for quantum advantage towards practical problems. 

\section*{Materials and Methods}

\subsection*{Conventions and notation}

\subsubsection*{Hamiltonian}

Our starting point is the Born-Oppenheimer Hamiltonian, written in second quantization using a basis of $\nmo$ orthonormal orbitals $\{ \varphi_p \}_{p=1}^{\nmo}$, as shown in Eq.~\ref{eq:es_ham}. We define the number operator $\hat{n}_{p \sigma} = \crt{p\sigma} \dst{p\sigma}$ which describes the number of electrons with spin $\sigma$ on orbital $p$. For non-relativistic all-electron calculations, the quantities
\begin{equation}
\begin{split}
E_{\mathrm{0}} &= \sum_{a<b}^{N_{nuc}} \frac{Z_a Z_b}{\| \posn{a} - \posn{b} \|} \\
h_{p r} &= \int d \pose{} \, \varphi^*_{p} (\pose{}) \, \left[ - \frac{1}{2} \, \frac{\partial^2}{\partial \vett{r}^2}  - \sum_{a=1}^{N_{nuc}} \frac{Z_a}{ \| \vett{r} - \posn{a} \| } \right] \, \varphi_{r}(\pose{}) \;,
\\
(pr|qs) &= \int d \pose{1} \int d \pose{2} \,  \frac{ \varphi^*_{p} (\pose{1}) \varphi_{r} (\pose{1}) \, \varphi^*_q (\pose{2}) \varphi_{s} (\pose{2}) }{ \| \pose{1} - \pose{2} \| } \;,
\end{split}
\end{equation}
describe the internuclear electrostatic interaction energy and the one-electron and two-electron parts of the Hamiltonian respectively (atomic units are used throughout, i.e. lengths and energies are measured in Bohr and Hartree units $a_B = \hbar^2/(m_e e^2)$ and $E_h = e^2/a_B$ respectively where $-e$ and $m_e$ are the electron charge and mass). The symbols $N_{nuc}$, $\posn{a}$, and $Z_a$ denote the total number of nuclei and their positions and atomic numbers respectively.

For relativistic and/or active-space calculations, the indices $p,r,q,s$ label active-space orbitals, and the quantities $E_0$, $h_{pr}$, and $(pr|qs)$ are modified to account for relativistic effects and/or the potential generated by the inactive-electron density.

In this work, we use orbitals from a restricted closed-shell Hartree-Fock calculation (also called molecular orbitals and denoted MOs) as the basis functions $\varphi_p$. Furthermore, we denote $N_\sigma$ the number of spin-$\sigma$ electrons in the exact ground state. Furthermore $N_\uparrow + N_\downarrow = N$, the toal number of electrons in the exact ground state.

\subsubsection*{Molecular species and active spaces}\label{sec: molecules and active spaces}

\begin{table}[b!]
\centering
\begin{tabular}{ccccc}
\hline
\hline
molecule & basis & $(N,\nmo)$ & $D$ \\
\hline
\nitrogen & 6-31G	 & (10e,16o) & \scientific{1.91}{7} \\
\nitrogen & cc-pVDZ & (10e,26o) & \scientific{4.32}{9} \\
\fetwostwo & TZP-DKH & (30e,20o) & \scientific{2.40}{8} \\
\fefoursfour & TZP-DKH & (54e,36o) & \scientific{8.86}{15} \\
\hline
\hline
\end{tabular}
\caption{Molecules studied in this work. For each molecule, we list the number of electrons and orbitals ($N$ and $\nmo$ respectively) studied on quantum hardware, along with the underlying basis set and the dimension $D = \binom{\nmo}{N/2}^2$ of the Hilbert space of $N_\uparrow=N_\downarrow=N/2$ electrons in $\nmo$ spatial orbitals (not considering molecular point-group symmetries).}
\label{tab:molecules}
\end{table}

The molecules simulated in this work are listed in Table \ref{tab:molecules}. For \nitrogen, we studied all non-core electrons and orbitals on quantum hardware. We computed the potential energy curve using (i) restricted Hartree-Fock (RHF) at 6-31G, and cc-pVDZ level \cite{hehre1969a,hehre1972a,dunning1989a} using \texttt{PySCF}~\cite{sun2018pyscf, sun2020recent} and enforcing $D_{\infty h}$ symmetry and, after projecting the non-relativistic Born-Oppenheimer Hamiltonian in the space spanned by all non-core RHF orbitals with standard functionalities, with
(ii) restricted and symmetry-preserving Moller-Plesset second-order perturbation theory (MP2), coupled cluster with singles and doubles (CCSD), complete active-space configuration interaction (CASCI), for 6-31G and selected configuration interaction, in its heat-bath flavor (HCI), for cc-pVDZ.

For \fetwostwo and \fefoursfour, we employed active spaces \cite{sharma2014low,li2017spin}, spanned by Fe[3d] and S[3p] orbitals, derived from a localized Density Functional Theory calculation with BP86 functional \cite{becke1988density,perdew1986density}, TZP-DKH basis \cite{jorge2009contracted}, and sf-X2C (spin-free exact two-component) Hamiltonian \cite{liu2010ideas,li2012spin} to include scalar relativistic effects. We computed approximations to the ground-state energy with restricted RHF, MP2, CCSD, and HCI.
The RHF, MP2, CCSD, CISD, CASCI and HCI calculations were carried out with the \texttt{PySCF} library~\cite{sun2018pyscf}. 

\subsubsection*{Electron configurations and qubit mapping} In this work, we represent many-electron states using qubit states with the Jordan-Wigner transformation, that maps an electronic configuration, i.e. a Slater determinant of the form
\begin{equation}
| \bts \rangle = \prod_{p\sigma} \left( \crt{p \sigma} \right)^{x_{p\sigma}} | \emptyset \rangle ,
\end{equation}
where $| \emptyset \rangle$ is the vacuum state (i.e., the state with zero electrons) and $x_{p\sigma} \in \{0,1\}$, onto an element of the computational basis, 
\begin{equation}
| \bts \rangle = \bigotimes_{p\sigma} |x_{p\sigma} \rangle ,
\end{equation}
labeled by a bitstring $\bts = (x_{\nmo-1 \downarrow} \dots x_{0 \downarrow} x_{\nmo-1 \uparrow} \dots x_{0 \uparrow} )$. The first half of the bitstring is denoted by  $\bts_\downarrow = (x_{\nmo-1 \downarrow} \dots x_{0 \downarrow} )$ and the second half of the bitstring is denoted by  $\bts_\uparrow = (x_{\nmo-1 \uparrow} \dots x_{0 \uparrow} )$. The JW mapping employs $M = 2\nmo$ qubits and allows computing the number of spin-$\sigma$ electrons for a given configuration $\bts$ as $N_{\bts \sigma} = \sum_{p} x_{p\sigma}$. The total number of electrons, $N_\bts = \sum_\sigma N_{\bts \sigma}$, is the Hamming weight of $\bts$. At the single-configuration level, any $\bts$ in the \textit{right particle sector} must satisfy: $N_{\bts \sigma} = N_\sigma$. An important example is the restricted Hatree-Fock (RHF) bitstring 
\begin{equation}
\label{eq:RHF_bitstring}
| \bts_{\mathrm{RHF}} \rangle = |
\underbrace{0 \dots 0}_{\nmo-N_\downarrow} \underbrace{1 \dots 1}_{N_\downarrow} 
\underbrace{0 \dots 0}_{\nmo-N_\uparrow} \underbrace{1 \dots 1}_{N_\uparrow} \;\; \rangle ,
\end{equation}
which by construction has $N_{\bts_{\mathrm{RHF}} \sigma} = N_\sigma$.

The Fock space is the vector space containing all possible electronic configurations for $\nmo$ orbitals with all possible filling factors for each spin sector. 
The terms \textit{determinant} and \textit{electronic configuration} are used interchangeably in the manuscript.

\subsection*{Sample-based Quantum Diagonalization}\label{Sec:S-CORE details}
This section provides a detailed description of the Sample-based quantum diagonalization (SQD) procedure. We describe the subspace projection and diagonalization, and the self-consistent configuration recovery scheme. We also motivate the energy-variance analysis presented on Fig.~\ref{fig03}.

In what follows, we use $\mathcal{X}$ and $\tilde{\mathcal{X}}$ to denote a set of configurations  sampled from probability distributions $P_\Psi(\bts) = | \langle \bts | \Psi \rangle |^2$ and $\tilde{P}_\Psi(\bts) = \langle \bts | \tilde{\rho} | \bts \rangle$, where $\tilde{\rho}$ is a density operator corresponding to a noisy counterpart of $| \Psi \rangle \langle \Psi |$.
We denote $\mathcal{X}_N \subset \tilde{\mathcal{X}}$ the subset of configurations  with the right particle number. The set of configurations  recovered by the configuration recovery procedure is denoted by $\mathcal{X}_{\rightarrow N}$ and  $\mathcal{X}_\mathrm{R} = \mathcal{X}_N \cup \mathcal{X}_{\rightarrow N}$ the set of configurations  output by self-consistent configuration recovery, and $\mathcal{S}^{(k)} \subset \mathcal{X}_\mathrm{R}$ with $k=1 \dots K$ a set of approximately $d$ configurations sampled from $\mathcal{X}_\mathrm{R}$. The wavefunction $|\psi^{(k)} \rangle$ in obtained by projection and diagonalization of $\hat{H}$ in the subspace spanned by the configurations in $\mathcal{S}^{(k)}$.

\subsubsection*{Eigenstate solver} \label{secS: eigenstate solver}
Given a set of $d$ electronic configurations, the many-electron Hamiltonian is projected and diagonalized in the subspace spanned by the single-particle states  defined by the electronic configurations, as proposed recently in~\cite{kanno2023qQSCI}. We begin  by considering a set of configurations  $\mathcal{X} = \{\bts_{(i)}\}$, all with the right-particle number for each spin sector. Sub-indices between parenthesis label configurations  in a set and not configuration components. We use the generic label $\mathcal{X}$ for a set of configurations  with the right particle number. In practice, the configuration recovery procedure will execute the eigenstate solver on the sets $\mathcal{X}_N$ and $\mathcal{X}_\textrm{R}$.

\paragraph{Conservation of spin.--}

In this study, we perform an approximate restoration of the total spin symmetry labelled by the  $\hat{S}^2$ quantum number. In particular, we are interested on  wavefunctions $| \psi^{(k)} \rangle$ that are as close as possible to singlet states, i.e., eigenfunctions of total spin with eigenvalue 0, $\hat{S}^2 | \psi^{(k)} \rangle = 0$. Conservation of symmetries is notoriously important because molecular eigenstates are joint eigenfunctions of $\hat{H}$ and its symmetries,
including $\hat{S}_z$, $\hat{N}$, molecular point-group symmetries (if any), and $\hat{S}^2$.
Conservation of $\hat{S}_z$ and $\hat{N}$ (and molecular point-group symmetries that are isomorphic to $\mathbb{Z}_2^{\times n}$) can be achieved in a relatively easy way, since the eigenfunctions of these operators are Slater determinants (SD) $| \bts \rangle$, e.g.
\begin{equation}
\hat{N} | \bts \rangle = (N_{\alpha\bts} + N_{\beta\bts} ) | \bts \rangle 
\;,\;
\hat{S}_z | \bts \rangle = (N_{\alpha\bts} - N_{\beta\bts} ) | \bts \rangle 
\;.
\end{equation}
Therefore, in this work, we ensure that the configurations used to span the subspace all have the desired eigenvalue of $\hat{S}_z$ and $\hat{N}$ as part of the configuration recovery procedure in the presence of noise.
Conservation of $\hat{S}^2$ is more difficult to achieve in CI methods since the eigenfunctions of $\hat{S}^2$ are not SDs.
The proper way of ensuring conservation of $\hat{S}^2$ is expanding $| \psi^{(k)} \rangle$ on a set of configuration state functions (CSFs), i.e. spin-symmetry-adapted linear combinations of SDs.
The use of CSFs was common in early CI codes since, in addition to the obvious benefits of reducing the memory footprint of the CI vector, automatic conservation of total spin enhanced the stability of the CI iterations.
Modern CI codes tend to employ SDs as opposed to CSFs, because the formation of the $\sigma$ vector (the most memory- and rate-limiting step of CI algorithms) is considerably more efficient and easily parallelized in SD-based algorithms, and in part because system memory and disk are more plentiful than in previous machines.
However, in SD-based CI codes, the conservation of total spin is no longer guaranteed. In the case of our HPC quantum estimator, sampling from a quantum computer may return sets of SDs that do not allow constructing eigenfunctions of total spin. For example, in a (2e,2o) system, one may sample the configuration $|1001 \rangle$ (having a single spin-down excitation over the RHF state $|0101\rangle$) which is a linear combination of the open-shell singlet and triplet states, respectively $\left( | 1001 \rangle \pm |0110 \rangle \right) / \sqrt{2}$. If the configuration $|0110 \rangle$ is not sampled, one can construct neither eigenfunction of total spin, leading to spin contamination or redundancy (i.e. the configuration $|1001 \rangle$ is involved in a CI calculation, but has coefficient $0$ in the CI vector).
In this work, to facilitate conservation of total spin, we relied on the following procedure: instead of collecting directly $d$ independent and identically distributed (i.i.d.)~samples from $\mathcal{X}$ to make the batch $\mathcal{S}^{(k)}$, we collect $\sqrt{d}/2$ samples and identify all unique configurations  $\bts^{\textrm{u}}$ ($\textrm{u}$ for unique) of length $M/2$ obtained from $\bts_{\uparrow (i)}$ and $\bts_{\downarrow (i)}$ for $1\leq i \leq \sqrt{d}/2$, forming the set $\mathcal{U}^{(k)} = \left\{\bts^{\textrm{u}}\right\}$. The size of the set is upper bounded by $\left|\mathcal{U}^{(k)} \right| \leq \sqrt{d}$. From $\mathcal{U}^{(k)}$ we obtain the batch set $\mathcal{S}^{(k)}$ as:
\begin{equation}
    \mathcal{S}^{(k)} = \left\{\bts \; \bigg| \; \bts = \bts^{\textrm{u}}_{(i)} \oplus \bts^{\textrm{u}}_{(j)} \textrm{ for  all } \bts^{\textrm{u}}_{(i)}, \bts^{\textrm{u}}_{(j)} \in \mathcal{U}^{(k)}\right\}.
\end{equation}
The size of the set above is upper bounded by $\left| \mathcal{S}^{(k)} \right|\leq d$.
This procedure facilitates total spin conservation (for example, from the configuration $|1001 \rangle$ one can build the set $\{ |1001 \rangle, |1010 \rangle, |0101 \rangle, |0110 \rangle \}$, which contains two closed-shell configurations and allows constructing an open-shell singlet state) but does not enforce it. Therefore, in combination with the sampling strategy mentioned above, we achieve the conservation of total spin by a soft constraint in the eigenstate solver, i.e. by adding a penalty term to mitigate spin contamination,
\begin{equation}
    \left(H + \lambda \big[ S^2 - s(s+1) \big]^2 \right) |\psi\rangle = E |\psi\rangle,
\end{equation}
where $\lambda$ can be understood as a Lagrange multiplier that penalizes contributions from $S^2 \neq s(s+1)$. In this work, we employed a soft constraint with $\lambda = 0.2$.

\paragraph{Projection and diagonalization.--}
For each subsampled set $\mathcal{S}^{(k)}$, the Hamiltonian is projected into the corresponding subspace spanned by the configurations in $\mathcal{S}^{(k)}$:
\begin{equation}\label{eqS:projection to subspace}
    \hat{H}_{\mathcal{S}^{(k)}} = \hat{P}_{\mathcal{S}^{(k)}} \hat{H}  \hat{P}_{\mathcal{S}^{(k)}} \textrm{, with } \hat{P}_{\mathcal{S}^{(k)}} = \sum_{\bts \in {\mathcal{S}^{(k)}}} |\bts \rangle \langle \bts|.
\end{equation}
We then diagonalize this projected Hamiltonian (solving $\hat{H}_{\mathcal{S}^{(k)}} |\psi^{(k)}\rangle = E^{(k)} |\psi^{(k)}\rangle$), and its ground state forms an approximation to the ground state of $\hat{H}$.
The approximate ground state $|\psi^{(k)} \rangle$ is defined by its amplitudes in the subspace:
\begin{equation}\label{eqS:SCIWaveFunctionAmplitudes}
    \left|\psi^{(k)} \right\rangle = \sum_{\bts\in \mathcal{S}^{(k)}} c^{(k)}_{\bts} |\bts\rangle.
\end{equation}
The $k^\textrm{th}$ estimate of the ground-state energy is given by:
\begin{equation}\label{eqS:sample energy}
    E^{(k)} = \left\langle \psi^{(k)} \right| \hat{H}_{\mathcal{S}^{(k)}} \left|\psi^{(k)} \right\rangle.
\end{equation}

\subsubsection*{Self-consistent configuration recovery}\label{sec: S-CORE details}

After a quantum state $\hat{\rho}$, corresponding to a noiseless state $|\Psi\rangle$, is prepared in our pre-fault-tolerant quantum processor, we measure it in the computational basis, obtaining the set of measurements
\begin{equation}\label{eqS:sampled_configurations}
    \tilde{\mathcal{X}} = \left\{\bts \; | \; \bts \sim \tilde{P}_\Psi \right\}.
\end{equation}
The class of noiseless states $|\Psi\rangle$ considered in this work are eigenstates to the total particle number operator and the total number operator for each spin species: 
\begin{equation}
\begin{split}
     \sum_{p = 1}^{\nmo}\sum_\sigma \hat{n}_{p\sigma} |\Psi \rangle &= N |\Psi \rangle, \quad
     \sum_{p = 1}^{\nmo} \hat{n}_{p\uparrow} |\Psi \rangle = N_\uparrow |\Psi \rangle, \quad
     \sum_{p = 1}^{\nmo} \hat{n}_{p\downarrow} |\Psi \rangle = N_\downarrow |\Psi \rangle. \\
\end{split}
\end{equation}
From the measurements on the quantum processors, we observe that there are a number of configurations in $\tilde{\mathcal{X}}$ whose $N_{\bts\uparrow}$ and $N_{\bts\downarrow}$ do not match the $N_\uparrow$ and $N_\downarrow$ of the ground state. In Table~\ref{tab:hardware details} we report typical values of the fraction of sampled configurations with the wrong particle number. Since the circuits we use to produce $|\Psi\rangle$ are particle-number preserving, we are certain that configurations with wrong particle numbers have been corrupted by noise. It is this subset of configurations that the configuration recovery scheme is applied to. The configurations in $\tilde{\mathcal{X}}$ whose $N_{\bts\uparrow} = N_{\uparrow}$ and $N_{\bts\downarrow} = N_{\downarrow}$ are not subject to the configuration recovery subroutine.

We then probabilistically flip bits and restore the correct particle number using information obtained from observables of the system. We use the spin-orbital occupancy averaged over all collected batches of subsamples $\vett{n}$, whose components are defined in Eq.~\ref{eq:nR_def}.

 Consider a configuration $\bts$ with $N_{\bts \uparrow }$ spin-up electrons, where $N_{\bts \uparrow }>N_\uparrow$.
From the set of occupied spin-orbitals in the first half of $\bts$, $|N_{\bts\uparrow}-N_\uparrow|$ bits are sampled to be flipped.
If instead $N_{\bts \uparrow }<N_\uparrow$, the bits to be flipped are instead sampled from the unoccupied spin-orbitals. The same procedure applies to the spin-down orbitals. 

The probability of flipping bit $x_{p\sigma}$ depends on the distance between the value of the bit and the reference orbital occupancy $n_{p\sigma}$.
The simplest approach would be to define the distribution proportional to $\left|x_{p\sigma} - n_{p\sigma}\right|$.
However, this introduces an undesirable effect: if for some spin-orbital $p\sigma$, $n_{p\sigma}\approx0.5$, then $\left|x_{p\sigma} - n_{p\sigma}\right|\approx0.5$ as well, i.e., we assign roughly $50\%$ probability weight to flip the bit, regardless of its initial value.
On the other hand, the initial value $x_{p\sigma}$ will in general retain some correlation with the other values in the bitstring, even in the presence of noise.
Hence a better approach is to de-weight the probability of flipping when $\left|x_{p\sigma} - n_{p\sigma}\right|$ is small by using a modified rectified linear unit (ReLU) function $w(\left|x_{p\sigma} - n_{p\sigma}\right|)$, defined as
\begin{equation}\label{Eq:flip relu}
    w(y)=
    \begin{cases}
        \delta\frac{y}{h}\quad\text{if $y\le h$},\\
        \delta+(1-\delta)\frac{y-h}{1-h}\quad\text{if $y>h$}.
    \end{cases}
\end{equation}
The parameter $h\in(0,1)$ defines the location of the ``corner'' of the ReLU function, while the parameter $\delta\in[0,c)$ defines the value of the ReLU function at the corner.
$w$ becomes a true ReLU function when $\delta=0$, and for values of $\delta>0$ the ReLU is modified so that it is not identically zero except at $y=0$
In the specific cases in this work, we chose the values $\delta= 0.01$ and $h=N/M$ (the filling factor) in all experiments.


We do not assume that we know $\vett{n}$ \textit{a priori}, and instead we compute it and improve it self-consistently, following the procedure:
\begin{enumerate}
    \item Setup phase:
    \begin{enumerate}
        \item Find the subset of configurations  of $\tilde{\mathcal{X}}$ that live in the correct particle sector for both spin species, which we denote by $\mathcal{X}_N$: $\mathcal{X}_N \subset \tilde{\mathcal{X}}$.
        \item Obtain batches of samples $\left( \mathcal{S}^{(1)}, \hdots, \mathcal{S}^{(K)} \right)$ from $\mathcal{X}_N$ as described in Sec.~\ref{secS: eigenstate solver}. 
        \item Run the eigenstate solver on the batches and obtain approximate eigenstates $\left| \psi^{(1)} \right\rangle, \hdots , \left| \psi^{(K)}\right\rangle$ (Eqs.\ref{eqS:projection to subspace} and~\ref{eqS:SCIWaveFunctionAmplitudes}).
        \item From the approximate eigenstates construct the first guess for $\vett{n}$, according to Eq.~\eqref{eq:nR_def}.
    \end{enumerate}
    \item Self-consistent iterations (repeat until stopping criterion is met):
    \begin{enumerate}
        \item  $\vett{n}$ is used to correct the configurations with the wrong particle number in $\tilde{\mathcal{X}}$ (we give this subset the label $\mathcal{X}_{/N}$). The resulting set of recovered configurations is labelled $\mathcal{X}_{\rightarrow N}$. 
        \item From $\mathcal{X}_\textrm{R} = \mathcal{X}_N \cup  \mathcal{X}_{\rightarrow N}$, batches of samples $\left( \mathcal{S}^{(1)}, \hdots, \mathcal{S}^{(K)} \right)$ are obtained as described in Sec.~\ref{secS: eigenstate solver}. 
        \item Run the eigenstate solver on the batches and obtain approximate eigenstates $\left| \psi^{(1)} \right\rangle, \hdots , \left| \psi^{(K)}\right\rangle$ (Eqs.\ref{eqS:projection to subspace} and~\ref{eqS:SCIWaveFunctionAmplitudes}).
        \item From the approximate eigenstates construct refined guess for $\vett{n}$, according to Eq.~\eqref{eq:nR_def}.
        \item If the stopping criterion is not met, go back to step 2.(a).
    \end{enumerate}
\end{enumerate}

We direct the reader to Fig. 2 in the main text for a numerical emulation of the effect of the configuration recovery on the accuracy of SQD in the presence of noise.

\subsubsection*{Energy-variance extrapolation}\label{Sec:energy-variance theory}
It is expected that the accuracy of SQD (with or without configuration recovery) will increase as the number of configurations used for the subspace expansion $d$ is increased. However, the convergence of the ground-state properties with $d$ is not expected to follow any specific functional relation. Therefore, attempting to analyze the convergence as a function of $d$ is not well motivated. Instead, the different eigenstate approximations obtained for different values of $d$ and different batches of samples $\mathcal{S}^{(k)}$ are used for an energy-variance extrapolation~\cite{Kashima2001Energy_Variance, Kwon1998Energy_Variance, Imada2000Energy_Variance, Nomura2017Energy_Variance, Sorella2001Energy_Variance, Mizusaki2002Energy_Variance}. Consider the approximate eigenstate $\left|\psi \right\rangle$, whose energy is given by $E = \langle \psi | \hat{H} | \psi \rangle$, and consider the exact eigenstate energy $E_\textrm{T}$. The difference between the approximate energy and $E_\textrm{T}$,
\begin{equation}
    \delta E = \langle \hat{H} \rangle - E_\textrm{T},
\end{equation}
vanishes linearly with the Hamiltonian variance divided by the square of the variational energy~\cite{Kashima2001Energy_Variance}
\begin{equation}
    \frac{\Delta H}{E^2} = \frac{\langle \psi | \hat{H}^2 | \psi \rangle - \langle \psi | \hat{H} | \psi \rangle ^2}{E^2}.
\end{equation}
This linear relation is satisfied as long as $|\psi \rangle$ is sufficiently close to an eigenstate of the Hamiltonian, as measured by the state fidelity. The least squares fit of a collection of energy-variance points yields an estimate of $E_\textrm{T}$, as the intersect of the fit with the ordinates. This point is the extrapolation of the estimate of the energy to the limit where $|\psi\rangle$ coincides with the exact eigenstate. Besides the energy extrapolation, the energy-variance analysis may also reveal the existence of multiple eigenstates close in energy to the ground state. 

In the main text and in this supplementary material we apply the energy-variance analysis to two different sets of energy-variance pairs. The first one is the energy-variance pairs obtained by the HCI procedure where different energies and variances are obtained by changing the cutoff parameter that indirectly controls the number of determinants in the subspace projection and diagonalization, resulting in different levels of accuracy. The second set of energy-variance pairs are those obtained from SQD for different numbers of configurations $d$ as well as for different batches of sampled configurations $\mathcal{S}^{(k)}$.

\subsection*{Experimental details}
In this section we describe the quantum circuits that are employed to produce the configurations  to which we apply the eigenstate solver. Additionally, we provide details on the setting of circuit parameters from an efficient classical CCSD calculation, and the mapping of the circuits to quantum processors with heavy-hex connectivity.

\subsubsection*{Quantum circuits}\label{secS:quantum circuits}

In this work, we used the local-unitary cluster Jastrow (LUCJ) ansatz \cite{motta2023bridging} to sample randomly distributed electronic configurations. LUCJ derives from the unitary cluster Jastrow (UCJ) ansatz, which has the form \cite{matsuzawa2020jastrow} of a product of $L$ layers, as defined in Eq.~\ref{equation:uCJ}. We recall that:
\begin{equation}
\hat{K}_\mu = \sum_{pq, \sigma} K_{pq}^\mu \, \crt{p \sigma} \dst{q \sigma}
\;,\;
\hat{J}_\mu = \sum_{pr, \sigma\tau} J_{p\sigma,q\tau}^\mu \, \numberop{p \sigma} \numberop{q \tau}
\;.
\label{equation:uCJ_KJ}
\end{equation}
In Eq. \eqref{equation:uCJ_KJ}, $p,q=0 \dots \nmo-1$ label molecular spatial orbitals and $\sigma,\tau$ label spin polarizations ($\alpha,\beta$ for spin-up and spin-down electrons respectively).  
$K_{pq}^\mu$/$J_{pq,\sigma\tau}^\mu$ has complex/real matrix elements and is anti-Hermitian/symmetric.
The UCJ ansatz can be derived from a twice-factorized low-rank decomposition of the qUCCD ansatz \cite{motta2021low,matsuzawa2020jastrow}, and the $L$-product form is such that the exact full configuration interaction wavefunction can be obtained via Eq.~\eqref{equation:uCJ} \cite{matsuzawa2020jastrow,evangelista2019exact}. 

The local UCJ or LUCJ \cite{motta2023bridging} introduces a ``local'' approximation of the UCJ ansatz, which makes the following modifications for opposite-spin and same-spin number-number terms: 
\begin{equation}
\begin{split}
\sum_{pq} J_{p\alpha,q\beta}\hat{n}_{p\alpha}\hat{n}_{q\beta} &\rightarrow \sum_{p \in S} J_{p\alpha,p\beta}\hat{n}_{p\alpha}\hat{n}_{p\beta}
\\
\sum_{pq} J_{p\sigma,q\sigma} \hat{n}_{p\sigma}\hat{n}_{q\sigma} &\rightarrow \sum_{pq \in S^\prime} J_{p\sigma,q\sigma} \hat{n}_{p\sigma}\hat{n}_{q\sigma},
\end{split}
\end{equation}
where $\sigma=\alpha, \beta$ and the sets $S,S^\prime$ are such that the quantum circuit implementing $e^{i \hat{J}_\mu}$ has depth $O(1)$  and only comprises $O(|S|+|S^\prime|)$ number-number ``nn gates,'' i.e. two-qubit unitaries of the form $U_{nn}(\varphi) = e^{- i \frac{\varphi}{4} (Z_p + Z_q - Z_p Z_q) }$, acting on adjacent qubits $p,q$ in the topology of a certain processor. For example, on a heavy-hex processor, $S = \{ 4k \;,\; k=0\dots (\nmo-1)/4 \}$ and $S^\prime = \{ (p,p+1) \;,\; p=0\dots N-2 \}$ \cite{motta2023bridging}.
The circuits $e^{\pm \hat{K}_\mu}$, on the other hand, can be implemented by a Bogolyubov circuit acting on $\nmo$ qubits and comprising $O(\nalpha (\nvir))$ gates and depth $O(\nmo)$ \cite{reck1994experimental,clements2016optimal,jiang2018quantum}.
Through the local approximation, the LUCJ Ansatz balances hardware friendliness and accuracy, the latter ultimately deriving from its connection to coupled cluster theory and adiabatic state preparation \cite{motta2023bridging}.

Unless otherwise specified,  we use the truncated LUCJ circuit $|\Psi\rangle = e^{\hat{K}_2}e^{-\hat{K}_1} e^{i \hat{J}_1} e^{\hat{K}_1} | \bts_\rhf \rangle$, which is the result of considering the two-layer LUCJ circuit and removing the last orbital rotation and last Jastrow operations. The resulting state is implemented by a circuit whose depth is identical to the single-layer LUCJ circuit. The addition of the $e^{\hat{K}_2}$ operation to the circuit can have a large impact on the configurations generated by the circuit when the parameters are set from the $t_2$ tensor from a classical restricted closed-shell CCSD calculation (as described in the next section). For dynamically correlated species, the $J_{p\sigma,q\tau}^\mu$ parameters obtained from $t_2$ can have a small amplitude, resulting in the approximate cancellation of the $\exp(-\hat{K}_1)$ and $\exp(\hat{K}_1)$ terms in the ansatz. Without $\exp(\hat{K}_2)$, the resulting wavefunction can be over-concentrated around the Hartree-Fock configuration.  Therefore, for dynamically correlated species the action of $\exp(\hat{K}_2)$ is to remove some of the excessive concentration of the $|\Psi\rangle$ wavefunction, when parameters are set from a restricted closed-shell CCSD calculation.

\subsubsection*{Initialization of LUCJ parameters}

In our experiments, we parametrize the LUCJ circuits using the following procedure:
\begin{enumerate}
\item First, we carry out a classical restricted closed-shell CCSD calculation, yielding amplitudes $t_{1,ai}$ and $t_{2,aibj}$, where $ij$/$ab$ labels occupied/unoccupied orbitals in the RHF state.
\item We reshape the $t_2$ tensor into the matrix $(t_2)_{ai,bj}$ and diagonalize it, $(t_2)_{ai,bj} = \sum_y \tau_y U_{ai,y} U_{bj,y}$, where the eigenvectors $\tau_y$ are sorted in decreasing order of absolute value. 
\item We extend the unitaries to the following matrices,
\begin{equation}
\tilde{U}_{y,pr} = \delta_{pa} \delta_{ri} U_{ai,y}
\;,
\end{equation}
i.e. matrices where only the occupied/unoccupied block is non-zero.
\item We define the Hermitian operators 
\begin{equation}
 X_{\pm,y} = \frac{1 \mp i}{2} \left( \tilde{U}_y \pm i \tilde{U}_y^T \right)
\end{equation}
and their eigenpairs
\begin{equation}
X_{\pm,y} V_{\pm,y} = g_{\pm,y} V_{\pm,y}~.
\end{equation}
\item We define the operators
\begin{equation}
\begin{split}
\left( J_{2y} \right)^{\sigma\tau}_{pr} &= \tau_y \left( g_{+,y} \right)_p \left( g_{+,y} \right)_r \\
\left( K_{2y} \right)^\sigma_{pr} &= \log(V_{+,y})_{pr} \\
\left( J_{2y+1} \right)^{\sigma\tau}_{pr} &= - \tau_y \left( g_{-,y} \right)_p \left( g_{-,y} \right)_r \\
\left( K_{2y+1} \right)^\sigma_{pr} &= \log(V_{-,y})_{pr}~. \\
\end{split}
\end{equation}
\item We retain the first $L$ matrices $J$, $K$. This allows us to refine a non-local UCJ wavefunction \cite{motta2021low,matsuzawa2020jastrow}.
\item We zero out the entries of the $J$ matrices leading to quantum gates acting on non-adjacent qubits. For a heavy-hex lattice, we retain the following elements:
\begin{equation}\label{eq: LUCJ heavy-hex connectivity}
\begin{split}
&\left( J_{\mu} \right)^{\alpha\alpha}_{p,p+1} \;,\; p=0 \dots \nmo-2 \\
&\left( J_{\mu} \right)^{\beta\beta}_{p,p+1} \;,\; p=0 \dots \nmo-2 \\
&\left( J_{\mu} \right)^{\alpha\beta}_{p,p} \;,\; p = 0 \dots \nmo-1 \;,\; p\%4 = 0~. \\
\end{split}
\end{equation}
\end{enumerate}
The final sparsification allows us to construct a LUCJ wavefunction. In a conventional LUCJ calculation, these parameters are the starting point of a variational optimization. For the hardware experiments reported in this study, we employed these parameters, without further optimization, to define a LUCJ circuit, that we used to sample randomly distributed configurations.

\subsubsection*{Mapping to heavy-hex processors}

The choice of retaining the elements $( J_{\mu} )^{\alpha\beta}_{p,p}$ with $p = 0 \dots \nmo-1 \;,\; p\%4 = 0$ when implementing LUCJ on a heavy-hex processor has an important technical motivation: on such devices, assuming a number of qubits greatly exceeding $\nmo$, spin-up and spin-down orbitals can be mapped on two segments of adjacent qubits forming a ``zig-zag'' pattern and connected through an auxiliary qubit for $p=0,4,8,\dots$ as shown in the three rightmost panels of Fig.~\ref{fig03}. Such a qubit layout allows implementing LUCJ with a minimal overhead of $\mathsf{SWAP}$ gates (two per auxiliary qubit and layer of LUCJ). 
However, on current processors with up to 133 qubits, for $\nmo > 21$ one cannot couple $\nmo/4$ spin-orbitals with opposite spins through auxiliary qubits without incurring a significant overhead of $\mathsf{SWAP}$ gates, for the simple reason that a chain of 22 or more qubits is longer than the ``diagonal'' of the processor.
In such a situation, as shown in the rightmost panel of Fig.~\ref{fig03}, the segments on which spin-up and spin-down qubits are mapped form two ``tails'' that are not connected by auxiliary qubits. This fact has two implications: (i) no more than 6 spin-orbitals with opposite spins can be coupled through auxiliary qubits, and (ii) if one retains the elements $( J_{\mu} )^{\alpha\beta}_{p,p}$ with $p = 0 \dots \nmo-1 \;,\; p\%4 = 0$ and $p \leq 16$, the largest elements of $J_\mu$ may be discarded, yielding a lower-accuracy wavefunction.
The first problem is a fundamental one, that can only be resolved with a substantially different mapping of fermionic degrees of freedom onto qubits and/or through the availability of larger processors.
The second problem, on the other hand, has a simple solution, that we now describe.
First, for any pair of spatial orbitals $p,r = 0 \dots \nmo-1$, the orbital rotation
\begin{equation}
\hat{S}_{pr} = e^{ - i \frac{\pi}{2} \sum_\sigma \left( \crt{p\sigma} \dst{r\sigma} + \crt{r\sigma} \dst{p\sigma} \right)}
\end{equation}
implements the permutation $S_{pr} \in \mathrm{S}_{\nmo}$ exchanging orbitals $p$ and $r$, in the sense that
\begin{equation}
\hat{S}^\dagger_{pr} \left( \sum_{qs,\tau} M_{qs} \crt{q\tau} \dst{s\tau} \right) \hat{S}_{pr} 
= \sum_{qs,\tau} M^\prime_{qs} \crt{q\tau} \dst{s\tau}
\;,\;
M^\prime_{qs} = M_{S_{pr}(q),S_{pr}(s)}
\;.
\end{equation}
Three immediate implications of this fact are: 
\begin{itemize}
\item $\hat{S}^\dagger_{pr} \numberop{q\tau} \hat{S}_{pr} = \numberop{S_{pr}(q)\tau}$~.
\item for any density-density operator $\hat{J}_\mu = \sum_{qs,\sigma\tau} ( J_{\mu} )^{\sigma\tau}_{q,s} \numberop{q\sigma} \numberop{s\tau}$, one has
\begin{equation}
\hat{S}^\dagger_{pr} \hat{J}_\mu \hat{S}_{pr} 
= \sum_{qs,\tau} ( J^\prime_{\mu} )^{\sigma\tau}_{q,s} \numberop{q\sigma} \numberop{s\tau} = \hat{J}^\prime_\mu
\;,\;
( J^\prime_{\mu} )^{\sigma\tau}_{q,s} = ( {J}_{\mu} )^{\sigma\tau}_{S_{pr}(q),S_{pr}(s)}
\;.
\end{equation}
\item that for any permutation $S \in \mathrm{S}_{\nmo}$ there exists an orbital rotation $e^{\hat{K}_S}$ implementing the permutation $S$ (this is true because permutations can be written as products of exchange permutations, an exchange permutation can be implemented by an orbital rotation, and orbital rotations are closed under multiplication).
\end{itemize}
Consider now the tensor $( J_{\mu} )^{\sigma\tau}_{q,s}$ resulting from the low-rank decomposition of the CCSD operator described in the previous Subsection. Let $p_0 \dots p_\ell$ be the $\ell$ elements of $( J_{\mu} )^{\alpha\beta}_{q,s}$ with the largest absolute values, and let $S \in \mathrm{S}_{\nmo}$ be the permutation such that $S(p_0) = 0 \dots S(p_\ell) = 4\ell$. Then,
\begin{equation}
e^{-\hat{K}_S} \hat{J}_\mu e^{\hat{K}_S} 
= \hat{J}^\prime_\mu
\end{equation}
where the elements of $( J^\prime_{\mu} )^{\alpha\beta}_{q,s}$ with the largest absolute values are at positions $0, \dots, 4\ell$. Before sparsifying the tensor $\tilde{J}_\mu$, one can use the identity
\begin{equation}
e^{\hat{K}_\mu} e^{i \hat{J}_\mu} e^{-\hat{K}_\mu} =
e^{\hat{K}_\mu} e^{\hat{K}_S} e^{i \hat{J}^\prime_\mu} e^{-\hat{K}_S} e^{-\hat{K}_\mu} =
e^{\hat{K}^\prime_\mu} e^{i \hat{J}^\prime_\mu} e^{-\hat{K}^\prime_\mu} 
\end{equation}
to obtain a UCJ operator with transformed orbital rotations $e^{-\hat{K}^\prime_\mu}$ and an opposite-spin density-density interaction whose largest elements in absolute value act on spatial orbitals $p=0,4,8,\dots,4\ell$. Sparsification of $( J^\prime_{\mu} )^{\alpha\beta}_{q,s}$ then leads to retaining the $l$ dominant opposite-spin density-density interaction terms (as many as allowed by the size and topology of the available heavy-hex processor) with a minimal overhead of $\mathrm{SWAP}$ gates.

In this work, we retained the elements $( J_{\mu} )^{\alpha\beta}_{p,p}$ with $p = 0 \dots \nmo-1 \;,\; p\%4 = 0$ and $p \leq 16$. In future work, the procedure described here could be used to modify the LUCJ wavefunction and the resulting probability distribution for electronic configurations.


\clearpage 

%
\bibliography{main} 
\bibliographystyle{sciencemag}

%
%
%
%
%
%


\section*{Acknowledgments}
We thank the IBM Quantum Service and Data team for help with the workflow execution. 
The authors acknowledge feedback and insightful conversations with 
Alberto Baiardi,
Sergey Bravyi,
Giuseppe Carleo,
Garnet Kin-Lic Chan,
Antonio C\' orcoles,
Oliver Dial,
Jay Gambetta,
Artur Izmaylov,
Caleb Johnson,
Yukio Kawashima,
David Kremer,
Isaac Lauer,
Peter Love,
Guglielmo Mazzola,
Takahito Nakajima,
Paul Nation,
Hanhee Paik,
Emily Pritchett,
Max Rossmannek,
Paul Schweigert,
James E.T. Smith, 
Miles Stoudenmire,
Nobuyuki Yoshioka, 
Andrew Wack, and
Christa Zoufal.
A part of this work is supported by project JPNP20017, funded by the New Energy and Industrial Technology Development Organization (NEDO), Japan, 
and by the RIKEN TRIP initiative (RIKEN Quantum).
\paragraph*{Author contributions:}
Design of the workflow and experiments: J. R.-M., M. M., W. K., K. S., M. C. T., A. M. Implementation of the workflow: J. R.-M., M. M., A. J.-A., S. M., S. S., T. S., I. S., R.-Y. S., K. J. S. Tuning and calibration of the Heron processor: H.H. Execution of the quantum part of the workflow on Heron: M. M., P. J., M. T.  Execution of the classical part of the workflow on Fugaku:
T. S., R.-Y. S., S. Y. Numerical benchmarks: J. R.-M., M.M., T. S., K. J. S.
Analytical derivations: J. R.-M., M.M., W. K., K. S., M. C. T., A. M. 
All authors contributed to the manuscript writing and data analysis. 
\paragraph*{Data and materials availability:}
All data supporting the findings is available in the main text and Supplementary materials. The code to reproduce all results reported is available as the open-source package qiskit-addon-sqd~\cite{sqd_addon}.



\subsection*{Supplementary materials}
Supplementary Text\\
Figs. S1 to S22\\
Tables S1 to S2\\
References \textit{(7-\arabic{enumiv})}\\ 


\newpage


\renewcommand{\thefigure}{S\arabic{figure}}
\renewcommand{\thetable}{S\arabic{table}}
\renewcommand{\theequation}{S\arabic{equation}}
\renewcommand{\thepage}{S\arabic{page}}
\setcounter{figure}{0}
\setcounter{table}{0}
\setcounter{section}{0}
\setcounter{equation}{0}
\setcounter{page}{1} 


\begin{center}
\section*{Supplementary Materials for\\ \scititle}

\author{
	Javier Robledo-Moreno$^{1\dagger}$,
	Mario Motta$^{1\ast}$,
	Holger Haas$^{1}$,
	Ali Javadi-Abhari$^{1}$,\and
	Petar Jurcevic$^{1}$,
	William Kirby$^{2}$,
	Simon Martiel$^{3}$,
	Kunal Sharma$^{1}$,
	Sandeep Sharma$^{4}$,\and
	Tomonori Shirakawa$^{5, 6, 7}$,
	Iskandar Sitdikov$^{1}$,
	Rong-Yang Sun$^{5, 6, 7}$,
	Kevin~J.~Sung$^{1}$,\and
	Maika Takita$^{1}$,
	Minh C. Tran$^{2}$,
	Seiji Yunoki$^{5, 6, 7, 8}$,
	Antonio Mezzacapo$^{1\perp}$.\and
	\small$^{1}$IBM Quantum, IBM T.J. Watson Research Center, Yorktown Heights, NY 10598, USA.\and
	\small$^{2}$IBM Quantum, IBM Research Cambridge, Cambridge, MA 02142, USA.\and
	\small$^{3}$IBM Quantum, IBM France Lab, Orsay, France.\and
	\small$^{4}$Department of Chemistry, University of Colorado, Boulder, CO 80302, USA.\and
	\small$^{5}$Computational Materials Science Research Team, RIKEN Center for Computational Science (R-CCS), \and \small Kobe, Hyogo, 650-0047, Japan.\and
	\small$^{6}$Quantum Computational Science Research Team, RIKEN Center for Quantum Computing (RQC),\and \small  Wako, Saitama, 351-0198, Japan.\and
	\small$^{7}$RIKEN Interdisciplinary Theoretical and Mathematical Sciences Program (iTHEMS), \and \small  Wako, Saitama 351-0198, Japan. \and
	\small$^{8}$RIKEN Center for Emergent Matter Science (CEMS), Wako, Saitama 351-0198, Japan.\and
	\small$^\dagger$Corresponding author. Email: j.robledomoreno@ibm.com\and
	\small$^\ast$Corresponding author. Email: mario.motta@ibm.com\and
	\small$^\perp$Corresponding author. Email: mezzacapo@ibm.com\and
}
\end{center}

\subsubsection*{This PDF file includes:}
Supplementary Text\\
Figures S1 to S22\\
Tables S1 to S2\\

\newpage

\tableofcontents

\section{Convergence properties of the components of sample-based quantum diagonalization}

\subsection{Wavefunction concentration and accuracy}\label{Sec: wavefunction concentraction}
\subsubsection{Sample complexity}

Under the assumption that the ground state is sufficiently concentrated, we show here that the energy obtained from the estimator converges to the ground-state energy exponentially fast in the number of samples.
Suppose we sample from the ground state $\ket{G} = \sum_{I} c_{I(\bts)} \ket{{I(\bts)}}$, where $I = 0,\dots,2^M-1$ label $2^M$ computational configurations  and $c_I$ are the corresponding coefficients.
In other words, the symbol $I(\bts)$ denotes the integer with binary representation $\bts$ for a given binary string $\bts$. To abbreviate the notation we drop the explicit dependence of $\bts$ on $I$.
Let $P_I = \abs{c_I}^2$ be the probability of each bitstring and, without loss of generality, assume the configurations  are ordered by $P_I$: $P_0 \geq P_1 \geq P_2 \geq \dots$.
To quantify the concentration of the configurations , we define parameters $\alpha_m$ and $\beta_m$ for a given $m$ such that
\begin{align}
    \sum_{I = 0}^{m-1} P_I \geq \alpha_m, \quad \text{and} \quad
    &P_1 \geq \dots \geq P_m \geq \beta_m.
\end{align}
The first condition imposes that the total probability of the first $m$ configurations  must be at least $\alpha_m$ while the second lower bounds the individual $P_I$ by $\beta_m$.
We first prove general statements using $\alpha_m$ and $\beta_m$ and, after that, apply these statements to different distributions of $P_x$.

If we draw $N_s$ samples from $\ket{G}$, we can show that the probability of \emph{not} obtaining all of the first $m$ configurations  among the $N_s$ samples decays exponentially with $N_s$.
The probability of not seeing a given bitstring $I$ in any of the $N_s$ samples is $(1 - P_I)^{N_s}$.
Therefore, the probability $p_{\text{fail}}$ of having at least one of the first $m$ configurations  not appearing in the $N_s$  samples is upper bounded by
\begin{align} 
	\sum_{I = 0}^{m-1}  (1 - P_I)^{N_s}
	\leq \sum_{I = 0}^{m-1} \left(1-\beta_m\right)^{N_s}
	\leq m e^{-N_s \beta_m}.
\end{align}
Choosing $N_s$ large enough so that $m \leq e^{N_s\beta_m /2}$, the bound further simplifies to $p_{\text{fail}} \leq e^{-N_s\beta_m/2}$.
Therefore, if we would like the probability of failure to be at most $\eta$, it is sufficient to choose
\begin{align} 
	N_s \geq \frac{2}{\beta_m} \log{\frac{1}{\eta}}.
\end{align}

Next, to understand how $m$ should scale with the system size, we look at the error in the energy of the approximate ground state constructed from diagonalizing the Hamiltonian in the sampled basis.
First, let $\mathcal S = \{\ket{I}:I = 0,\dots,m-1\}$ be the truncated basis consisting of exactly the first $m$ configurations .
Let 
\begin{align} 
	\ket{G_m} = \frac{1}{\sqrt{\mathcal N}} \sum_{I = 0}^{m-1} c_I \ket{x} 
\end{align}
be the normalized state constructed by truncating $\ket{G}$ to the subspace $\mathcal S$. Here, $\mathcal N = \sum_{I = 1}^m P_I $ is a normalization constant.
We have
\begin{align} 
	&\norm{\ket{G} - \ket{G_m}}^2 
	= 2 - \innerprod{G}{G_m}-\innerprod{G_m}{G}
	= 2 - 2\sqrt{\sum_{I = 0}^{m-1} P_I} 
	\leq 2 - 2\sqrt{\alpha_m}.
\end{align}
Intuitively, if $\alpha_m \approx 1$, $\ket{G}$ is close to $\ket{G_m}$. In particular, it implies an upper bound on the energy of $\ket{G_m}$:
\begin{align} 
	\bra{G_m} \hat{H} \ket{G_m} &\leq \bra{G} \hat{H} \ket{G} + 2\|\hat{H}\| \norm{\ket{G} - \ket{G_m}}\nonumber\\
	&\leq \bra{G} \hat{H} \ket{G} + 2\sqrt{2}\|\hat{H}\|(1-\sqrt{\alpha_m})^{1/2}.  \label{eq:bound1}
\end{align}

Now, let $\tilde H$ be the $m\times m$ matrix that represents the Hamiltonian in the truncated subspace $\mathcal S$, i.e. the top left $m\times m$ block of $H$, and let $\ket{\tilde G}$ be the ground state of $\tilde H$. 
We have
\begin{align} 
	\bra{\tilde G} \hat{H} \ket{\tilde G}
	= \bra{\tilde G} \tilde H \ket{\tilde G}
	&\leq \bra{G_m} \tilde H \ket{G_m} 
	= \bra{G_m} \hat{H} \ket{G_m},  	 \label{eq:bound2}
\end{align}  
where the inequality follows from $\ket{\tilde G}$ being the ground state of $\tilde H$.
Combining Eqs.~(\ref{eq:bound1}) and (\ref{eq:bound2}), we have a bound on the difference between the ground-state energy of $\tilde H$ and the original Hamiltonian $H$:
\begin{align} 
	\bra{\tilde G} \tilde H \ket{\tilde G}
	- \bra{G} \hat{H} \ket{G} \leq 2\sqrt{2}\|\hat{H}\|(1-\sqrt{\alpha_m})^{1/2}.
\end{align}

Earlier, we assumed that $\mathcal S$ consists of exactly the first $m$ bit strings.
Note that $\bra{\tilde G} \tilde H \ket{\tilde G}$ cannot increase if we add more configurations  to $\mathcal S$.
Therefore, combining with the earlier statement about the probability of seeing all $m$ bit strings in $N$ samples, we have that
for any $\eta \in (0,1)$, choosing
\begin{align} 
	 N_s \geq \max\left\{\log\frac{2m}{\beta_m},\frac{2}{\beta_m} \log{\frac{1}{\eta}}\right\}
\end{align}
guarantees
\begin{align} 
	\bra{\tilde G} \tilde H \ket{\tilde G}
	- \bra{G} \hat{H} \ket{G} \leq 2\sqrt{2}\|\hat{H}\|(1-\sqrt{\alpha_m})^{1/2}
\end{align}
with probability at least $1-\eta$.
This result allows us to estimate the sufficient number of samples $N_s$ given a distribution of the configurations  in the ground state.
For example, if $P_I \propto e^{-I}$ decays exponentially, we have $1- \sqrt{\alpha_m} \propto e^{-m}$ and $\beta_m \propto e^{-m}$.
At $m \approx 2 c \log M$, where $c$ is a parameter to be determined,  choosing $N_s = \Omega(M^{2c})$ is sufficient to guarantee an energy error at most $\mathcal{O}\left({\|\hat{H}\|/M^c}\right)$. We have the freedom to choose $c$ large enough so that $\|\hat{H}\|/M^c \ll 1$. In particular, if $\|\hat{H}\| = \mathcal{O}\left({M^4}\right)$, we would choose $c > 4$.
In short, the sufficient number of samples $N_s$ scales only polynomially with the number of qubits $M$.

This favorable scaling persists even if $P_I \propto 1/I^\gamma$ decays algebraically for some constant $\gamma > 1$.
In this case, we have $1- \sqrt{\alpha_m} \propto 1/m^{\gamma-1}$ and $\beta_m \propto 1/m^\gamma$.
Therefore, choosing $N_s = \Omega(m^{\gamma})$, we can guarantee the error is at most $O\big({\|\hat{H}\|/m^{(\gamma-1)/2}}\big)$. Again, we have the freedom to choose $m \propto M^c$ for $c$ large enough so that $\|\hat{H}\|/M^{c(\gamma-1)/2}\ll 1$.
With this choice of $m$, the number of samples only needs to scale polynomially with $M$ as $O\left(M^{c\gamma}\right)$.

Recall that in our procedure, further subsample from the set of samples $\mathcal S$.
If the size of the subsamples is large enough, one can also prove that at least one of them contains all the $m$ configurations .
For simplicity, we assume that we will draw $K$ subsamples uniformly from $\mathcal S$.
This subsampling probability is the worst case for the success probability and, in practice, we would instead draw these subsamples based on their frequencies in the $N_s$ samples.
Assume that each subsample has $d$ configurations .
Additionally, we are interested in the limit $m \ll d \ll N_s$.
We will calculate the probability that none of these $K$ subsamples has all of the $m$ correct configurations 
and show that this probability decays exponentially with $K$.

First, consider a subsample of $d$ samples.
There are $\binom{N_s}{d}$ ways to choose this subsample.
Among them, $\binom{N_s-m}{d-m}$ contain the $m$ configurations .
So the probability of \emph{not} having all $m$ configurations  is 
\begin{align} 
	 	p_{\mathrm{fail}}^{(1)} \equiv 
	 	  1 - \frac{d!(N_s-m)!}{N_s!(d-m)!}
	 	\leq 1 - \left(\frac{d-m}{N_s}\right)^m.
\end{align} 
So the probability that none of the $K$ subsamples has all of the $m$ configurations  is bounded by
\begin{align} 
	p_{\mathrm{fail}}^{(K)} \leq \left[1 - \left(\frac{d-m}{N_s}\right)^m\right]^K,
\end{align}
which indeed decays exponentially with $K$.

In the first example above where $P_I \propto e^{-I}$, if we choose $d \propto M^c$ such that $N_s/d = \kappa$ is a constant, we have
\begin{align} 
     p_{\mathrm{fail}}^{(K)} \lesssim \left(1 - \frac{1}{\kappa^{c \log M}}\right)^K 
     = \left(1 - \frac{1}{M^{c\log \kappa}}\right)^K.
\end{align}
Therefore, a polynomially large number of subsamples $K \propto M^{c\log \kappa}$ would ensure that the failure probability is at most a constant.
Although the scaling of $d$ with $M$ is the same as that of $N_s$, choosing a large $\kappa$ significantly reduces the dimension of the matrices that we need to diagonalize. 
Instead of diagonalizing a large $N_s \times N_s$ matrix, we diagonalize $K$ different $d\times d$ matrices, which can be performed in parallel.

\subsubsection{Assessing wavefunction concentration}
\textit{A-priori} it is not possible to know if the ground-state wavefunction of a given system is concentrated. In this subsection, we present a number of tests that can be carried out to study the concentration of the ground state wavefunction, having access only to the eigenstates $|\psi^{(k)} \rangle$ produced by the estimator with or without configuration recovery. 

Assume that the distribution over the space of electronic configurations generated by the wave function $|\Psi \rangle$ coincides with that of the ground state. The study of the convergence of the approximate ground state properties obtained from  $|\psi^{(k)} \rangle$, as a function of $d$, can reveal if the ground state wavefunction is concentrated. When the ground state is not concentrated, the convergence of observables like the energy with $d$ can be slow, or not converge at all. In the particular case of the energy, as the number of configurations $d$ is increased, the value for the approximate ground-state energy decreases. If the approximate ground state energy has not converged when reaching the largest $d$ amenable by the available computational resources, it is clear that the size of the support of the ground-state wavefunction exceeds the value of $d$.

Furthermore, when the ground state wave function is not concentrated, different batches of sampled configurations $\mathcal{S}^{(k)}$ will likely share only a few common configurations, resulting in a large variance between the properties extracted from the different $|\psi^{(k)} \rangle$. 

The expectation values of observables being close to their extremal values can also be a good indicator  of the concentration. 
For example, the closer the occupations are to a corner of the hypercube $[0,1]^M$, the more concentrated the wavefunction is on the bitstring that corresponds to the corner. 

We remark that while these tests can indicate the presence of a non-concentrated ground state wavefunction, they do not guarantee a definitive answer to whether the ground-ground state wavefunction was indeed concentrated.  

\subsubsection{A numerical study of the accuracy: Hubbard model}

To illustrate the dependence of the accuracy of the Selected Configuration Interaction (SCI)-based eigensolver as a function of the ground-state wavefunction concentration we consider the $\nmo$-site Hubbard model in a fully-connected lattice with random hopping amplitudes:
\begin{equation}
    \hat{H} = -\frac{1}{\sqrt{L}}\sum_{\substack{p, q = 1 \\ p \neq q}}^{\nmo}
    \sum_{\sigma}\,
    t_{pq}\,\hat{a}^{\dagger}_{p\sigma}\hat{a}_{q\sigma} + U \sum_{p = 1}^{\nmo} \hat{n}_{p \uparrow} \hat{n}_{p\downarrow}.
    \label{Eq:Hubbard Hamiltonian}
\end{equation}
 The hopping amplitudes $t_{pq} = t_{qp}$ are independent random variables drawn  from the Gaussian distribution 
with mean $0$ and standard deviation $1$, i.e. 
$\overline{t_{pq}}=0$ and $\overline{t^2_{pq}}=t^2=1$. The results presented in this section are obtained from the average of results obtained from twenty disorder realizations. In the infinite-volume, this model is known to possess {\it self-averaging} properties~\cite{SYK_review}. However, on finite-size systems, there are sample-to-sample fluctuations. The numerical study in this section does not consider the effect of noise, so we run the estimator without configuration recovery. For the discussion that follows, we span the wavefunction amplitudes in terms of electronic configurations in the localized basis, i.e. the basis used to define the Hamiltonian in Eq.~\eqref{Eq:Hubbard Hamiltonian}, contrary to the molecular systems considered where the reference basis is that of Hartree-Fock orbitals.

We choose the Hubbard model for this analysis because the ground-state properties, and the wavefunction concentration, depend on a single parameter: $U$. Furthermore, the study of the Hubbard model with various flavors of Configuration Interaction approaches has been considered by previous works~\cite{Friedman1997CIHubbard, Louis1990CiHubbard, Schwarz2015CIHubbard, Dobrautz2019CIHubbard}. The onsite Hubbard repulsion $U$ controls the nature of the ground state of the system. In the $\nmo \rightarrow \infty$ limit, and for $0 \leq U < U_c$, the ground state is a Fermi liquid metal. For $U>U_c$, the ground state is a Mott insulator. The metal to insulator transition in this case is of second order~\cite{Rozenberg_1994,Georges_1996}. \textit{Dynamical Mean Field Theory} studies report a transition point of $U_{c}=5.82...$~\cite{Lee_2017}. The transition on finite-size systems is harder to identify~\cite{gauvinndiaye2023mott}. The change in the nature of the ground state is reflected in the concentration of the wavefunction amplitudes.  In the Mott phase, electrons tend to localize and there is a preference for configurations with low double occupancy~\cite{gauvinndiaye2023mott}. For $U = 0$ and small $U$, and given the disordered nature of the model, it is expected that all electronic configurations will be equivalent, especially after considering a number of disorder realizations. It is therefore expected that the eigenstate solver will show better accuracy for larger values of $U$.

Since the goal of this numerical experiment is to isolate the performance of SQD as a function of the wavefunction concentration, the electronic configurations used to run the SCI eigenstate solver are the $d = 10^3$ configurations of highest amplitude in the exact ground state. This choice removes any effects in the accuracy coming from the choice of the quantum circuit generating the configurations. This procedure is equivalent to the QSCI proposal~\cite{kanno2023qQSCI}, where the configurations are sampled from the exact ground-state wavefunction. As a measure of concentration, we consider the entropy of the distribution defined by the wavefunction amplitudes:
\begin{equation}
\label{eq:entropy distribution}
    S = -\sum_\bts |c_\bts|^2 \log_2 |c_\bts|^2 \;.
\end{equation}
To measure the accuracy in the ground state we consider the relative error in the ground-state energy, a common metric in variational many-body calculations. The relative error is refined as: $\left|E_\textrm{SQD} - E_\textrm{exact} \right|/E_\textrm{exact}$. For the numerical experiments we consider $\nmo = 10$, and 31 equally-spaced values of $U$, between $U = 1$ and $U = 16$, both included.

\begin{figure*}
    \centering
    \includegraphics[width=1\linewidth]{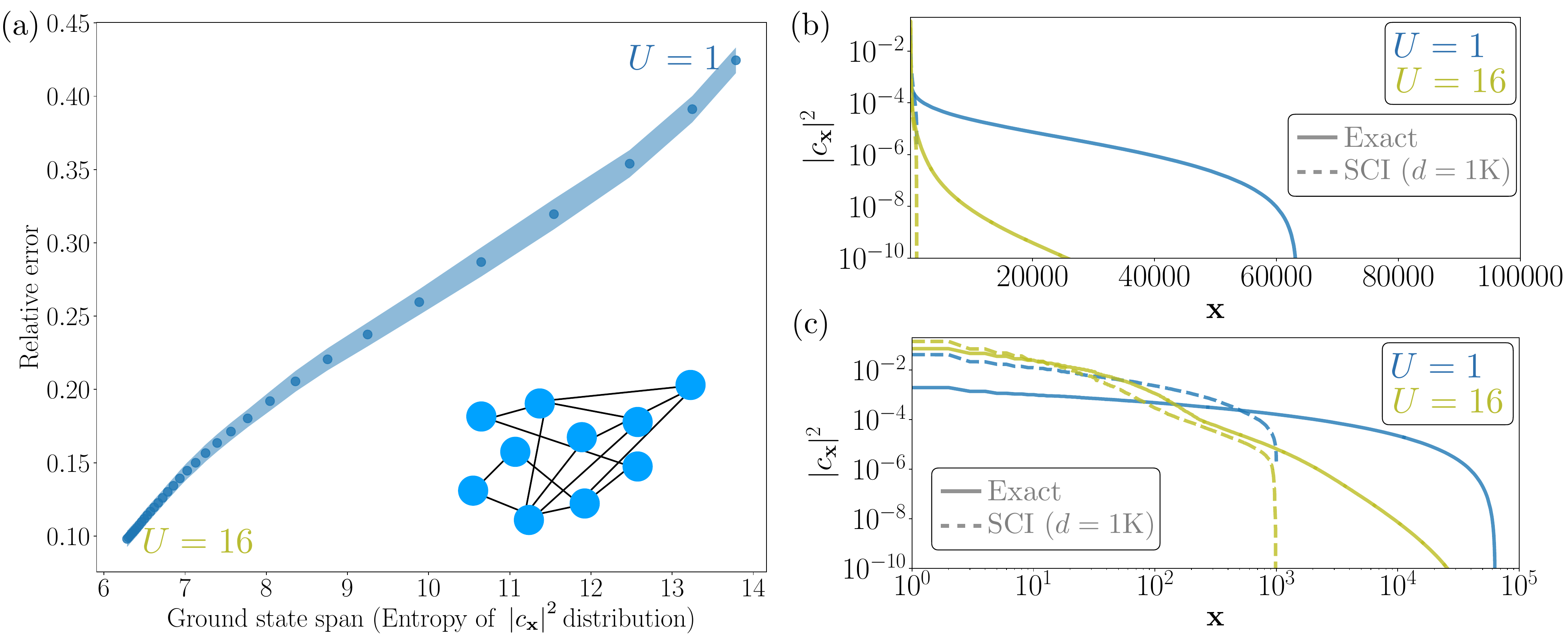}
    \caption{Numerical study on the dependence of the performance of SCI with the ground-state wavefunction concentration in the fully connected Hubbard model with random hopping amplitudes as depicted in the inset of panel (a). Quantities are averaged over twenty disorder realizations. \textbf{(a)} Relative error in the ground-state energy as a function of the exact ground-state wavefunction span, as defined in Eq.~\eqref{eq:entropy distribution}. Different points correspond to different values of the onsite interaction between $U = 1$ and $U = 16$ as indicated. The shaded region indicates the standard error in the mean. \textbf{(b)} Wavefunction amplitudes in decreasing order of magnitude for two values of $U$ as indicated in the legend. Solid lines show the exact ground amplitudes while the dashed lines show the approximate ground state amplitudes obtained by SCI with $10^3$ configurations (see supplementary text). \textbf{(c)} Same as \textbf{(b)} with a logarithmic scale in the horizontal axis to better show the structure of the SCI-constructed amplitudes.}
    \label{fig:Hubbard concentration}
\end{figure*}

Fig.~\ref{fig:Hubbard concentration} (a) shows the relative error in the ground-state energy obtained by SQD as a function of the entropy of the distribution of the wavefunction amplitudes. The more concentrated the ground-state wavefunction (lower entropy), the lower the relative error becomes. As expected, SQD performs better for large $U$ deep in the Mott phase. For small $U$ in the metallic phase, the SCI trial wavefunction with $d = 10^3$ does not have enough expressive power to accurately represent the ground state of the system, whose amplitudes anti-concentrate.

Panels (b) and (c) in Fig.~\ref{fig:Hubbard concentration} show a comparison between the exact and SQD wavefunction amplitudes ordered by their magnitude for $U = 1$ (metal) and $U = 16$ (Mott insulator). For more concentrated wavefunctions ($U = 16$) the SQD amplitudes are in better agreement with the exact amplitudes. For $U = 1$ the number of relevant electronic configurations in the exact wavefunction greatly exceeds the $d = 10^3$ configurations that SQD is allowed to use, explaining the poor performance of SCI in the prediction of the ground-state energy as shown in Fig.~\ref{fig:Hubbard concentration} (a). 

\subsection{An analytical lower bound to the probability of recovering families of configurations}

Without loss of generality, we reorder the orbital labels such that $n_1 \leq n_2\leq \hdots \leq n_M$. Note that in the previous expression we have combined the spin-orbital multi index $(p \sigma)$ in to a single index $p$ running from $1\leq p \leq M$. 
In the following we assume that there exists a good reference configuration $\vett{x}_r$, obtained by assigning value 1 to the $N$ bits of  $\vett{n}$ which have the largest magnitude, and zero otherwise. The average deviation of $\vett{x}_r$ from $\vett{n}$ is quantified by $\varepsilon$: $|\vett{x}_r - \vett{n}|_1 = \varepsilon M $. $\varepsilon$ represents the average distance between each bit in $\vett{x}_r$ and the corresponding entry in $\vett{n}$. The bits $p$ for which $(x_r)_p = 1$ are referred to as the \textit{1-sector} of $\vett{x}_r$, while the bits for which $(x_r)_p = 0$ are referred to as the \textit{0-sector} of $\vett{x}_r$.  For this analysis we take the probability of flipping a bit in $\vett{x}$ proportional to $|x_p-n_p|$, which on average takes the value:
\begin{itemize}
    \item $\varepsilon$; if $p$ is in the 1-sector of $\vett{x}_r$ and $x_p = 1$.
    \item $1-\varepsilon$; if $p$ is in the 1-sector of $\vett{x}_r$ and $x_p = 0$.
    \item $\varepsilon$; if $p$ is in the 0-sector of $\vett{x}_r$ and $x_p = 0$.
    \item $1-\varepsilon$; if $p$ is in the 0-sector of $\vett{x}_r$ and $x_p = 1$.
\end{itemize}
We use these average values for our analytical derivations.

\medskip 

 Let $\mathcal{U} = \mathcal{U}_L \cdots \mathcal{U}_2 \cdot \mathcal{U}_1$ denote the ideal unitary channel that produces $\psi = \mathcal{U}(|0\rangle \langle 0|
)$. Here, $\mathcal{U}_i$ denotes a two qubit unitary. Let us analyze the effect of noise on $\mathcal{U}$. Suppose that each two-qubit gate $\mathcal{U}_i$ is followed by a Pauli noise channel that preserves the state with probability $p$, i.e, $\mathcal{P}(\rho) = p \rho + (1-p)P\rho P$ with $P$ denoting a Pauli operator. Then after $L$ quantum operations, the noisy output state $\widetilde{\rho}$ is given by  
\begin{align}\label{seq:noisy-state}
    \widetilde{\rho} = p^L \rho + \sigma~,
\end{align}
where $\sigma$ denotes the component of the state affected from non-identity Pauli paths. 

From Eq.~\eqref{seq:noisy-state}, it follows that the number of experiments needed to sample from the ideal state $\psi$ scales as $1/p^L$. Let $p = e^{-\lambda}$, where $\lambda$ denotes the noise rate. Then the number of shots needed for sampling bitstrings from $\psi$ scales as $e^{\lambda L}$, which is exponential in the number of noisy operations and the noise rate. In general, without performing quantum error correction, any quantum algorithm employed on noisy devices is limited to a maximum number of operations $L$ that scales inversely proportional to the noise rate $\lambda$.  
Note that for our application, one might also get correct bitstrings from $\sigma$, as defined in Eq.~\eqref{seq:noisy-state}, which can decrease the number of experiments needed to obtain good bitstrings.

Above we discussed the effect of noise in obtaining samples from the ideal state. We now analyze the probability of recovering configurations using the configuration recovery scheme described earlier, as a function of their Hamming distance from $\vett{x}_r$. In general, this analysis depends on the noise model considered. To simplify the analysis, we consider a simple global depolarizing noise model, whose effect on the probability of sampled configurations is described by:
\begin{equation}
    \widetilde{P}_\Psi (\bts) = \alpha P_\Psi(\bts) + (1-\alpha) \frac{\mathbb{I}}{2^M},
\end{equation}
where $\alpha \in [0, 1]$ is a parameter that quantifies the amount of quantum signal. Note that at $\alpha = 1$, we exactly sample from the ideal state and, therefore, we do not need to perform the recovery process. While we use for simplicity a global depolarizing noise model, we expect that the analysis can be generalized to account for more realistic noise models, including the one defined in Eq.~\eqref{seq:noisy-state}.

The configurations that can be sampled from such a probability distribution fall into three categories: configurations with the right particle number that are in the support of the ideal distribution $P_\Psi(\bts)$, configurations with the right particle number that are not the support of $P_\Psi(\bts)$, and configurations with the wrong particle number.
Configuration recovery does not apply to the first two categories. The third category, configurations with incorrect particle numbers, are guaranteed to have come from the noisy part of $\widetilde{P}_\Psi (\bts)$.
We now derive a lower bound on the probability of drawing a configuration with wrong number of particles from $\widetilde{P}_\Psi (\bts)$ and then converting it (via the configuration recovery scheme) to a particular configuration $\vett{x}_\text{target}$ with the right particle number, whose Hamming distance to $\vett{x}_r$ is $2b$.

Since $\vett{x}_\text{target}$ is Hamming distance $2b$ from $\vett{x}_r$, it contains $b$ $1$s in the $0$-sector and $b$ $0$s in the $1$-sector.
There are two cases for initial configurations with wrong number of particles: those that have too many 1s or too many 0s.
\begin{enumerate}
\item 
Too many 1s. In this case only bits with the value 1 will be flipped by the configuration recovery. Consider the set of initial noisy configurations  with Hamming weight $N+g+h$. Here, $g$ is the number of 1s in the 0-sector of $\vett{x}_r$ in the initial configuration that need to be flipped to reach $\vett{x}_\text{target}$, while $h$ is the number of 1s in the 1-sector that need to be flipped to reach $\vett{x}_\text{target}$. All possible combinations of $g$ and $h$ are considered in this analysis. Starting from the initial noisy configuration, the probability of flipping one of the $g$  $1$s that need to be flipped to reach $\vett{x}_\text{target}$ (which we will call a ``successful'' bit-flip) is given by 
\begin{equation}
    \frac{g(1-\varepsilon)}{(1-\varepsilon)(b+g)+\varepsilon (N-b+h)};
\end{equation}
the denominator reflects the total number of $1$s weighted by their probabilities of being flipped.
After flipping the first 1 in the 0-sector, the probability of another successful bit-flip in the the same sector is similarly
\begin{equation}
    \frac{(g-1)(1-\varepsilon)}{(1-\varepsilon)(b+g-1)+\varepsilon (N-b+h)}.
\end{equation}
Thus, the probability of flipping all $g$ of the incorrect $1$s in the 0-sector is
\begin{equation}
    \prod_{i=0}^{g-1}\frac{(g-i)(1-\varepsilon)}{(1-\varepsilon)(b+g-i)+\varepsilon (N-b+h)}.
\end{equation}
The same argument can be applied to the bit-flips performed in the 1-sector. The probability of a successful first bit-flip in the 1-sector is given by:
\begin{equation}
    \frac{h\varepsilon}{(1-\varepsilon)b+\varepsilon (N-b+h)},
\end{equation}
by essentially the same argument as above.
The probability of a successful second flip is
\begin{equation}
    \frac{(h-1)\varepsilon}{(1-\varepsilon)b+\varepsilon (N-b+h-1)}.
\end{equation}
One can apply the same iterative argument in this case. Hence, the overall probability of obtaining the target bitstring $\vett{x}_\text{target}$ via configuration recovery is 
\begin{equation}
\begin{split}
    &\prod_{i=0}^{g-1}\left(\frac{(g-i)(1-\varepsilon)}{(1-\varepsilon)(b+g-i)+\varepsilon (N-b+h)}\right)\prod_{j=0}^{h-1}\left(\frac{(h-j)\varepsilon}{(1-\varepsilon)b+\varepsilon (N-b+h-j)}\right)\\
    &=\prod_{i=1}^{g}\left(\frac{i(1-\varepsilon)}{(1-\varepsilon)(b+i)+\varepsilon (N-b+h)}\right)\prod_{j=1}^{h}\left(\frac{j\varepsilon}{(1-\varepsilon)b+\varepsilon (N-b+j)}\right).
\end{split}
\end{equation}

\item
For the set of configurations  with Hamming weight $N-g-h$ that came from $g$ and $h$ instances of $1 \to 0$ bit-flip errors in the one- and zero-sectors of the $\vett{x}_r$ state (respectively), a symmetric argument applies. We omit the derivation for this case.
\end{enumerate}

The above gives the probabilities of recovering configurations  with excess 1s or 0s back to a particular bitstring $\vett{x}_\text{target}$ of Hamming distance $2b$ from $\vett{x}_r$.
We now integrate this probability over different values of $g$ and $h$ to obtain the probability of recovering a particular bitstring of Hamming distance $2b$ to $\vett{x}_r$, from a sample drawn from the uniform distribution. We obtain the lower bound
\begin{equation}\label{eq:lower bound p_recovery}
\begin{split}
    & P_\textrm{recovery}(M, N, b, \varepsilon) \geq \\ 
    &\sum_{g = 0}^{M-N-b}\sum_{h=0}^{N-b}\frac{\binom{M-N-b}{g}\binom{b}{h}}{2^M} \prod_{i=1}^{g}\left(\frac{i(1-\varepsilon)}{(1-\varepsilon)(b+i)+\varepsilon (N-b+h)}\right)
     \prod_{j=1}^{h}\left(\frac{j\varepsilon}{(1-\varepsilon)b+\varepsilon (N-b+j)}\right)\\
    &+ \sum_{g=0}^{N-b}\sum_{h=0}^{M-N-b} \frac{\binom{N-b}{g}\binom{b}{h}}{2^M} \prod_{i=1}^{g}\left(\frac{i(1-\varepsilon)}{(1-\varepsilon)(b+i)+\varepsilon (M-N-b+h)}\right)\prod_{j=1}^{h}\left(\frac{j\varepsilon}{(1-\varepsilon)b+\varepsilon (M-N-b+j)}\right)\\
    & \equiv F(M,N,b,\epsilon).
\end{split}
\end{equation}
The two terms correspond to the two cases above, with the combinatorial fractions in each giving the probabilities of obtaining a configuration equivalent to the particular bitstring $\vett{x}_\text{target}$ plus $g$ random $0\rightarrow1$ bit-flips in the 0-sector of $\vett{x}_r$ and $h$ random $0\rightarrow1$ bit-flips in the 1-sector (first term), or $g$ random $1\rightarrow0$ bit-flips in the one-sector of the HF state and $h$ random $1\rightarrow0$ bit-flips in the zero-sector (second term).
In summary, the probability of configuration recovery to output a configuration without error is lower bounded by $\alpha + (1-\alpha)F(M,N,b,\epsilon)$,
where $F(M,N,b,\epsilon)$ is defined in Eq.~\eqref{eq:lower bound p_recovery}.
Note that this lower bound approaches one as $\alpha\rightarrow 1$.

\begin{figure*}
    \centering
    \includegraphics[width=.7\linewidth]{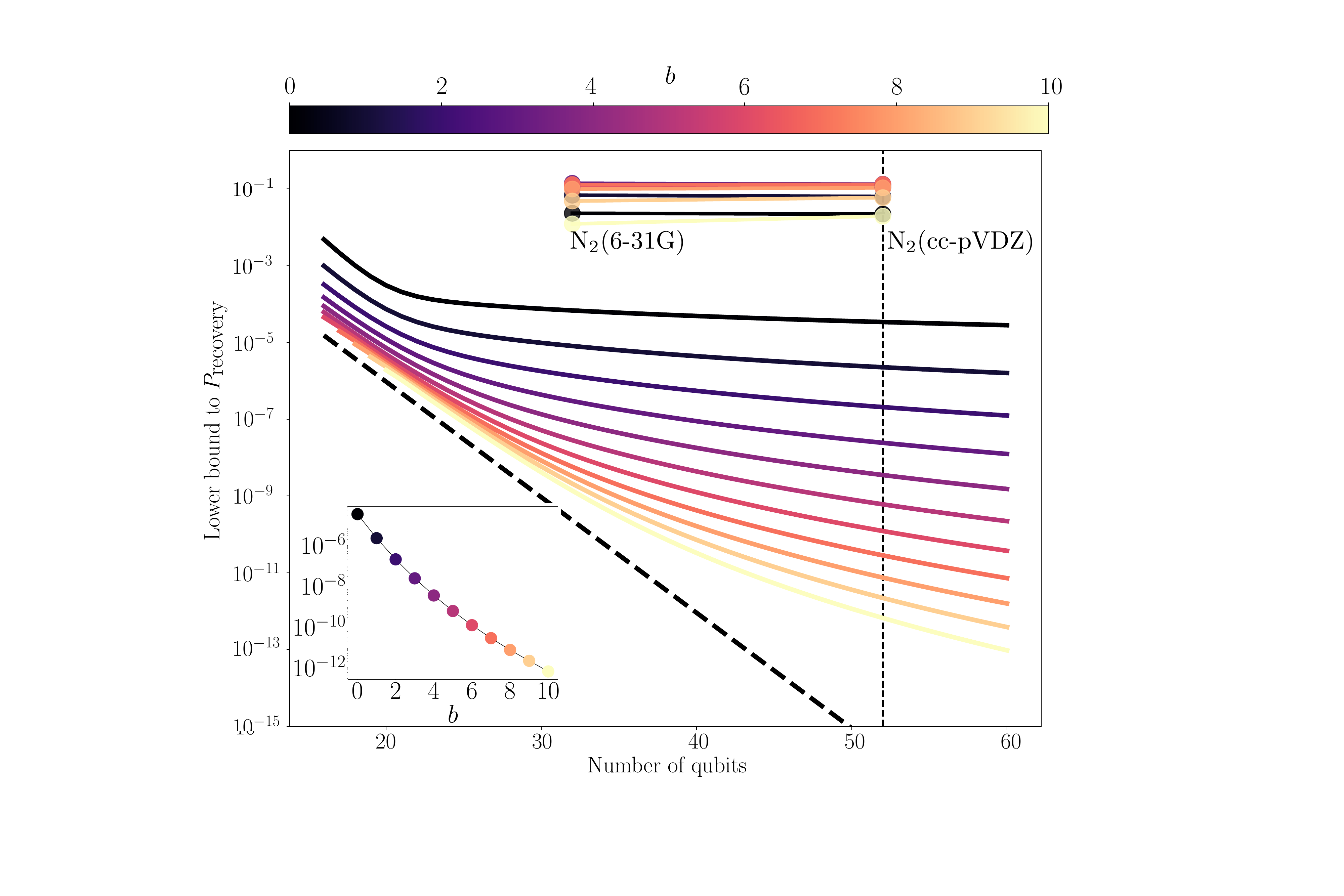}
    \caption{Lower bound to the probability of recovering a particular bitstring with Hamming distance $2b$ from a reference approximation $\vett{x}_r$ bistring from a bistring sampled from the uniform distribution. The lower bound is shown as a function of the number of qubits. Different values of $b$ are shown, as indicated by the colorbar. The dashed black curve shows $1/2^M$, which is the probability of obtaining the desired bitsting directly from the uniform distribution.  The circles connected by lines show the fraction of configurations for different families of bitstrings (as indicated by the colorbar) obtained by running the configuration recovery procedure on the samples from the quantum processor on the \nitrogen molecule for the 6-31-G and cc-pVDZ bases sets. The inset shows the lower bound as a function of $b$ for $M = 52$ qubits.}
    \label{fig:p_recovery}
\end{figure*}

Fig.~\ref{fig:p_recovery} shows the lower bound in Eq.~(\ref{eq:lower bound p_recovery}) to the recovery probability, for $\varepsilon = 0.1$ and $N = 10$, as a function of the number of qubits $M$. We observe that the lower bound decays as an exponential of the number of qubits and that the recovery of configurations with small values of $b$ is more likely than the recovery of configurations defined by large $b$ values. We remark that the configuration recovery probability for configurations close to $\vett{x}_r$ is substantially larger than the probability of obtaining the same samples by sampling from the uniform distribution, which would be the limit of no quantum signal. We argue that this feature helps in improving molecular energies over using simple post-selection over correct particle sector. Additionally, Fig.~\ref{fig:p_recovery} shows the fraction of configurations obtained from the configuration recovery procedure for different values of $b$ on actual hardware data on the \nitrogen molecule on different bases sets and correspondingly different qubit numbers. The fractions are orders of magnitude larger than the lower bound and do not appear to have a strong dependence on the system size. This observation highlights that the lower bound is quite loose and in practice the performance is better.

\section{Optimization of the circuit parameters}\label{Sec: optimization of S-CORE}

The optimization of circuit parameters to produce better distributions in the space of electronic configurations is an avenue of improvement of the SQD framework. 
We develop a formalism for parameter optimization using the subspace energy as the cost function, in contrast to previous experiments that rely on standard VQE optimization frameworks~\cite{kanno2023qQSCI, nakagawa2023adaptqsci}.
The HPC quantum estimator is constructed by running a diagonalization procedure in $K$ subspaces defined by different batches of sampled configurations, as described in the main text. The number of subspaces considered $K$ can go from $K = 1$ to $K = K_\textrm{max}$, where $K_\textrm{max} = \binom{D}{d}$. Recall that $D$ is the dimensionality of the subspace of the Fock space spanned by configurations with the correct particle number and $d$ is the dimension of the SCI subspace. Note that $K_\textrm{max}$ is double-combinatorial in the number of spin-orbitals and electrons. Therefore, computing the HPC quantum estimator (with or without configuration recovery) for all possible sets of $d$ configurations is not efficient. Instead, we propose the use of a Monte Carlo estimator over batches of configurations $\mathcal{S}^{(k)}$ to optimize the energy of the HPC quantum estimator averaged over different sets of configurations.

Recall that each configuration $\bts$ is sampled with probability $P_\Psi (\bts) = \left| \langle \bts | \Psi \rangle \right|^2$, in the noiseless case. The following discussion applies to the noisy distribution $\tilde{P}_\Psi$. The method presented in this section can be applied also in conjunction with the configuration recovery technique. Since each configuration is i.i.d. sampled, the joint distribution that describes the probability of sampling the configurations in a batch $\mathcal{S}^{(k)}$ with $d$ configurations is given by 
\begin{equation}\label{Eq:batch probability}
    P_\Psi \left(  \mathcal{S}^{(k)}  \right) = \prod_{\bts_{(i)} \in  \mathcal{S}^{(k)}} P_\Psi \left(\bts_{(i)} \right).    
\end{equation}
We consider the situation where the quantum circuit that prepares $|\Psi\rangle$ is characterized by a set of variational parameters $\theta$. The explicit dependence on variational parameters is denoted by $\Psi_\theta$. A cost function is formulated for the circuit optimization:
\begin{equation}\label{eqS:S-CORE cost function}
    \mathcal{E} (\theta) = \sum_{k = 1}^{K_\textrm{max}} P_{\Psi_\theta} \left(  \mathcal{S}^{(k)}  \right) E^{(k)},
\end{equation}
where $E^{(k)}$ is defined in the main text, Eq.~\eqref{eqS:sample energy}, as the SQD ground-state energy for the configurations in $\mathcal{S}^{(k)}$.  An estimator to the double combinatorially-large summation in $\mathcal{E} (\theta)$ can be obtained by the Monte-Carlo unbiased estimator:
\begin{equation}
    \mathcal{E} (\theta) \approx  \frac{1}{K} \sum_{k = 1}^K  E^{(k)} 
\end{equation}
with $K \ll K_\textrm{max}$ and where the $E^{(k)}$ are obtained from $\mathcal{S}^{(k)}$ sampled according to $P_{\Psi_\theta}\left(\mathcal{S}^{(k)}\right)$.
The gradient of the cost function with respect to the variational parameters is given by:
\begin{equation}
     \partial_\theta \mathcal{E} (\theta) = 
    \sum_{k = 1}^{K_\textrm{max}} \left\{ \sum_{\bts_{(i)} \in \mathcal{S}^{(k)}} \left[ \partial_\theta P_{\Psi_\theta} \left(\bts_{(i)} \right) \right] \left( \prod_{\substack{\bts_{(j)} \in \mathcal{S}^{(k)}\\ \bts_{(j)} \neq \bts_{(i)}}} P_{\Psi_\theta} \left(\bts_{(j)}\right) \right) \right\} \cdot E^{(k)}
\end{equation}
Multiplying each term in the sum from the product rule by $1 = P_{\Psi_\theta}\left(\bts_{(i)}\right)/P_{\Psi_\theta}(\bts_{(i)})$ we obtain:
\begin{equation}\label{eqS:S-CORE cost function gradient}
     \partial_\theta \mathcal{E} (\theta) = 
     \sum_{k = 1}^{K_\textrm{max}}  P_{\Psi_\theta} \left(\mathcal{S}^{(k)} \right)  \left(\sum_{\bts_{(i)} \in \mathcal{S}^{(k)}} \frac{\left[ \partial_\theta P_{\Psi_\theta} \left(\bts_{(i)}\right) \right]}{P_{\Psi_\theta} \left(\bts_{(i)}\right)}  \right)  E^{(k)},
\end{equation}
whose Monte-Carlo estimator is given by:
\begin{equation}
\label{eq: gradient MC estimator}
    \partial_\theta \mathcal{E} (\theta) \approx \frac{1}{K} \sum_{k = 1}^K \left(\sum_{\bts_{(i)} \in \mathcal{S}^{(k)}} \frac{\left[ \partial_\theta P_{\Psi_\theta} \left(\bts_{(i)}\right) \right]}{P_{\Psi_\theta} \left(\bts_{(i)}\right)}  \right)  E^{(k)}
\end{equation}
with $K \ll K_\textrm{max}$ and where again the $E^{(k)}$ are obtained from $\mathcal{S}^{(k)}$ sampled according to $P_{\Psi_\theta}\left(\mathcal{S}^{(k)}\right)$.
The evaluation of $\left[ \partial_\theta P_{\Psi_\theta} (\bts_{(i)}) \right]$ may be achieved by:
\begin{equation}
     \partial_\theta P_{\Psi_\theta} \left(\bts_{(i)}\right)  = \partial_\theta \left(\left\langle \bts_{(i)} | \Psi_\theta \right\rangle \left\langle \Psi_\theta | \bts_{(i)} \right\rangle \right) = 2\textrm{Re} \left\{ \left\langle \bts_{(i)} | \Psi_\theta \right\rangle  \left\langle\partial_\theta \Psi_\theta | \bts_{(i)} \right\rangle \right\}.
\end{equation}
We remark that the limit where $K = 1$ is equivalent to the stochastic gradient-descent optimization technique.

For most classes of variational circuits, the gradient of the circuits $| \partial_\theta \Psi_\theta \rangle$ may be implemented by parameter-shift rules or similar techniques. The reader may have noticed that the estimator for the gradient in the cost function of Eq.~\eqref{eq: gradient MC estimator} is biased if the support of $\left[ \partial_\theta P_{\Psi_\theta} \right]$ does not coincide with the support of $P_{\Psi_\theta}$, as a direct consequence of writing  $1 = P_{\Psi_\theta}\left(\bts_{(i)}\right)/P_{\Psi_\theta}(\bts_{(i)})$ for configurations outside the support of  $P_{\Psi_\theta}$, where that expression is ill-defined.

Another practical issue with the gradient-based optimization of $\mathcal{E}(\theta)$ is the requirement to evaluate $\left\langle \bts_{(i)} | \Psi_\theta \right\rangle$ and $\left\langle\partial_\theta \Psi_\theta | \bts_{(i)} \right\rangle$, which rely in the Hadamard test for their evaluation. The implementation of the Hadamard test requires controlled versions of the circuits that realize the variational states. The implementation of such controlled unitaries requires circuit depths beyond the reach of current quantum processors.

An alternative approach is to optimize $\mathcal{E}(\theta)$ by explicit-gradient-free methods like COBYLA~\cite{cobyla} or simulated annealing. Gradient-free optimization of the circuit parameters is used in the numerical study of Sec.~\ref{Sec: effect of jastrow}.

The optimization of the circuit parameters on quantum experiments to minimize the estimator energy will be the subject of future studies, which should consider  the effect of noise in different gradient-free optimizers. Furthermore, the goal of this work is not to show that the optimization of the circuit parameters can improve the accuracy of the estimator, but to show that the quantum centric supercomputing estimator allows to tackle electronic structure problems described by an unprecedented number of qubits. Moreover, we show that the estimator run on circuits with fixed parameters already achieves good levels of accuracy.

\subsection{Orbital optimization}
Orbital or single-particle basis rotations are a common approach to improve the accuracy of variational calculations of electronic structure systems. Orbital optimizations can be applied in conjunction with a wide variety of electronic structure methods, including complete active space diagonalizations~\cite{roos1980complete, head1988optimization, werner1985second, olsen2011casscf}, the \textit{density matrix renormalization group}~\cite{zgid_density_2008,wu_disentangling_2022, wouters_density_2014}, \textit{variational Monte Carlo}~\cite{robledomoreno2023OrbitalRotations} techniques, and quantum computing variational approaches~\cite{sokolov_quantum_2020, mizukami_orbital_2020, bierman_improving_2022}. Moreover, orbital optimizations are common practice on SCI-based approaches as well~\cite{smith2017cheap}. More generally, orbital optimizations make variational approaches invariant under orbital rotations. For notation clarity we replace in this section the notation of the two-body integral $(pr|qs)$ by the abbreviation $h_{pqrs}$.

The implementation of the orbital optimizations in conjunction with SQD (with or without configuration recovery) follows the procedure presented in Ref.~\cite{robledomoreno2023OrbitalRotations}. Orbital rotations consist of the application of the similarity transformation:
\begin{equation}\label{eq:similary Transform}
\hat{\tilde{H}}= \hat{U}^\dagger(\kappa)\hat{H} \hat{U}(\kappa), 
\end{equation}
where
\begin{equation}\label{eq:rotation parametrization}
    \hat{U}(\kappa) = \exp\left( \sum_{\substack{pq \\ \sigma}} \kappa_{pq} \crt{p\sigma} \dst{q\sigma}\right).
\end{equation}
In this work, we choose the matrix that parametrizes the rotation to be real: $\kappa \in \mathbb{R}^{\nmo \times \nmo}$. To enforce unitarity in $\hat{U}(\kappa)$ we require that $\kappa_{pq} = -\kappa_{qp}$. According to Thouless's Theorem~\cite{Thouless1960}, the action of the similarity transformation defined by $\hat{U}(\kappa)$ transforms the creation operators according to:
\begin{equation}
\crt{p\sigma} \mapsto \hat{U}(\kappa)^\dagger \crt{p\sigma} \hat{U}(\kappa) =  \sum_{t} \Omega_{t p} \crt{t\sigma},
\end{equation}
where $\Omega=\exp(\kappa)\in \mathbb{R}^{\nmo \times \nmo} $. The Hamiltonian in the rotated basis can therefore be written in terms of the reference creation and annihilation operators:
\begin{equation}
 \hat{\tilde{H}} =  \sum_{\substack{pq \\ \sigma}} \tilde{h}_{pq} \crt{p\sigma} \hat{a}_{q\sigma} + \sum_{\substack{pqrs \\ \sigma \tau}}  \frac{\tilde{h}_{pqrs}}{2} \crt{p\sigma} \crt{q\tau} \dst{s\tau}  \dst{r\sigma}
\end{equation}
where the one- and two-body integrals have been transformed according to the tensor transformations:
\begin{equation}\label{eq:RotatedIntegrals}
    \begin{split}
        \tilde{h}_{pq}  & = h_{tu} \Omega_{t p} \Omega_{u q} , \\
        \tilde{h}_{pqrs}  & = h_{tuvw} \Omega_{t p} \Omega_{u q} \Omega_{v r} \Omega_{w s},
    \end{split}
\end{equation}
using Einstein's summation convention.

A variational procedure is used to search for the single-particle basis that yields the optimal description of the ground state, given the collection of all possible $K_\textrm{max}$ \textit{bare} variational trial states $\left|\psi_\Theta^{(k)} \right\rangle$. The variational procedure is similar to the method presented in Sec.~\ref{Sec: optimization of S-CORE}. In this setting, $\left|\psi_\Theta^{(k)} \right\rangle$ is the SCI ansatz given as the linear combination of the $d$ determinants in $\mathcal{S}^{(k)}$ (see Eq.~\eqref{eqS:SCIWaveFunctionAmplitudes}). The set of variational parameters $\Theta$ is composed of the circuit parameters $\theta$ defining $\left|\Psi_\theta \right\rangle$ together with the wavefunction amplitudes $c_\bts^{(k)}$ in the subspace defined by the SCI configurations. Each variational state is \textit{dressed} by the same single-particle orbital rotation 
\begin{equation}\label{eq:variationalRotated}
    \left|\psi_{\{ \kappa, \Theta\}}^{(k)}\right\rangle = \hat{U}(\kappa) \left|\psi_\Theta^{(k)} \right\rangle.
\end{equation}
The loss function to be optimized in the variational setting is defined by the average, over batches of configurations, of the Rayleigh quotient:
\begin{equation}\label{eqS:rotatedCostFunction}
\mathcal{E} (\kappa, \Theta) = \sum_{k = 1}^{K_\textrm{max}} P_{\Psi_\theta}\left(\mathcal{S}^{(k)} \right) \left\langle \psi^{(k)} | \hat{U}^\dagger(\kappa) \hat{H} \hat{U}(\kappa) | \psi^{(k)} \right\rangle = \sum_{k = 1}^{K_\textrm{max}} P_{\Psi_\theta}\left(\mathcal{S}^{(k)} \right) \left\langle \psi^{(k)} |  \hat{\tilde{H}}(\kappa) | \psi^{(k)} \right\rangle .
\end{equation}
Gradient descent and its variants can be used to minimize both $\kappa$ and $\theta$ in $\mathcal{E} (\kappa, \Theta)$, while the optimization of $c_{\bts}^{(k)}$ is carried out by the diagonalization procedure (see Sec.~\ref{secS: eigenstate solver}). Note that the sum in the average contains double-combinatorially many terms. Consequently, a Monte-Carlo based estimator is used to obtained an unbiased estimate (see Sec.~\ref{Sec: optimization of S-CORE}). Gradients with respect to $\theta $ have the same expression as in Eq.~\eqref{eqS:S-CORE cost function gradient} replacing $\hat{H}$ by $\hat{\tilde{H}}$. Gradients with respect to the orbital rotations can be computed by the contraction of the bare one- and two-body reduced density matrices (1- and 2-RDMs) with the gradients of the one- and -two body integrals with respect to $\kappa_{pq}$ in Eq.~\eqref{eqS:rotatedCostFunction}:
\begin{equation}
    \partial_\kappa \mathcal{E}(\kappa,\Theta) = \sum_{k = 1}^{K_\textrm{max}} P_{\Psi_\theta}\left(\mathcal{S}^{(k)} \right) \left[\sum_{\substack{pq \\ \sigma}} \tilde{h}_{pq}' \Gamma_{pq; \sigma}^{(k)} + 
     \sum_{\substack{pqrs \\ \sigma}}  \frac{\tilde{h}_{pqsr}'}{2}\left( \Gamma_{pqrs; \sigma, \sigma}^{(k)} +\Gamma_{pqsr; \sigma, -\sigma}^{(k)}\right) \right].
\end{equation}
In the previous expression the gradients of the integrals in the rotated basis are given by
\begin{equation}
    \begin{split}
        \tilde{h}_{pq}'  & = h_{tu}\frac{\partial}{\partial \kappa} \left( \Omega_{t p} \Omega_{u q} \right)\\
        \tilde{h}_{pqrs}'  & = h_{tuvw} \frac{\partial}{\partial \kappa} \left( \Omega_{t p} \Omega_{u q} \Omega_{v r} \Omega_{w s}\right),
    \end{split}
\end{equation}
and the bare 1-RDMs and 2-RDMs are defined as
\begin{equation}\label{eq: density matrices}
    \begin{split}
        \Gamma_{pq; \sigma}^{(k)} & = \left\langle \psi^{(k)} |\crt{p \sigma} \dst{q \sigma}| \psi^{(k)} \right\rangle \\
        \Gamma_{pqsr; \sigma, \tau}^{(k)} & = \left\langle \psi^{(k)} |\crt{p \sigma} \crt{q \tau} \dst{s \tau} \dst{r \sigma} | \psi^{(k)} \right\rangle. 
    \end{split}
\end{equation}
The gradients of the integrals with respect to the rotation parameters are computed using the automatic differentiation (AD) tools of the software package Jax~\cite{jax2018}. Since the evaluation of the gradients with respect to $\kappa$ requires the evaluation of an intractable sum, we use an unbiased Monte-Carlo estimator for its evaluation:
\begin{equation}
    \partial_\kappa \mathcal{E}(\kappa,\Theta) \approx \frac{1}{K}\sum_{k = 1}^{K}\left[\sum_{\substack{pq \\ \sigma}} \tilde{h}_{pq}' \Gamma_{pq; \sigma}^{(k)} + 
     \sum_{\substack{pqrs \\ \sigma}}  \frac{\tilde{h}_{pqsr}'}{2}\left( \Gamma_{pqrs; \sigma, \sigma}^{(k)} +\Gamma_{pqsr; \sigma, -\sigma}^{(k)}\right) \right],
\end{equation}
with $K \ll K_\textrm{max}$ and where the batches of configurations labelled by $(k)$ are sampled according to $P_{\Psi_\theta}\left(\mathcal{S}^{(k)} \right)$ (see Eq.~\eqref{Eq:batch probability}).

For a full orbital optimization calculation, all sets of variational parameters $\kappa$, $\theta$, and $c_\bts^{(k)}$ should be updated. Since $\kappa$ and $\theta$ sets rely on gradients for their optimization, they can be updated simultaneously. However, the wavefunction amplitudes are not updated by gradients. In this case, the optimization strategy consists of the alternation of a number of gradient-descent steps optimizing  $\kappa$ and $\theta$ followed by the eigenstate solver that now uses samples from an updated distribution according to the change in $ \theta$ with a rotated Hamiltonian according to $\kappa$. This alternation is repeated for a number of iterations $N_\textrm{SCF}$. In this work, we do not consider the optimization of circuit parameters $ \theta$ and therefore the effect of the orbital optimization is not maximized, since the quantum circuit is not allowed to respond to the change in the Hamiltonian.

See Sec.~\ref{Sec:orbital optimizations N2} to see the effect of orbital optimizations in the accuracy of SQD (with configuration recovery). The section shows results run on measurement outcomes obtained from the Heron processor, to study the dissociation of \nitrogen at cc-pVDZ level of theory.

\section{Additional information about experimental details}
In this section we show examples of the structure of the compiled circuits. We show numerical experiments comparing the performance of the circuits we use against the performance of a classically efficient reduction of the LUCJ circuits.

We also describe the setup for all of the experiments conducted in this work, highlighting the quantum and classical hardware used. We show the quantum processor mappings, depth and number of gates in the circuits, and the number of measurement outcomes sampled. We also provide details about the classical computing resources.

\subsubsection{Compiled circuits}

\begin{figure*}[t!]
    \centering
    \includegraphics[width=0.8\linewidth]{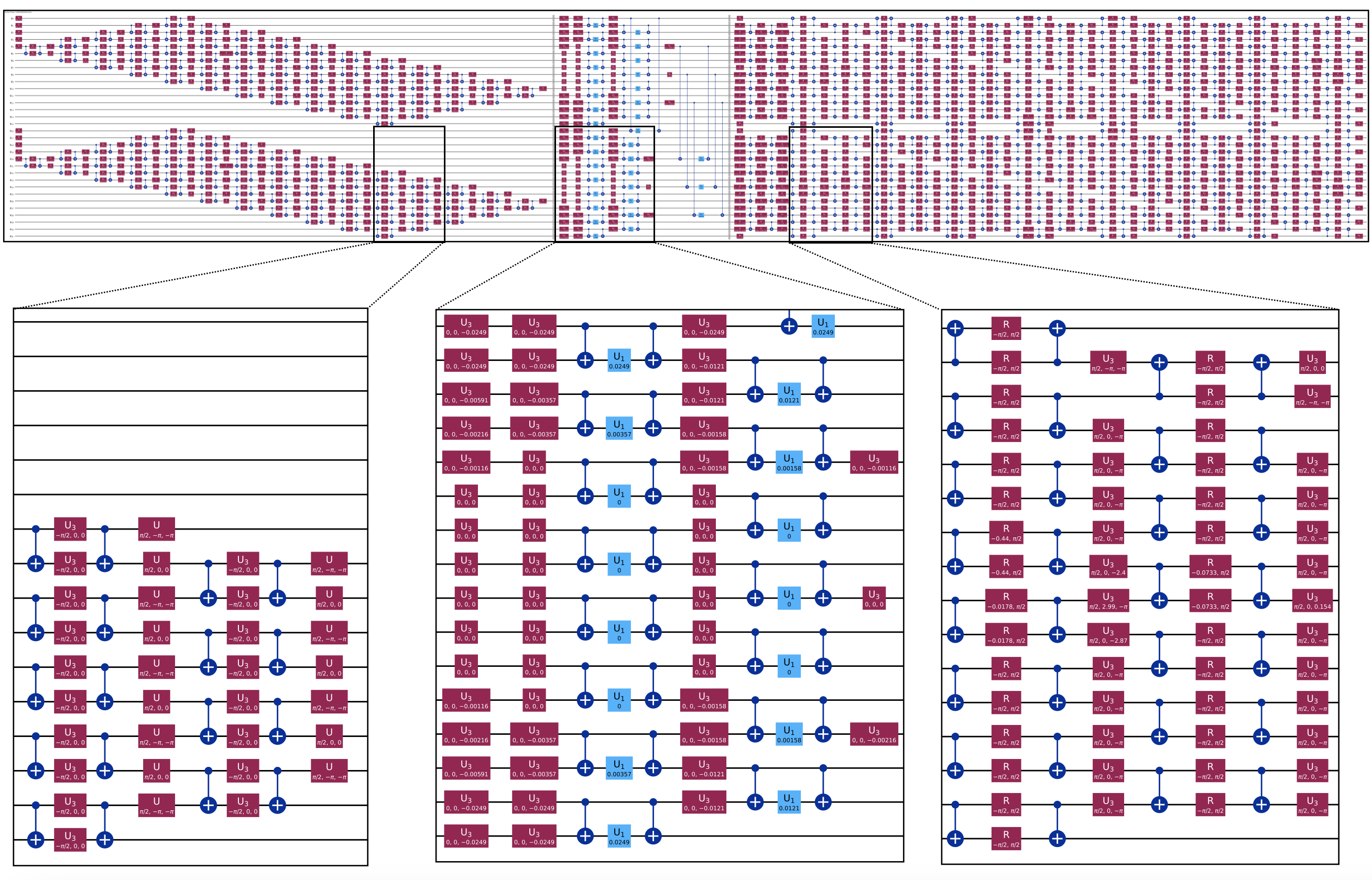}
    \caption{Top: LUCJ circuit used to simulate the ground state of N$_2$/6-31G at the equilibrium bondlength, compiled into single-qubit (red, purple blocks) and $\mathsf{CNOT}$ (light blue symbols) gates. Bottom: zoomed-out views of gates from the first orbital rotation, density-density interaction, and second orbital rotation (left to right).}
    \label{fig:circuit_n2_631g}
\end{figure*}

\begin{figure*}[t!]
    \centering
    \includegraphics[width=0.8\linewidth]{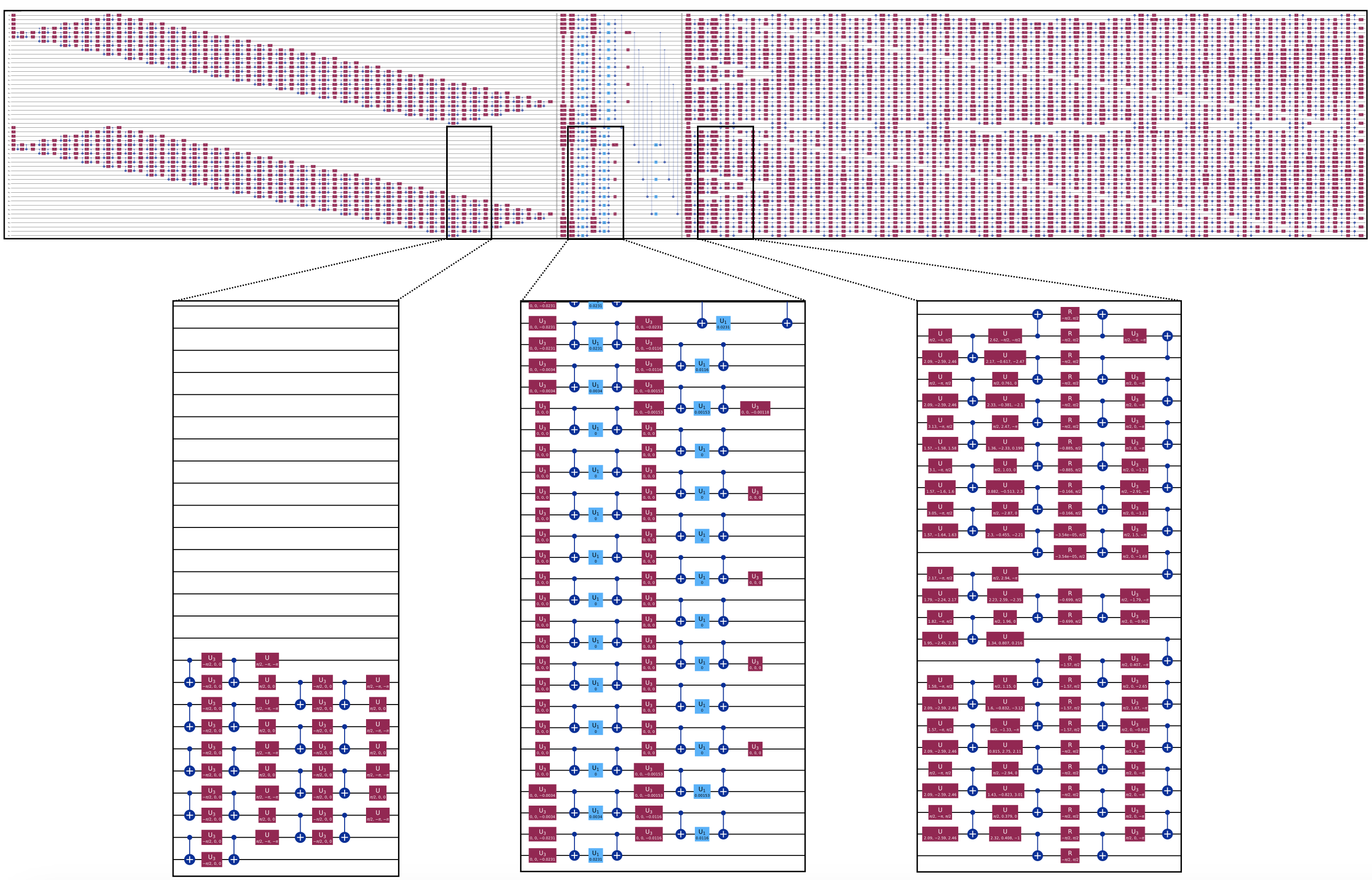}
    \caption{Top: LUCJ circuit used to simulate the ground state of N$_2$/cc-pVDZ at the equilibrium bondlength, compiled into single-qubit (red, purple blocks) and $\mathsf{CNOT}$ (light blue symbols) gates. Bottom: zoomed-out views of gates from the first orbital rotation, density-density interaction, and second orbital rotation (left to right).}
    \label{fig:circuit_n2_ccpvdz}
\end{figure*}

\begin{figure*}[t!]
    \centering
    \includegraphics[width=0.8\linewidth]{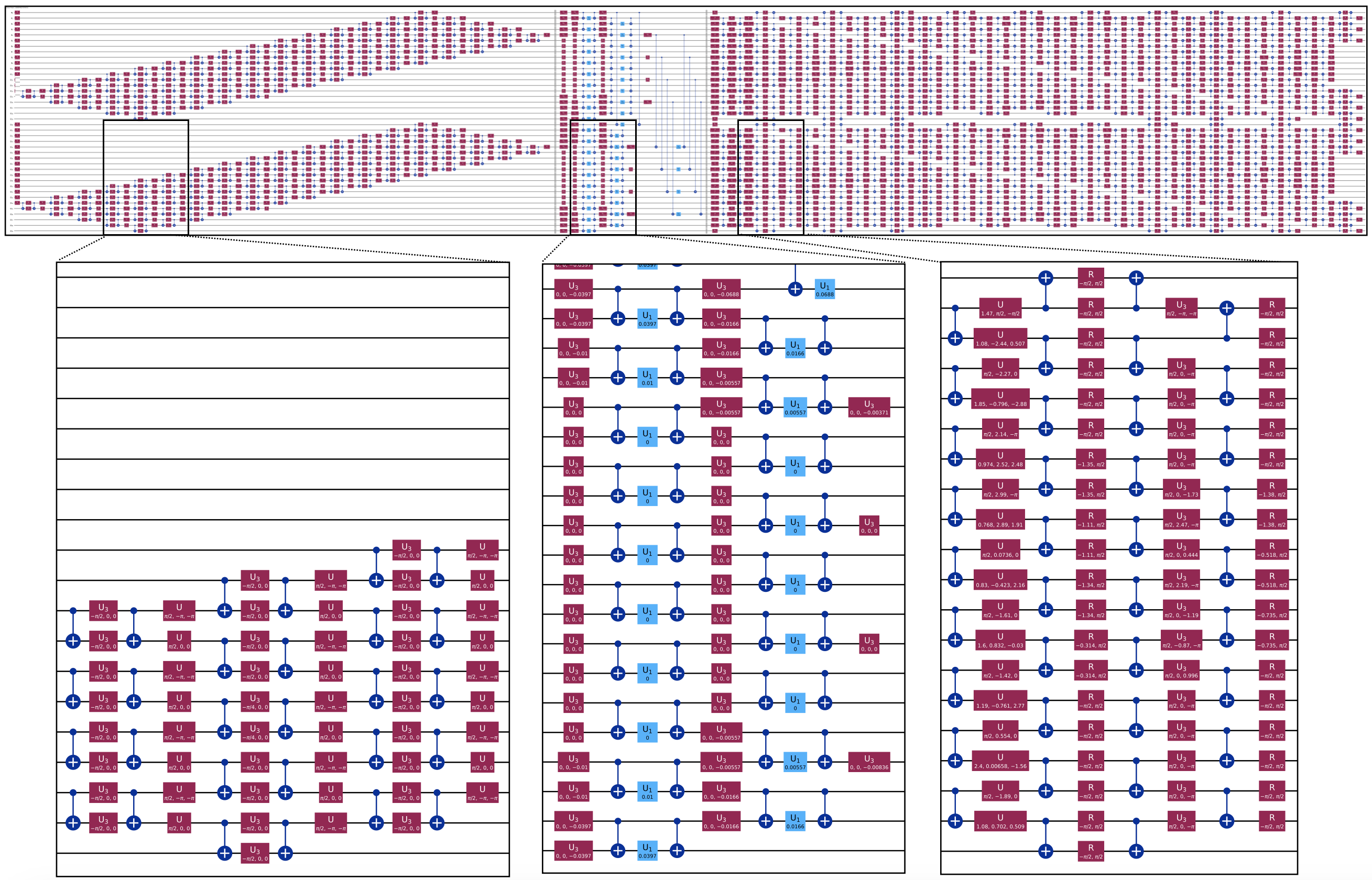}
    \caption{Top: LUCJ circuit used to simulate the ground state of the [2Fe-2S] cluster, compiled into single-qubit (red, purple blocks) and $\mathsf{CNOT}$ (light blue symbols) gates. Bottom: zoomed-out views of gates from the first orbital rotation, density-density interaction, and second orbital rotation (left to right).}
    \label{fig:circuit_fe2s2}
\end{figure*}

\begin{figure*}[t!]
    \centering
    \includegraphics[width=0.8\linewidth]{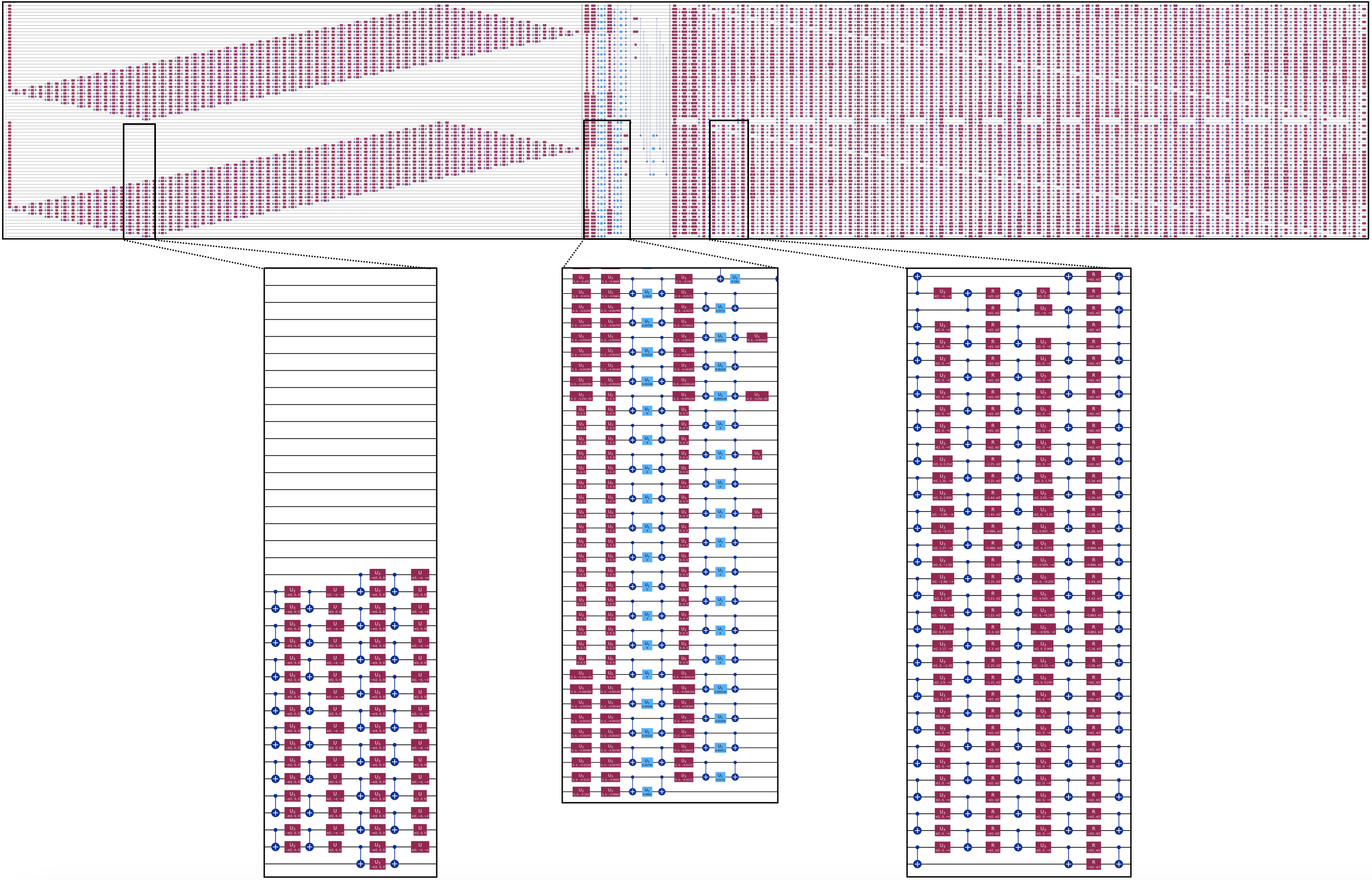}
    \caption{Top: LUCJ circuit used to simulate the ground state of the [4Fe-4S] cluster, compiled into single-qubit (red, purple blocks) and $\mathsf{CNOT}$ (light blue symbols) gates. Bottom: zoomed-out views of gates from the first orbital rotation, density-density interaction, and second orbital rotation (left to right).}
    \label{fig:circuit_fe4s4}
\end{figure*}

In Figures~\ref{fig:circuit_n2_631g}, \ref{fig:circuit_n2_ccpvdz}, ~\ref{fig:circuit_fe2s2}, and ~\ref{fig:circuit_fe4s4}, we show the specific LUCJ circuits considered in this work. Each circuit is compiled into single-qubit 
\begin{equation}
\mathsf{U}_3(\theta,\phi,\lambda) = 
\left( 
\begin{array}{rr}
\cos(\theta/2) &  -e^{i\lambda} \sin(\theta/2) \\
e^{i\phi}  \cos(\theta/2) & e^{i (\phi+\lambda)} \cos(\theta/2) \\
\end{array} \right)
\end{equation}
and two-qubit $\mathsf{CNOT}$ gates. For readability, the displayed circuits contain barriers separating orbital rotations and density-density interactions, which we removed in the final compilation to \device{nazca}/\device{torino} to reduce circuit depth. It is worth noting that the leftmost part of the quantum circuit does not implement the Bogolyubov transformation $\exp(\hat{K}_1)$, but prepares the Slater Determinant $\exp(\hat{K}_1) | \bts_\rhf \rangle$. In other words, it reproduces the action of $\exp(\hat{K}_1)$ on a specific Slater determinant, which allows for a drastic reduction in circuit size without information loss. Subsequent orbital rotations are not amenable to such a circuit simplification.

\subsubsection{Effect of the Jastrow terms in the accuracy}
\label{Sec: effect of jastrow}

In this section we study the impact of the two-body terms, $\exp(i \hat{J}_\mu)$, of the LUCJ ansatz in the performance of SQD. We compare the case where the two-body parameters are zero, ${J_{p\alpha, p\beta} = J_{p\alpha, p\alpha} = 0}$, against the most general case where they can take arbitrary real values. Without the density-density interactions, the remaining circuit is a free-fermion evolution, i.e. the wavefunction amplitudes remain those of a single Slater determinant~\cite{Thouless1960}, so it can be efficiently simulated by a classical computer. 

We present a numerical study of that emulates the procedure without taking into account the effect of noise. Thus, there is no configuration recovery. For the quantum circuit simulations, we used the software library \texttt{ffsim}~\cite{ffsim}.  \texttt{ffsim} performs the state vector simulation within a subspace of fixed particle number and total $\hat{z}$ component of the spin, making the simulations much more efficient than a generic quantum circuit simulator.

Three circuits are considered: LUCJ($L = 1$), LUCJ($L = 4$), and the LUCJ circuit without two-body terms, which we refer to as ``Determinant''. We study the accuracy of the estimator in the dissociation of \nitrogen (6-31G). For the LUCJ circuits, we consider those with heavy-hex-like connectivity in the density-density interactions; see Eq.~\eqref{eq: LUCJ heavy-hex connectivity}. The circuits are optimized to minimize the estimator energy, as described in Sec.~\ref{Sec: optimization of S-CORE} using the COBYLA~\cite{cobyla} optimizer. At each optimization step we consider $K = 1$ batches of samples obtained from the set of sampled configurations  $\mathcal{X}$, where $|\mathcal{X}| = 10^7$. The value of $d$ at each point in the dissociation curve is chosen to coincide with the number of configurations resulting from a converged HCI calculation, carried out on \texttt{PySCF}, in the same system.

\begin{figure*}
    \centering
    \includegraphics[width=1\linewidth]{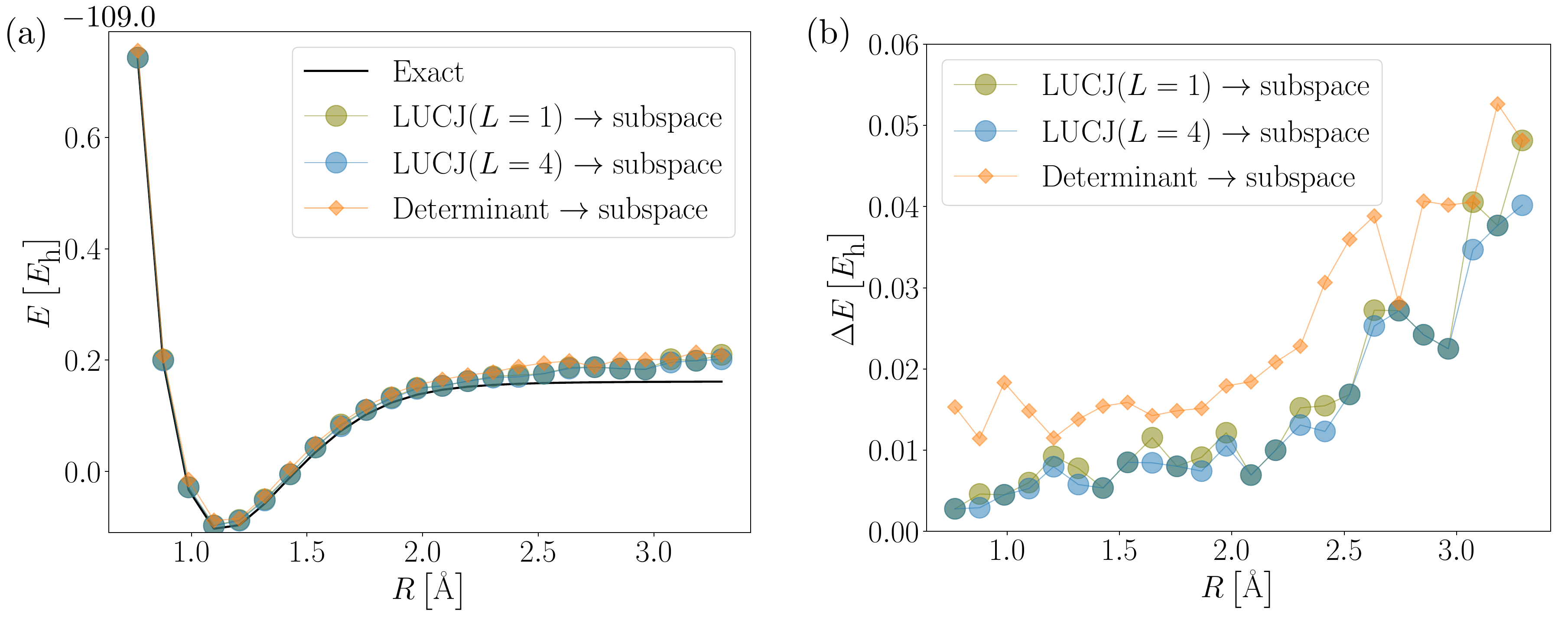}
    \caption{Numerical and noiseless study of the accuracy of the SCI-based HPC  quantum estimator obtained from different levels of approximation of the LUCJ circuit and samples obtained from an Slater determinant (limit when the sampling from the LUCJ circuit can be efficiently simulated). The system considered is the \nitrogen dissociation in the 6-31G basis. The circuits are optimized to minimize the estimator energy. In the legend, $L$ refers to the number of layers of the LUCJ ansatz. \textbf{(a)} ground-state energy as a function of the bond length. \textbf{(b)} Difference between the exact diagonalization energy and the SQD energy as a function of the bond length.}
    \label{fig:N2_LUCJ_vs_determinant}
\end{figure*}

Fig.~\ref{fig:N2_LUCJ_vs_determinant} (a) shows the potential energy surface obtained from SQD run on configurations  generated by LUCJ($L = 1$), LUCJ($L = 4$) and the determinant circuits. Panel (b) in Fig.~\ref{fig:N2_LUCJ_vs_determinant} shows the energy error from the three different circuits. For most bond lengths, the error of SQD is largest for the determinant circuit, showing that the density-density interactions play a crucial role for the generation of relevant electronic configurations. This claim is also supported by the decrease of the error (on average) when increasing the number of LUCJ layers.  We note that the optimization of the LUCJ circuits takes more iterations to converge than the optimization of the determinant circuits.

\subsection{Quantum and classical computational resources}

\subsubsection{processor layouts}

This section provides technical details about the experiments carried out on the quantum processors. The circuits that we use come from the truncation of two layers of the LUCJ circuits (as described in Sec.~\ref{secS:quantum circuits}). For the quantum runs, the circuit parameters are not optimized and instead are obtained from a classical CCSD calculation, as described in Sec.~\ref{secS:quantum circuits}. Note that the circuits themselves are still classically challenging to simulate. 

\begin{figure*}
    \centering
    \includegraphics[width=1\linewidth]{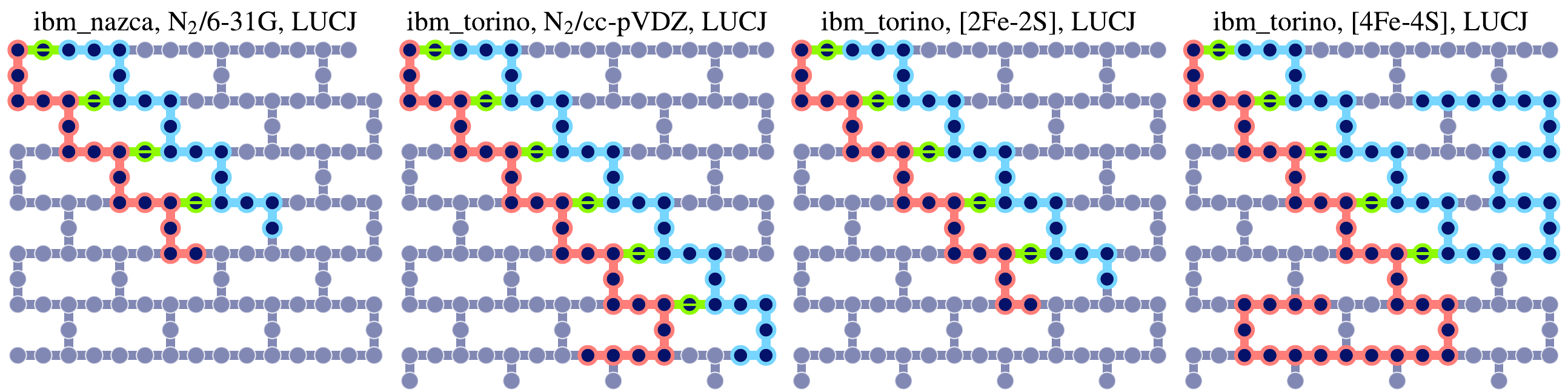}
    \caption{Left to right: schematics of the processors used to carry out experiments for the N$_2$ molecule with 6-31G basis, the N$_2$ molecule with cc-pVDZ basis, the [2Fe-2S] cluster, and the [4Fe-4S] cluster, using local unitary cluster Jastrow (LUCJ) quantum circuits. Qubits used in the calculation are shown in red (for qubits associated with $\alpha$ spin-orbitals), blue for qubits associated with $\beta$ spin-orbitals), and green (for auxiliary qubits).}
    \label{figS_N2_circiuts}
\end{figure*}

Table~\ref{tab:hardware details} lists the different molecular species, together with the number of qubits used to encode the $|\Psi\rangle$ that generates samples of electronic configurations. The same table also provides details about the total circuit depths, and numbers of one- and two-qubit gates, as well as the IBM quantum processor that was used to run the experiments. The specific qubit layouts are also provided in Table~\ref{tab:hardware details} and depicted in Fig.~\ref{figS_N2_circiuts}.

In the experiments that study the dissociation of \nitrogen, the number of measurement outcomes collected is $|\tilde{\mathcal{X}}| =$ \scientific{1}{5} and \scientific{9.8304}{4} per point in the dissociation curve at 6-31G and cc-pVDZ level of theory respectively. In the experiments to study the ground-state properties of [2Fe-2S] and [4Fe-4S], the number of measurement outcomes collected is $|\tilde{\mathcal{X}}| =$\scientific{2.4576}{6}. 
Out of the total number of measurement outcomes, a fraction of the configurations live in the subspace of the Fock space with the correct particle number, defined as
\begin{equation}
p_N^{\mathrm{hw}} = \frac{1}{N_s} \sum_{\ell=1}^{N_s} 
\delta_{N_{\bts_\ell\alpha}, N_\alpha} 
\delta_{N_{\bts_\ell\beta}, N_\beta} 
\;,\;
N_s = |\tilde{\mathcal{X}}|
\;.
\end{equation}
Table~\ref{tab:hardware details} shows the 95\% confidence interval for that fraction. Recall that the measurement outcomes that correspond to configurations with the correct particle number are used in the setup phase of the configuration recovery procedure. Table~\ref{tab:hardware details} also shows the fraction of sampled configurations  with the correct particle number if samples were collected from the uniform distribution over configurations  of length $M$,
\begin{equation}
p_N^{\mathrm{unif}} = \binom{\nmo}{N_\alpha} \binom{\nmo}{N_\beta} 2^{-2\nmo}
\;.
\end{equation}
As shown in the Table, the two probabilities are statistically distinguishable for all the experiments we carried out.
Our experiments used twirled readout error mitigation (ROEM) \cite{nation2021scalable} to mitigate errors arising from qubit measurement, and dynamical decoupling (DD) \cite{viola1998dynamical,kofman2001universal,biercuk2009optimized,niu2022effects} to mitigate errors arising from quantum gates. We employed the implementation of ROEM and DD available on the \texttt{Runtime} library of Qiskit~\cite{Qiskit}, through the \texttt{Sampler} primitive.
DD is implemented by sequences of $X$ control pulses, whose effect is to protect qubits from decoherence due to low-frequency system-environment coupling. Here, we applied sequences of two $X$ pulses (as in  Ramsey echo experiments) to idle qubits.

\begin{landscape}
\begin{table*}
\centering
\begin{tabular}{cccccccc}
\hline
\hline
\footnotesize
system & \footnotesize q (JW, Tot.) & \footnotesize ($d,\mathsf{CNOT},\mathsf{u})$ & \footnotesize device & \footnotesize layout & \footnotesize $p_N^{\mathrm{hw}}$ [95$\%$ c.i.] & \footnotesize $p_N^{\mathrm{unif}}$ & \footnotesize $  \footnotesize |\tilde{\mathcal{X}}|$ \\
\hline
\footnotesize N$_2$, (10e,16o) & \footnotesize (32, 36) &\footnotesize (148,762,1408) & \footnotesize \device{nazca} &\footnotesize [0$\rightarrow$82]+[2$\rightarrow$73] & \footnotesize (0.0069,0.0071) & \footnotesize 0.0044 &  \footnotesize \scientific{1}{5} \\
\hline
\footnotesize N$_2$, (10e,26o) & \footnotesize (52, 58) & \footnotesize (223,1792,3412) & \footnotesize \device{torino} & \footnotesize [0$\rightarrow$121]+[2$\rightarrow$127] & \footnotesize (0.0016,0.0017) & \footnotesize \scientific{9.6}{-7} & \footnotesize \scientific{9.8304}{4} \\
\hline
\footnotesize [2Fe-2S], (30e,20o) & \footnotesize (40, 45) & \footnotesize (173,1100,2070) & \footnotesize \device{torino} & \footnotesize [0$\rightarrow$104]+[2$\rightarrow$94] & \footnotesize (0.0044,0.0045) & \footnotesize 0.00022 &\footnotesize \scientific{2.4576}{6} \\
\hline
\footnotesize [4Fe-4S], (54e,36o) & \footnotesize (72, 77) & \footnotesize (301,3590,6980) & \footnotesize \device{torino} & \footnotesize [0$\rightarrow$100]+[2$\rightarrow$28] & \footnotesize (\scientific{4.69}{-5},\scientific{6.61}{-5}) & \footnotesize \scientific{1.88}{-6}  & \footnotesize \scientific{2.4576}{6} \\
\hline
\hline
\end{tabular}
\caption{Details of hardware simulations. For each active space (column 1 from the left), we list the details of the LUCJ quantum circuit executed on hardware, specifically: its number of qubits (q) both for the Jordan-Wigner encoding (JW) and the total number of qubits used including auxiliary ones (Tot.) (column 2) and count of quantum operations (circuit depth $d$, number of $\mathsf{CNOT}$ and single-qubit $\mathsf{u}$ gates, column 3), the device used (column 4) and the qubit layout chosen (column 5, where $[a\rightarrow b]$ indicates the shortest path between qubits $a$ and $b$ in the device topology, also visible in Fig.~\ref{figS_N2_circiuts}). In columns 6 and 7 we report a 95\% confidence interval for the fraction $p_N^{\mathrm{hw}}$ of configurations with the correct particle number and the corresponding value for configurations  with uniform probability distribution (see the text for the definition of these quantities). In column 8, we list the number of noisy configurations sampled from quantum hardware.}
\label{tab:hardware details}
\end{table*}
\end{landscape}

\subsubsection{Classical resource details}
The SCI-based eigenstate solver used by SQD is the one implemented in \texttt{PySCF}~\cite{sun2018pyscf,sun2020recent} for the \nitrogen (6-31G, cc-pVDZ), and [2Fe-2S] systems. For the [4Fe-4S] molecule we use the SCI-based eigenstate solver implemented in \texttt{DICE}~\cite{sharma2017semistochastic, holmes2016heat} because of its ability to run across multiple computer nodes.

For the [4Fe-4S] experiments, the eigenstate solver can be distributed across a number of classical nodes. The workflow for [4Fe-4S] is run on the supercomputer Fugaku. The supercomputer Fugaku has a total of 158,976 nodes with Armv8.2-A SVE 512 bit architecture. Each node has 48 compute  cores and 32 GiB of memory. The largest calculation in this work uses 6400 nodes of the Fugaku supercomputer running in parallel.

\section{Additional experimental results}
This section contains the results from the experiments that are run on the quantum processors, as described in the previous section. We show a collection of investigations that are complementary to those shown in the main text. 

We first visualize the prediction individual wave function components obtained by SQD with configuration recovery in a system size where we have access to the exact ground state wave function amplitudes (\nitrogen in the 6-31G basis). 

We then study the scaling and runtime of the estimator. In particular, we study the effect and cost that more intensive classical resources have on the quality of the estimator. We choose the \nitrogen (cc-pVDZ) and [2Fe-2S] molecules for this analysis.

We also study the effect of orbital optimizations in conjunction with the HPC quantum eigensolver in the quality of the predictions. We choose \nitrogen in the cc-pVDZ basis as the case of study. 

The results reported in Figure 4 of the main text, for the [2Fe-2S] cluster show the resolution of three different eigenstates. In this section we  provide additional information about their nature. In particular, we study the average orbital occupancy  for the three eigenstates in the MO and localized bases. 

Lastly, we investigate the amount of signal that can be extracted and amplified from our experiments. For this study we consider the dissociation of  \nitrogen (cc-pVDZ) as well as the study of the low-energy spectrum of the [2Fe-2S] molecule. The same study for the [4Fe-4S] molecule is in the main text.

\subsection{Visualization of the wave function amplitudes}

The aim of this subsection  is to show the accuracy of individual wavefunction amplitudes obtained with SQD (with configuration recovery). Configuration recovery is run on measurement outcomes obtained from a quantum processor. We consider the dissociation of \nitrogen in a basis set that is amenable to exact diagonalization techniques but larger than a minimal basis set: the 6-31G basis. The exact diagonalization wavefunction serves as the reference to asses the accuracy of the amplitudes produced by the estimator. For each point of the dissociation curve we collect $|\tilde{\mathcal{X}}| = 100 \cdot 10^3$ measurement outcomes and run the configuration recovery procedure with $K = 10$ batches of samples. The configuration recovery method is run for ten sefl-consistent iterations. The value of the subspace dimension for the SCI eigenstate solver is $d = \textrm{4M}$. For the potential energy surface we report $\min_k \left(E^{(k)} \right)$. We show the amplitudes of the wavefunction from the batch of configurations whose energy is the lowest.

\begin{figure*}
    \centering
    \includegraphics[width=1\linewidth]{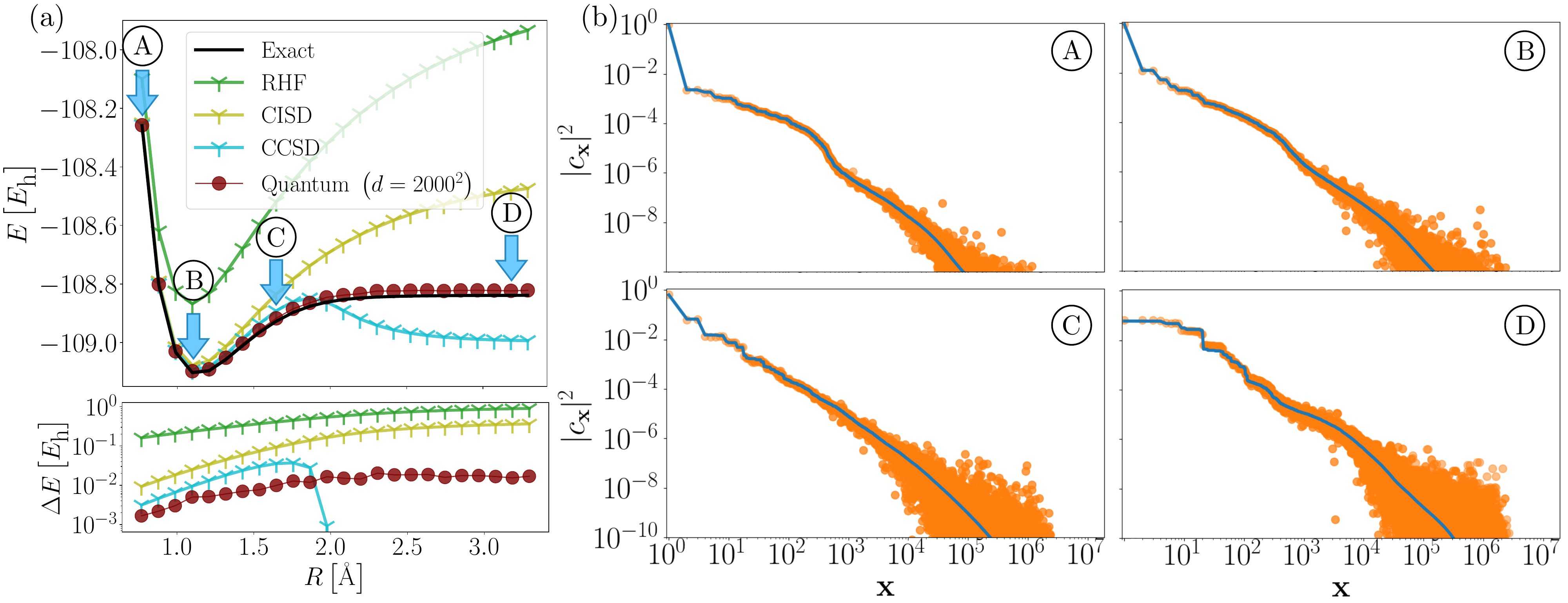}
    \caption{Visualization of the wavefunctions produced by SQD in the dissociation of  \nitrogen (6-31G), and the comparison against the exact wavefunctions. \textbf{(a)} Shows the energy and energy error as a function of the bond length for the estimator with configuration recovery run from measurement outcomes from the quantum processor. Classical approximate methods are shown for reference. \textbf{(b)} Shows the wavefunction amplitudes $|c_{\bts}|^2$ for all possible electronic configurations $\bts$. The configurations in the horizontal axis are ordered according to the magnitude of the amplitudes in the exact wavefunction. The blue solid lines correspond to the exact ground state amplitudes while the orange dots correspond to the wavefunction amplitudes obtained from the estimator with configuration recovery. Different points in the dissociation curve are shown and labelled A, B, C, D, following the same encoding as in panel \textbf{(a)}. }
    \label{fig:N2_wave_function}
\end{figure*}

Panel (a) in Fig.~\ref{fig:N2_wave_function} shows the potential energy surface of the molecule obtained by the estimator as well as the error in the estimation of the ground-state energy, as a function of the bond length. The error in the estimation of the ground-state energy remains below $10 \; mE_\textrm{h}$ for all points in the dissociation curve, thus showing good agreement with the exact results.

Panel (b) in Fig.~\ref{fig:N2_wave_function} shows a comparison between the exact ground state wavefunction amplitudes and the wavefunction amplitudes obtained by the estimator, for different points in the dissociation curve. The agreement is exceptional for the larger wavefunction amplitudes and slowly deteriorates in the description of the tails of the wavefunction. We also observe that for larger bond lengths, where higher levels of static correlation are present, the agreement in the tails is worse than for bond lengths closer to the equilibrium geometry. This is to be expected for two reasons. First, for larger bond lengths the wavefunction amplitudes are less concentrated than close to equilibrium. As discussed in Sec.~\ref{Sec: wavefunction concentraction}, the accuracy of the estimator using a SCI-based eigenstate solver deteriorates as the wavefunction becomes less concentrated. The second reason is that for bond lengths close to equilibrium, the wavefunction has a strong mean field character, but not for larger bond lengths. This directly impacts the shape of the reference spin-orbital occupancy $\vett{n}$ that is used to recover configurations. When the wavefunction  is mean-field dominated, $\vett{n}$ takes the shape of a sharp step function shape with all spin-orbital average occupancies close to 0 or 1. One possible effect of static correlations is to soften the step function, moving average occupancies of some spin-orbitals away from 0 or 1. The recovery of configurations used in this work is more effective when the components of $\vett{n}$ are close to 0 or 1.

In summary, SQD with configuration recovery produces accurate representations of the ground state wavefunction (on experimental data), even at an individual amplitude level. Less concentrated wavefunctions tend to challenge the procedure more, both because the accuracy of the SCI solver is decreased, and because the configuration recovery becomes less effective due to softer profiles of $\vett{n}$.

\subsection{Scaling and runtime}

We begin this section by studying the improvement of the accuracy of SQD with the number $d$ of determinants used for the subspace expansion and diagonalization. It is clear that larger $d$ yields more expressivity in the variational wavefunction, since we are increasing the possible size of the support of the state. It is also clear that increasing $d$ increases the HPC quantum estimator runtime since both the projection and diagonalization processes become more computationally intensive. To show the improvement of the accuracy with $d$ we consider the dissociation of \nitrogen (cc-pVDZ), where the measurement outcomes are generated by the quantum processor.  The estimator is applied to $10^5$ measurement outcomes and we take $K = 10$ batches of samples and report $\min_k\left(E^{(k)} \right)$ for the potential energy surface for different values of $d$. Ten self-consistent iterations are considered for the recovery of configurations.

\begin{figure*}
    \centering
    \includegraphics[width=.5\linewidth]{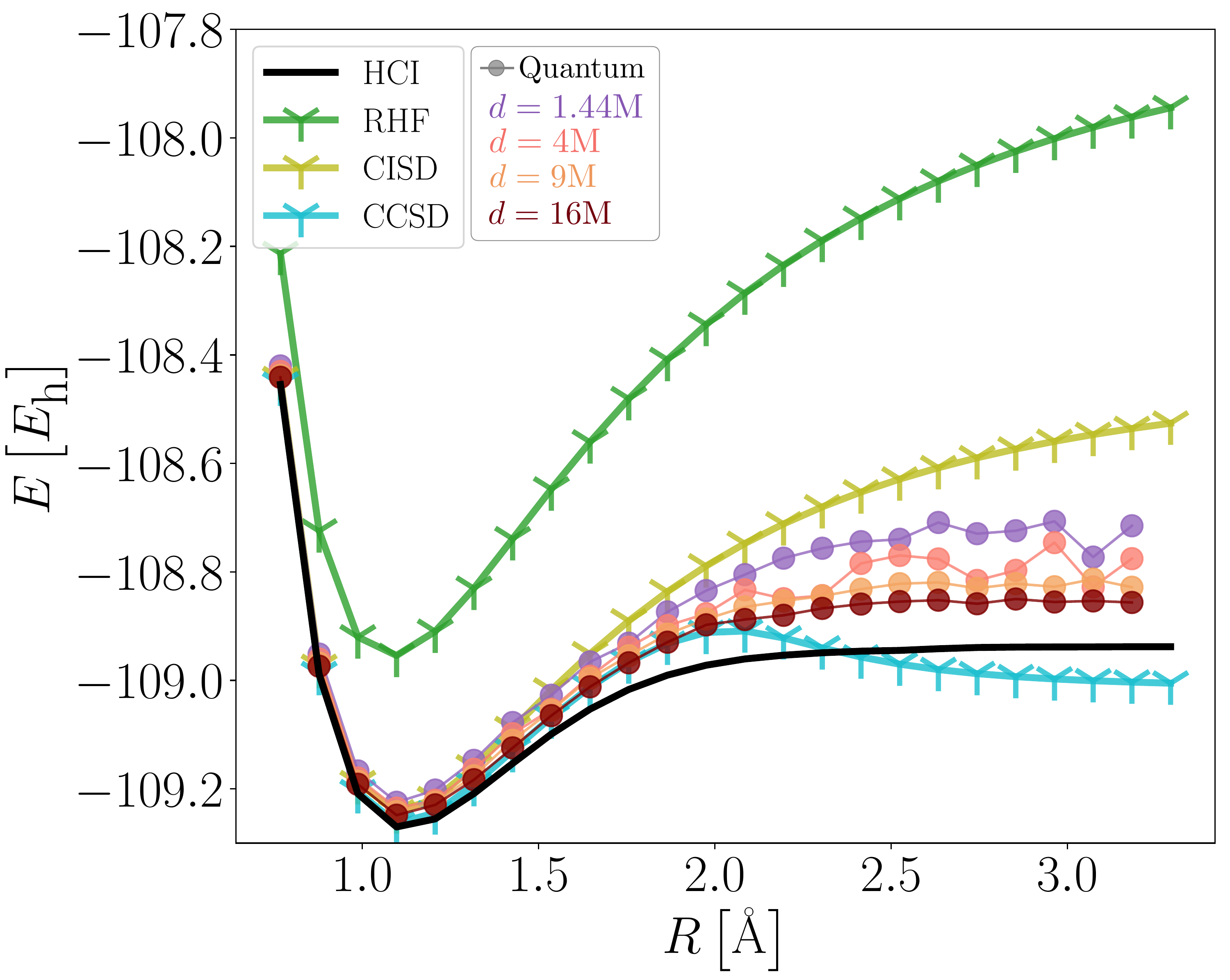}
    \caption{Improvement of the estimator accuracy (with configuration recovery) by increasing the number of determinants for the subspace expansion and diagonalization. Potential energy surface of the \nitrogen molecule (cc-pVDZ) obtained from the HPC quantum estimator estimator using measurements outcomes from the quantum processor. The energies reported correspond to the lowest energy amongst the K batches of configurations, i.e. ${\min_k \left(E^{(k)}\right)}$. Different subspace dimensions $d$ are considered, as indicated in the legend. Different classical methods are shown for reference.}
    \label{figS:N2_improvement_with_d}
\end{figure*}

Fig.~\ref{figS:N2_improvement_with_d} shows the potential energy surfaces obtained by the estimator with different values of $d$, the number of determinants used for the subspace expansion and diagonalization. Increasing the value of $d$ improves the results both qualitatively and quantitatively. For the smaller values of $d$, the potential energy surface shows unphysical oscillations when the bond length is large. For larger values of $d$, not only the energy values are lower, but also the oscillations in the energy as a function of the bond length are significantly decreased in amplitude.
Improvement of the ground state approximation with an increasing amount of computational resources is also observed in the energy-variance analyses of the main text (Fig. 4), and the supplementary materials (Figs.~\ref{fig:Fe2S2_SCORE-vs-UniformRP} and~\ref{fig:Fe2S2_SCORE-vs-Uniform}).

\begin{figure*}
    \centering
    \includegraphics[width=1\linewidth]{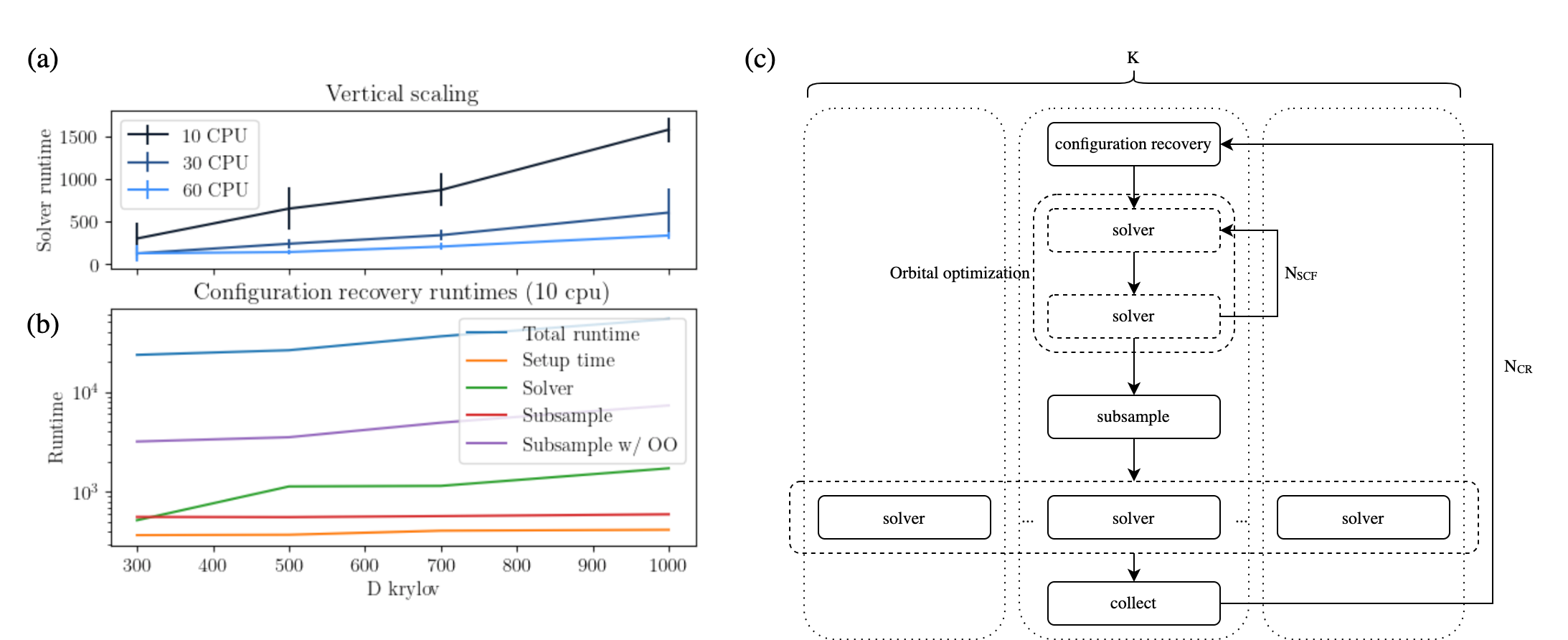}
    \caption{Classical-resource scaling of the HPC estimator with configuration recovery and orbital optimizations. \textbf{(a)} Vertical scaling of the solver (increase number of CPUs per node) for [2Fe-2S]. \textbf{(b)} Relative runtimes of blocks of the estimator for [2Fe-2S]. \textbf{(c)} Horizontal scaling of the estimator. Total runtime without horizontal scaling is $\mathrm{S} \cdot (N_\textrm{SCF} + K) \cdot N_\textrm{CR}$. The total runtime with horizontal scaling is $\mathrm{S} \cdot (N_\textrm{SCF} + 1) \cdot N_\textrm{CR}$, where $\mathrm{S}$ is the runtime of the solver, $\mathrm{K}$ is the number of nodes in a cluster, $\mathrm{N_\textrm{CR}}$ is the number of iterations, and $\mathrm{N_\textrm{SCF}}$ is the number of orbital optimization steps. 
    }
    \label{fig:Scaling}
\end{figure*}

We analyze the vertical and horizontal scaling of the workflow, in the case where the subspace diagonalization is run on a single computer node, using the solver provided in the \texttt{PySCF} library. The [2Fe-2S] iron-sulfur cluster is used as the system of reference. Fig.~\ref{fig:Scaling} shows vertical scaling (increasing resources within a single node) of the eigenstate solver, and relative runtimes and schematics of horizontal scaling (adding nodes to cluster) of different blocks of the estimator (with configuration recovery). The eigenstate solver contributes the most to total runtime of the algorithm. Vertical scaling of the eigenstate solver shows noticable decrease in runtime up to a point. Experiments showed that for [2Fe-2S], 30+ CPUs per node is the optimal configuration. A good strategy for classical scaling would be to find the optimal vertical configuration and then scale horizontally for better utilization of resources. 

To conclude this section, we study the dependence of the runtime and accuracy of the estimator in the [4Fe-4S] system as a function of the amount of classical compute. The classical compute can be increased by increasing the size of the subspace dimension $d$, or by collecting more batches of configurations $K$. The \texttt{DICE} library allows to distribute the subspace diagonalization across a number of independent computer nodes. It is expected that more classical nodes will yield faster runtimes per diagonalization. For this experiment the Heron processor Montecarlo is used, with $\left|\mathcal{X} \right| = 3163742$ collected measurement outcomes, and $p_N^{hw} = 5.4 \cdot 10^{-4}$. 

\begin{figure*}
    \centering
    \includegraphics[width=1\linewidth]{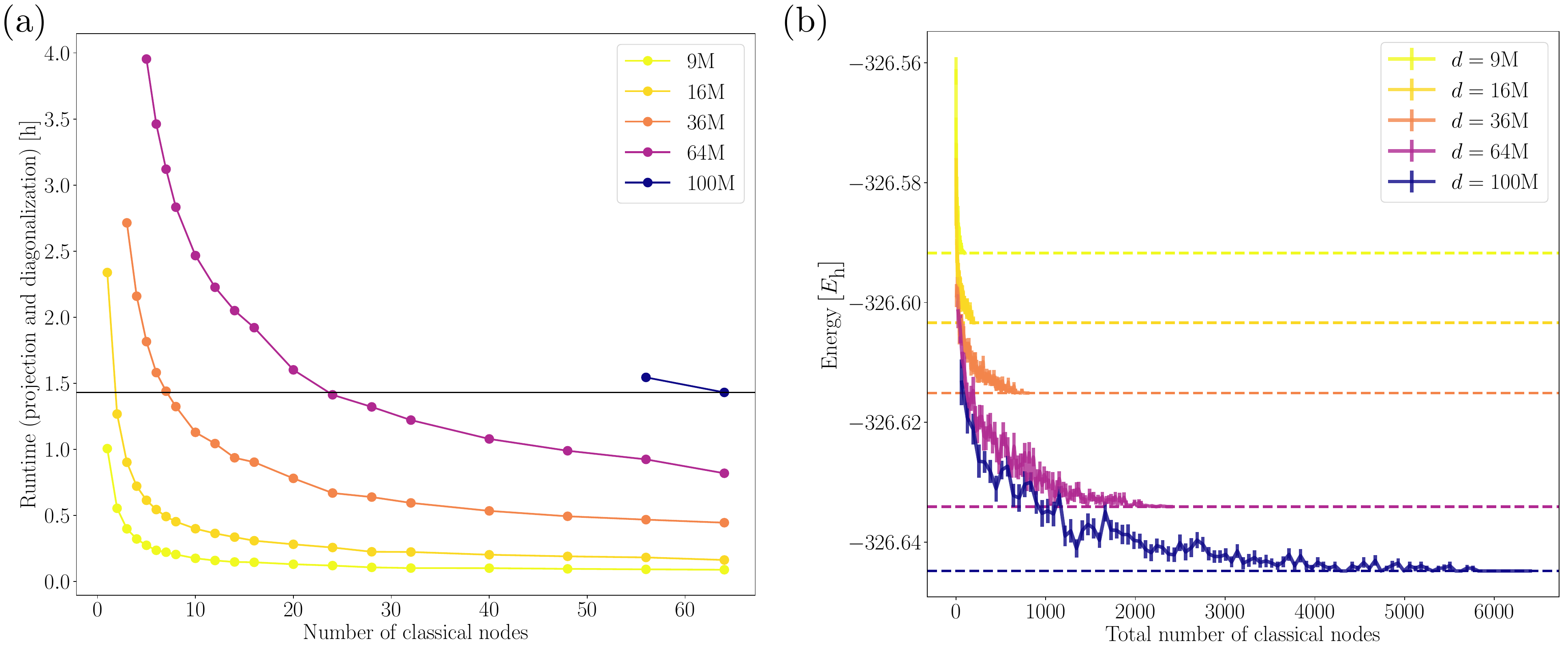}
    \caption{Accuracy and runtime of SQD for the [4Fe-4S] iron-sulfur cluster. \textbf{(a)} Runtime of a single subspace projection and diagonalization as a function of the number of nodes that are used to distribute the calculation. Different colors correspond to different subspace dimensions. The smallest number of nodes per curve corresponds to the limit where the calculation runs out of memory on a single node. The subspace projection and diagonalization are handled by the \texttt{DICE} library. The horizontal line shows a fixed runtime of 1.44 hours for the different subspace sizes. \textbf{(b)} Statistical convergence of SQD as the number of batches of samples $K$ is increased. The vertical axis shows the the average minimum energy obtained by a fixed number of batches of samples. Each batch of samples is run on fixed number of nodes and the diagonalization for the different batches are run in parallel. The number of nodes for the diagonalization of a single batch is chosen to obtain a fixed runtime of 1.44 hours for all of the different subspace dimensions $d$. The horizontal axis shows the total number of nodes corresponding to different numbers of batches. The largest number of batches corresponds to $K = 100$. }
    \label{figS:runtime Fe4S4}
\end{figure*}

Fig.~\ref{figS:runtime Fe4S4} (a) shows the runtime of the diagonalization of a single batch of configurations as a function of the number of classical nodes. Different subspace dimensions are considered. As expected, with an increased number of nodes, the runtime of the diagonalization decreases. After a certain number of nodes, the speedup saturates, due to the overhead in inter-node communications. This panel allows us to identify the number of classical nodes necessary to have a fixed runtime for the diagonalization for different subspace dimensions. To study improvement of the accuracy of the estimator with the number of batches of configurations $K$, we run the workflow with $K = 100$. From the $K = 100$, batches, we subsample the energy of smaller groups of batches and collect the minimum energy for each subsample. The subsampled minimum energies are averaged. Different sizes of the smaller groups of batches are considered until reaching the group of $K = 100$ batches of configurations. Panel (b) in Fig.~\ref{figS:runtime Fe4S4} shows the average of the minimum energies as a function of the total number of nodes for different values of $d$, such that the runtime of each digonalization is fixed to 1.44 hours. The total number of classical nodes is given by the product of  the size of the groups of batches of configurations and the number of nodes required to obtain a runtime of 1.44 hours per subspace projection and diagonalization.

\subsection{Effect of orbital optimizations in the dissociation of \nitrogen (cc-pVDZ)}\label{Sec:orbital optimizations N2}

We apply orbital optimizations (OO) to improve the accuracy of SQD,  in the study of the dissociation of \nitrogen (cc-pVDZ). The estimator with configuration recovery and its orbital-optimized counterpart are applied to the same set of noisy measurement outcomes $\tilde{\mathcal{X}}$ obtained from the quantum processor. We use $K = 10$ batches of samples in each case and $N_\textrm{SCF} = 10$ iterations of the alternation between the optimization of $\kappa$ and running of the eigenstate solver to update $c_\bts^{(k)}$. After each $c_\bts^{(k)}$ update, $5000$ iterations of gradient descent with momentum are used to optimize $\kappa$. Before applying orbital optimizations, the configuration recovery procedure is run for 10 iterations on the reference basis (molecular orbitals). Various sizes $d$ of the batches are considered. 

\begin{figure*}
    \centering
    \includegraphics[width=1\linewidth]{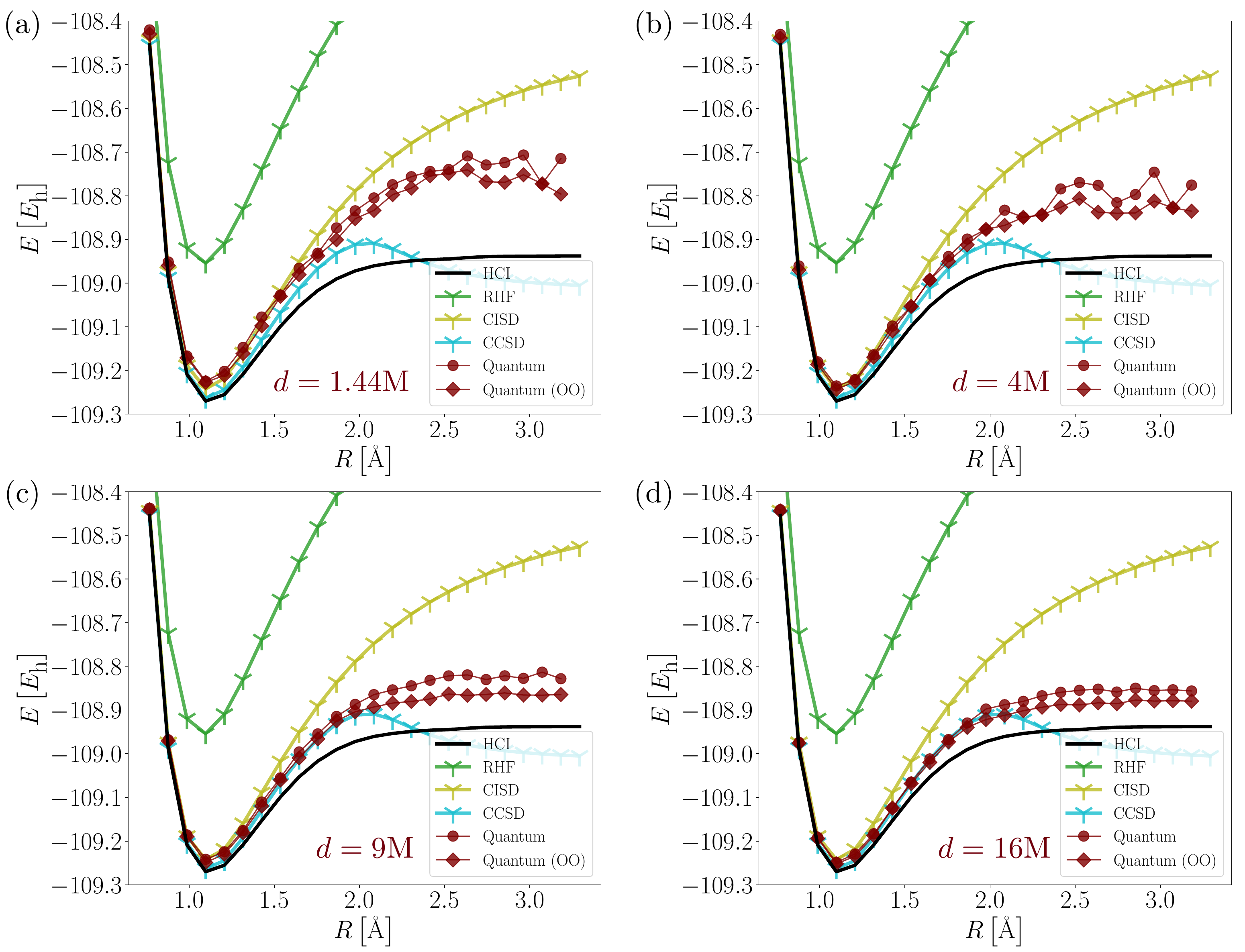}
    \caption{Effect of orbital optimizations (OO) in the performance of SQD with configuration recovery in the dissociation of \nitrogen (cc-pVDZ). The ground-state energy is shown as a function of the bondlength. The estimator is run on measurement outcomes from a quantum processor. The energies shown correspond to the lowest energy amongst the $K$ batches of configurations, i.e. ${\min_k \left(E^{(k)}\right)}$. Energies from different classical methods are shown for reference, as indicated in the legend. Results are shown for the estimator run with different numbers of configurations $d$, as indicated in each panel \textbf{(a)}-\textbf{(d)}.}
    \label{fig:N2_MO-vs-OO}
\end{figure*}

Fig.~\ref{fig:N2_MO-vs-OO} compares the ground-state energy obtained by the estimator without and with orbital optimizations. We observe that close to the equilibrium bond length the effect of the orbital optimizations is negligible. However, upon dissociation, the estimator with orbital optimizations obtains noticeably lower ground state energies. For smaller values of $d$, the estimator without orbital optimizations shows unphysical oscillations in the potential energy surface. The orbital optimizations decrease the amplitude of those oscillations. 

In summary, orbital optimizations allow finding the single-particle basis in which the estimator is most accurate, thus improving the quality of its predictions. Furthermore, the optimization of the circuit parameters $\theta$ would allow the circuit to respond to the change of basis to produce electronic configurations better suited for the new basis. As mentioned earlier, the optimization of the circuit parameters will be the subject of follow-up projects. Note that the unrestricted optimization of $\kappa$ can break symmetries that may be desirable to preserve, so it is up to the user to decide whether to use orbital optimizations or not. 

\begin{figure*}
    \centering
    \includegraphics[width=.5\linewidth]{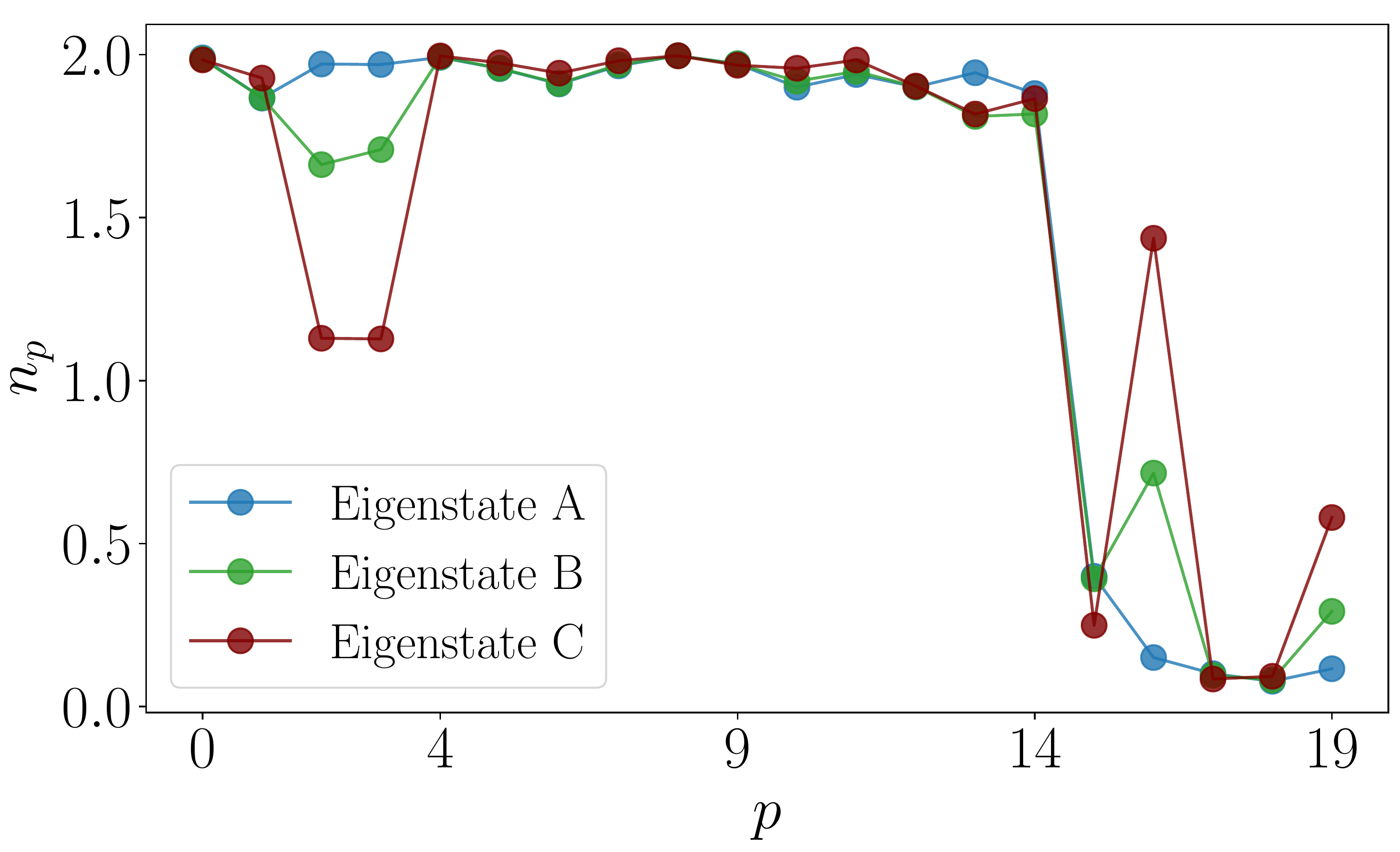}
    \caption{Orbital occupancy profile $\vett{n}$ for the [2Fe-2S], produced by SQD with configuration recovery. The increasing index $p$ labels molecular orbitals of increasing energy. The occupancy profile is shown for the three eigenstates identified in the energy-variance analysis (see Fig.~4 (b) in the main text).}
    \label{fig:Fe2S2_occupancies}
\end{figure*}
\begin{figure*}[t!]
    \centering
    \includegraphics[width=1\linewidth]{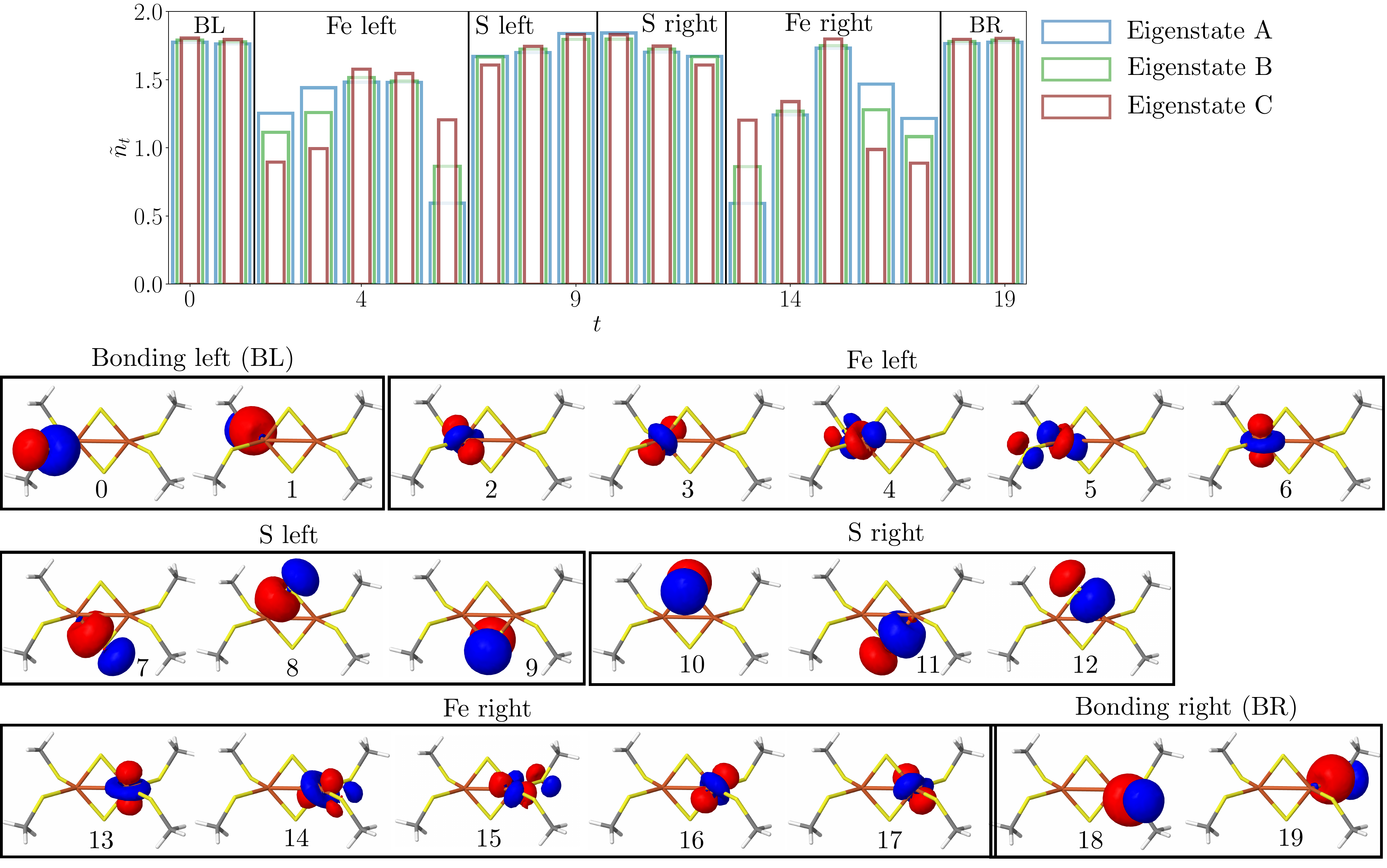}
    \caption{Orbital occupancy profile $\tilde{\vett{n}}$ for the [2Fe-2S], produced by SQD with configuration recovery. The increasing index $t$ labels localized orbitals as shown in the bottom of the panel. The occupancy profile is shown for the three eigenstates identified in the energy-variance analysis (see Fig.~4 (b) in the main text). The representation of the orbitals is obtained from the repository~\cite{FeSRepo} from Ref.~\cite{li2017spin}.}
    \label{fig:Fe2S2_occupancies_localized}
\end{figure*}

\subsection{Orbital occupation numbers for the low-energy spectrum of [2Fe-2S]}

The calculations reported in Figure 4 of the main text, for the [2Fe-2S] cluster, identified three distinct eigenstates that we labeled A, B, and C. 
In this section, we provide additional information about the nature of the three identified eigenstates. The purpose of this analysis is to illustrate the robust classification of the wavefunctions produced by our estimator into three families (A, B, and C) but also to shed light on the nature of the corresponding eigenstates and the ability of our method to accurately capture a multireference character in electronic eigenstates, pinpointing areas of possible future improvement.
In particular, we show the spatial orbital occupancy for the three eigenstates that SQD finds for [2Fe-2S] both in the MO basis and in a basis of spatially local orbitals.
The MO basis helps highlight deviations from the mean-field states or lack thereof, and the basis of localized orbitals (identifiable with e.g. Fe $3d$ and S $2p$ orbitals) helps characterize our method's ability to capture local antiferromagnetic correlations between transition-metal atoms.

The spatial orbital occupancy is given by $n_p = \sum_\sigma n_{p\sigma}$. Therefore $n_p\in[0, 2]$. For Eigenstate A, the occupancy comes from the estimator run with $d = \textrm{9M}$. For Eigenstates B and C, the occupancy comes from the estimator run with $d = \textrm{16M}$. We cannot compare the occupancy for the same value of $d$ since the description of eigenstates A and B require different numbers of configurations (see Sec.~\ref{sec:quantum signal 2Fe-2S} and Fig.~\ref{fig:Fe2S2_SCORE-vs-Uniform}). For Eigenstates B and C, we use the same number of determinants to make the comparison as even as possible. The number of batches of configurations is $K = 10$.

Fig.~\ref{fig:Fe2S2_occupancies} shows the $n_p$ profile for the three eigenstates identified from the energy-variance analysis, in the basis of MOs. The occupancy profile for Eigenstate A is the typical profile of an eigenstate with a strong mean-field character. The occupancy of the lowest-energy $N_\sigma = 15$ molecular orbitals is close to its maximum value, while the remaining orbitals have lower occupancy. The character of the occupancy profile for Eigenstate B is different from that of Eigenstate A. The occupancy of the third and fourth orbitals is depleted to give a more prevalent presence of electrons in orbitals $p = 16$ and $p = 19$. This effect is even more noticeable for eigenstate C, where the occupancy of the third and fourth orbitals is depleted by half and the occupancy of orbital $p = 16$ is now over the half-filling factor.

To further understand the nature of Eigenstates B and C, we study the occupancy in the basis of localized orbitals $\tilde{\vett{n}}$ (see Sec.~\ref{sec: molecules and active spaces} for details about the basis of localized orbitals).
The reported occupancies are constructed as follows: first, we obtain the one-body density matrix for the approximate ground state obtained from each batch of configurations, expressed in the basis of MOs: $\Gamma_{pq, \sigma}^{(k)}$ (see Eq.~\ref{eq: density matrices}). We then rotate the density matrix to the basis of localized orbitals with a similarity transformation denoted by $\Omega$,
\begin{equation}
    \tilde{\Gamma}_{tu, \sigma}^{(k)} = \Gamma_{pq, \sigma}^{(k)} \Omega_{pt} \Omega_{qu},
\end{equation}
where $\tilde{\Gamma}_{tu, \sigma}^{(k)}$ is the one-body density matrix in the basis of localized orbitals. The occupancy of localized orbital $t$ is given by
\begin{equation}
    \tilde{n}_t = \frac{1}{K} \sum_{k = 1}^K \sum_\sigma \tilde{\Gamma}_{tt, \sigma}^{(k)}.
\end{equation}
Fig.~\ref{fig:Fe2S2_occupancies_localized} shows a comparison of the value of $\tilde{\vett{n}}$ for Eigenstates A, B, and C. The reflection symmetry along the plane that separates the left and right sides of the molecule is broken for the three approximate eigenstates. This is highlighted by the observation that the occupancy of the equivalent orbitals for the two Fe atoms are not the same. We also observe that the occupancy of the bonding orbitals of the SCH$_3$ groups and the occupancy of the $3p$ orbitals of the S atoms is similar for the three eigenstates. The occupancy of the $3d$ orbitals of Fe has the largest differences among the three different eigenstates. The more notable discrepancy in the d-orbitals of Fe may be a consequence of the fact that the low-energy physics of this molecule is dominated by the antiferromagnetic coupling of the electrons in the Fe $3d$ orbitals~\cite{li2017spin, sharma2014low}.

\subsection{Quantum signal in the experiments}

In this section we assess the quality of the quantum signal for all our experiments, going beyond the [4Fe-4S] analysis presented in the main text.. Given the large circuit sizes of our experiments:
\begin{itemize}
    \item $[$2Fe-2S$]$: 40 qubits (not including auxiliary), 1100 two-qubit gates, 3170 total gates,
    \item \nitrogen: 52 qubits (not including auxiliary), 1792 two-qubit gates, 5204 total gates,
    \item $[$4Fe-4S$]$: 72 qubits (not including auxiliary), 3590 two-qubit gates, 10570 total gates,
\end{itemize}
one should verify  whether there is a useful signal coming out of the quantum circuits, comparing samples from $\tilde{P}_\Psi$ and uniform random samples. We first investigate the fraction of sampled configurations  that live in the correct particle sector $p_s^{\textrm{hw}}$ and compare it to the fraction that would be obtained if the samples came from the uniform distribution $p_s^\textrm{unif}$. The values for $p_s^\textrm{hw}$ alongside their 95\% confidence interval, and $p_s^\textrm{unif}$ are reported in Table~\ref{tab:hardware details}. For the 40-, 52-, and 72-qubit experiments, we observe that $p_s^\textrm{hw}$ is $\sim 20$, $\sim 10^3$, and $\sim 40$ times larger than $p_s^\textrm{unif}$ respectively. This provides a first indication that configurations  sampled from the quantum processor contain a signal that is distinguishable from white noise.

The second set of tests that we conduct is the comparison of the ground-state energy obtained by SQD with and without applying the configuration recovery procedure. Note that not applying configuration recovery makes the estimator equivalent to the QSCI framework~\cite{kanno2023qQSCI} . Having reasonable energies (lower than or in the vicinity of the RHF energy) without applying configuration recovery reveals that the set $\mathcal{X}_N$ already contains reasonable configurations to construct a non-random ground state.

The third set of tests involves comparing the performance of SQD with configuration recovery run on configurations  obtained from the quantum processor and samples obtained from the uniform distribution. We consider the uniform distribution over all possible configurations in the Fock space, as well as the uniform distribution over configurations with the correct particle number, as in Fig. 4 (c) in the main text~\footnote{When considering samples from the uniform distribution over electronic configurations on the correct particle sector, the configuration recovery method cannot be applied since it only recovers configurations with the wrong particle number.}. The results for these last two sets of tests are shown in the coming subsections for the 40-, 52- and 72-qubit experiments. 

\subsubsection{N$_2$ experiments}

\paragraph{Effect of configuration recovery on the accuracy of the HPC quantum estimator.--}
In this section, we compare the accuracy of the estimator without configuration recovery (applied to $\mathcal{X}_N \subset \tilde{\mathcal{X}}$) to the accuracy of the estimator with configuration recovery (applied to $\tilde{\mathcal{X}}$). As for all \nitrogen experiments, $|\tilde{\mathcal{X}}| = 100 \cdot 10^3$. From the experimental measurements, we observe that the average (over the bondlengths considered in this study) size of the $\mathcal{X}_N $ set is  $|\mathcal{X}_N| = 165$. The maximum number of determinants that can be extracted to perform the projection and diagonalization is $d^{hw}_N = 110 \textrm{K}$. Since typical values of $d$ used for the \nitrogen molecule exceed $d^{hw}_N$ we perform the projection and diagonalization using all the configurations obtained from $\mathcal{X}_N$ (number of batches of configurations is $K =1$). For the estimator run with configuration recovery  we use $K = 10$. In both cases, we report the $\min_k\left(E^{(k)} \right)$ energy for the potential energy surface. When running the estimator with configuration recovery we use 10 self-consistent recovery iterations.

\begin{figure*}
    \centering
    \includegraphics[width=.5\linewidth]{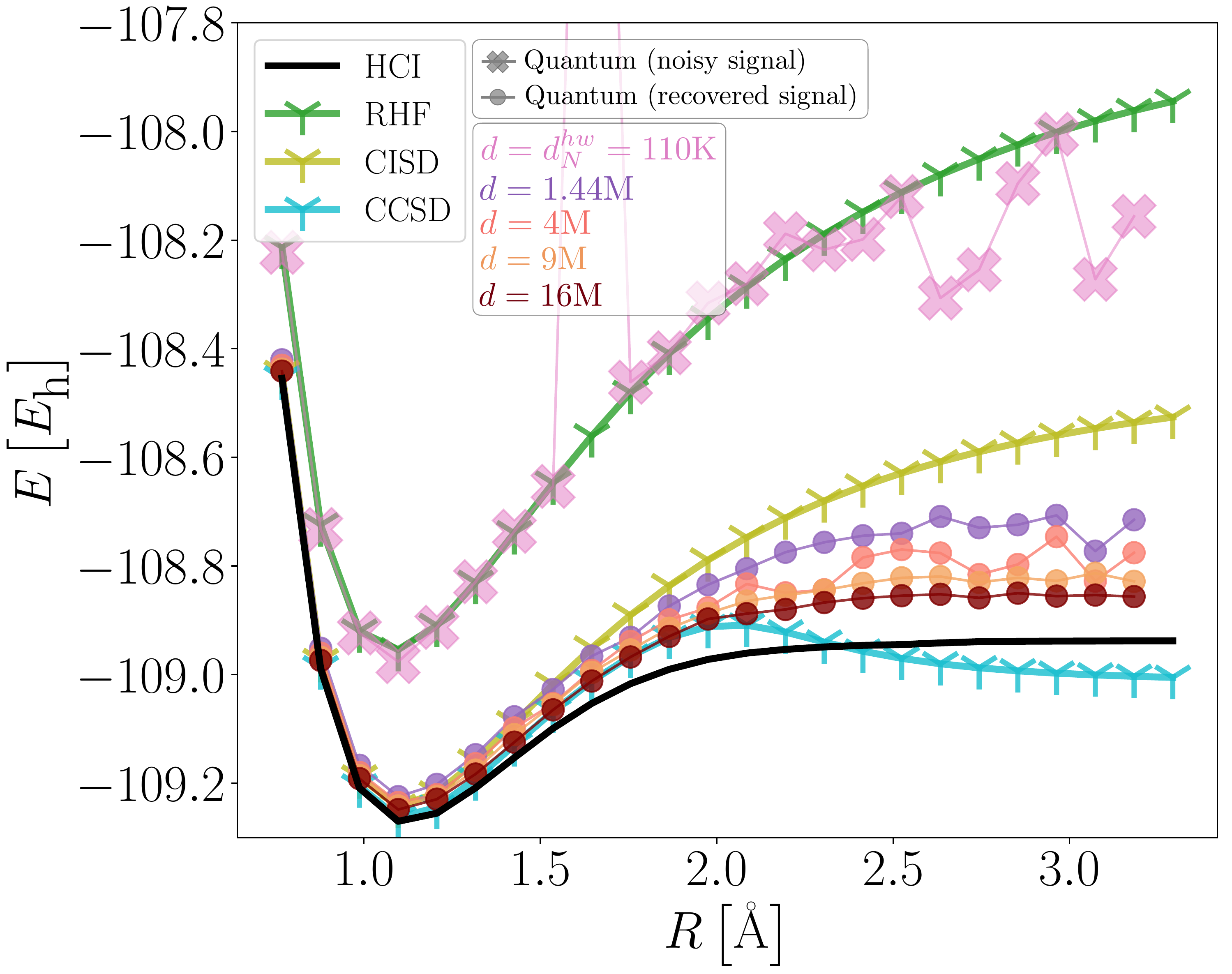}
    \caption{Comparison of the performance of SQD, with and without configuration recovery, run on the same set of measurement outcomes from the quantum processor. The dissociation of \nitrogen (cc-pVDZ) is considered. We compare the performance of the estimator with no configuration recovery (crosses) to the performance of the estimator with configuration recovery (points). Different numbers of configurations $d$ are considered, labeled by different colors. $d_N^{hw}$ refers to the number of configurations on the right particle sector of $N$ electrons extracted from the measurement outcomes, used for the estimator without configuration recovery (see the text for more details). We note that with current noise rates, $d_N^{hw}< 1.44 \textrm{ M}$. Different classical methods are shown for reference, as indicated in the legend.}
    \label{figS:N2_QSCI-vs-SCORE}
\end{figure*}

Fig.~\ref{figS:N2_QSCI-vs-SCORE} shows the comparison between SQD run with and without configuration recovery. The ground-state energy is shown as a function of the bond length. The estimator run on $\mathcal{X}_N$, i.e. without configuration recovery, obtains energies comparable to RHF energies. A level of accuracy comparable to RHF is not desirable for correlated electronic structure methods. However, since no mitigation techniques have been used in the large circuits producing the samples, this is an indication that a useful quantum signal (mostly of mean-field character) is present in $\tilde{\mathcal{X}}$. As expected, the estimator with configuration recovery shows higher levels of accuracy as compared to not using configuration recoivery, obtaining a qualitatively correct dissociation curve upon increasing $d$.

The accuracy of the energies obtained by SQD without configuration recovery could be improved by obtaining more measurement outcomes and thus more configurations on the right particle sector. The typical values of $d$ used for the estimator with configuration recovery are larger than $d^{hw}_N $. With current noise rates, obtaining a larger size of $|\mathcal{X}_N|$ to run the estimator without configuration recovery, requires the collection of many more measurement outcomes, at an efficiency where approximately only 1 in 1000 measured configurations has the right particle number (see Tab~\ref{tab:hardware details} for the precise fraction). Configuration recovery allows improving the accuracy of SQD without the collection of more measurement outcomes from the quantum processor. This is desirable as it reduces the overall cost of the calculation. 

\begin{figure*}
    \centering
    \includegraphics[width=1\linewidth]{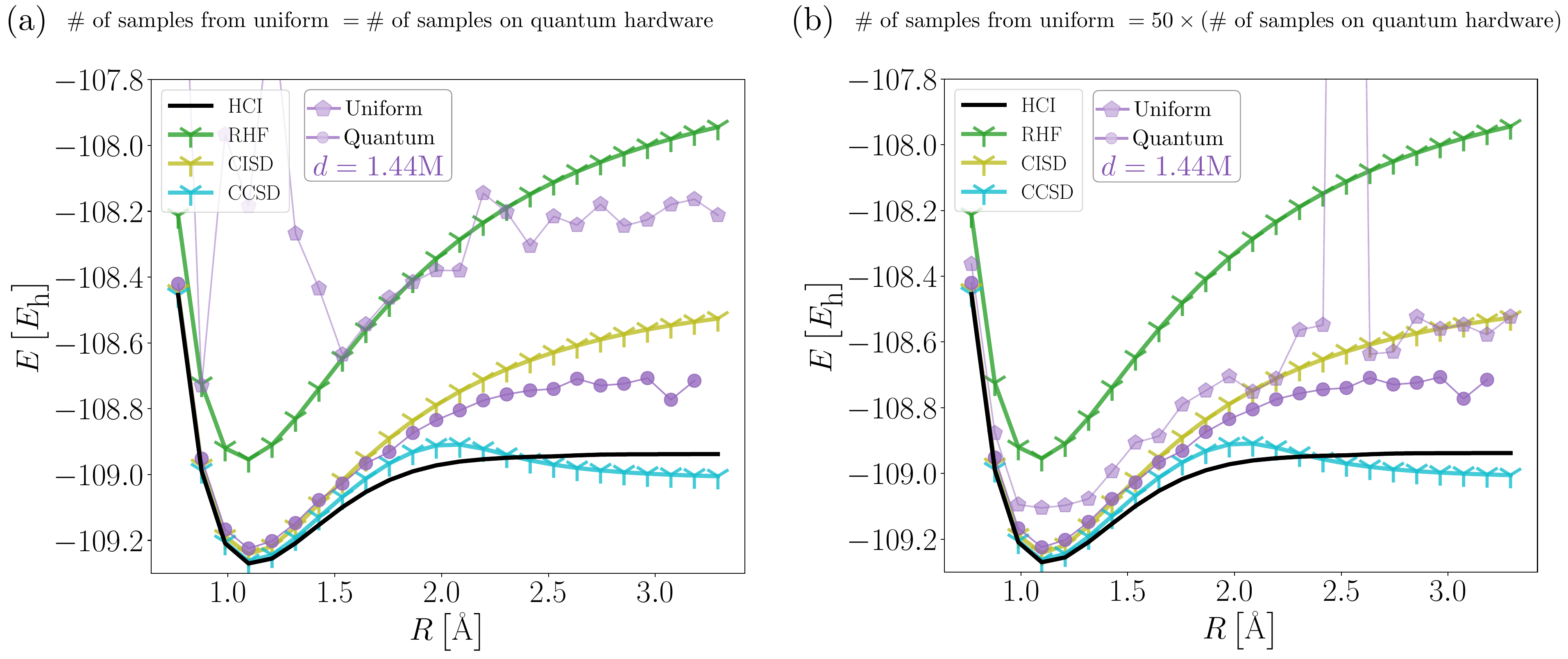}
    \caption{ Comparison of the estimator with configuration recovery applied to measurement outcomes from the quantum processor and configurations  drawn from the uniform distribution, in the dissociation of \nitrogen (cc-pVDZ). Pentagons correspond to the energies obtained with the configurations  sampled from the uniform distribution while dots correspond to the energies obtained from configurations  sampled from a quantum processor. Different classical methods are shown for reference, as indicated in the legend. \textbf{(a)} Uniform samples of configurations in the right particle sector. \textbf{(b)} Uniform samples of configurations over Fock space. }
    \label{fig:N2_SCORE-vs-Uniform}
\end{figure*}

\paragraph{Comparison of the quality of the estimator with configuration recovery, applied to quantum data and configurations drawn from the uniform distribution.--}
We first compare the performance of the estimator with configuration recovery applied to measurement outcomes from the quantum processor versus its application to configurations  sampled from the uniform distribution in the right particle sector. The estimator is applied to $10^5$ samples in both scenarios and we use the same value of $d$. In both cases, we take $K = 10$ batches of samples and report $\min_k\left(E^{(k)} \right)$ for the potential energy surface. When running the estimator with configuration recovery we use 10 self-consistent iterations. Recall that the configuration recovery procedure recovers configurations that have the wrong particle number due to noise, thus the application of the estimator to configurations sampled from the uniform distribution in the right particle sector is equivalent to applying the SCI eigenstate solver to the random configurations. Panel (a) in Fig.~\ref{fig:N2_SCORE-vs-Uniform} shows the comparison. We observe that the potential energy surface obtained from the measurement outcomes from the quantum processor is both qualitatively and quantitatively more accurate than the one obtained from the uniform distribution over configurations  in the right particle sector. The potential energy surface obtained from the configurations sampled from the uniform distribution shows a very prominent non-smooth behavior with energies worse than the RHF energies for a number of bond lengths. 

The comparison with samples obtained from the configurations uniformly sampled over the full Fock space and the quantum samples is shown in Fig.~\ref{fig:N2_SCORE-vs-Uniform} (b). We remark that $10^5$ samples over the uniform distribution in the Fock space of electronic configurations are unlikely to reveal any configurations in the correct particle sector, since the probability of a uniformly random bitstring falling in the correct particle sector is $\binom{\nmo }{N_\uparrow}\binom{\nmo }{N_\downarrow}/2^{M} = 9.6\cdot 10^{-7}< 10^{-5}$. Recall that the configurations on the right particle sector are required to start the configuration recovery procedure by providing the initial average occupancies. Therefore, we take $50 \cdot 10^6$ configurations  sampled from the uniform distribution over configurations  of length $M$ for this comparison. Note that the use of more samples from the uniform distribution as compared to the number of measurement outcomes from quantum constitutes a comparison that favors the uniform distribution scenario.

In spite of this, Fig.~\ref{fig:N2_SCORE-vs-Uniform} (b) shows a higher level of accuracy obtained by the estimator with configuration recovery run using measurement outcomes from the quantum processor.
These results indicate that the quantum processor has a signal that is amplified by the configuration recovery procedure and used by SQD to procedure more accurate ground-state representations than running the same estimator on uniform noise.

\subsubsection{[2Fe-2S] experiments}\label{sec:quantum signal 2Fe-2S}

\paragraph{Effect of configuration recovery on the accuracy of the HPC quantum estimator.--}
In this section, we compare the accuracy of the estimator without configuration recovery (i.e., applied to $\mathcal{X}_N \subset \tilde{\mathcal{X}}$) to the accuracy of the estimator with configuration recovery (applied to $\tilde{\mathcal{X}}$). As for all [2Fe-2S] experiments, $|\tilde{\mathcal{X}}| = 2.4576 \cdot 10^6$. In both cases, we take $K = 10$ batches of samples. When running the configuration recovery procedure, we use 10 self-consistent recovery iterations.

\begin{figure*}
    \centering
    \includegraphics[width=.7\linewidth]{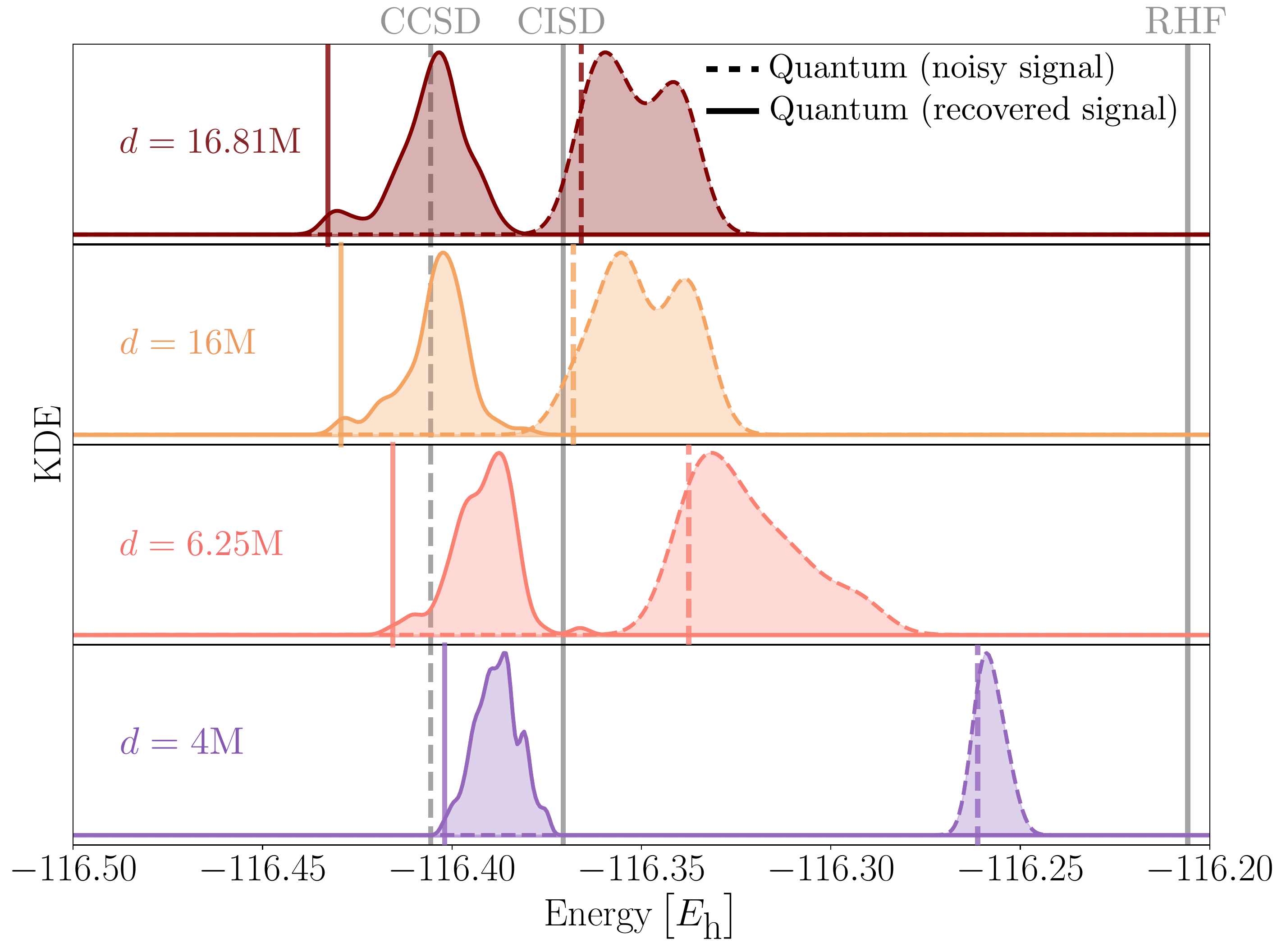}
    \caption{Comparison of the performance of SQD, with and without configuration recovery, run on the same set of measurement outcomes for the ground state of [2Fe-2S]. The \textit{kernel density estimation} (KDE) of the distribution of ground state energies obtained from the different batches of configurations is shown.  The performance of the estimator with no configuration recovery (dashed lines) is compared to the performance of the estimator with configuration recovery (solid lines). Different panels correspond to different subspace dimensions $d$, as indicated in each panel. In each panel, the solid and dashed vertical lines matching the color of the KDE shows the lowest value of the energy amongst the different batches of configurations. The grey vertical lines show the value of the ground-state energy obtained from different classical methods, as indicated at the top of the panel.}
    \label{fig:Fe2S2_QSCI-vs-SCORE}
\end{figure*}

Fig.~\ref{fig:Fe2S2_QSCI-vs-SCORE} shows the \textit{Kernel Density Estimation} (KDE) of the $E^{(k)}$ distribution running SQD both with and without configuration recovery for different values of $d$. The qualitative behavior of the KDE for the estimator with and without vonfiguration recovery is similar. Upon increasing the value of $d$, the distributions shift towards lower values of the energy. However, the distributions of $E^{(k)}$ for the estimator with configuraiton recovery have lower energy than those without. Without configuration recovery, the best ground-state energy obtained is between the RHF and CISD energies, whereas when using the configuration recovery procedure, the energy is decreased even below the CCSD estimate. Again, note that the fact that the estimator without configuration recovery still has energies below RHF is a promising indication that the raw data from the processor contains a signal that the configuration recovery procedure can amplify.

\begin{figure*}
    \centering
    \includegraphics[width=1\linewidth]{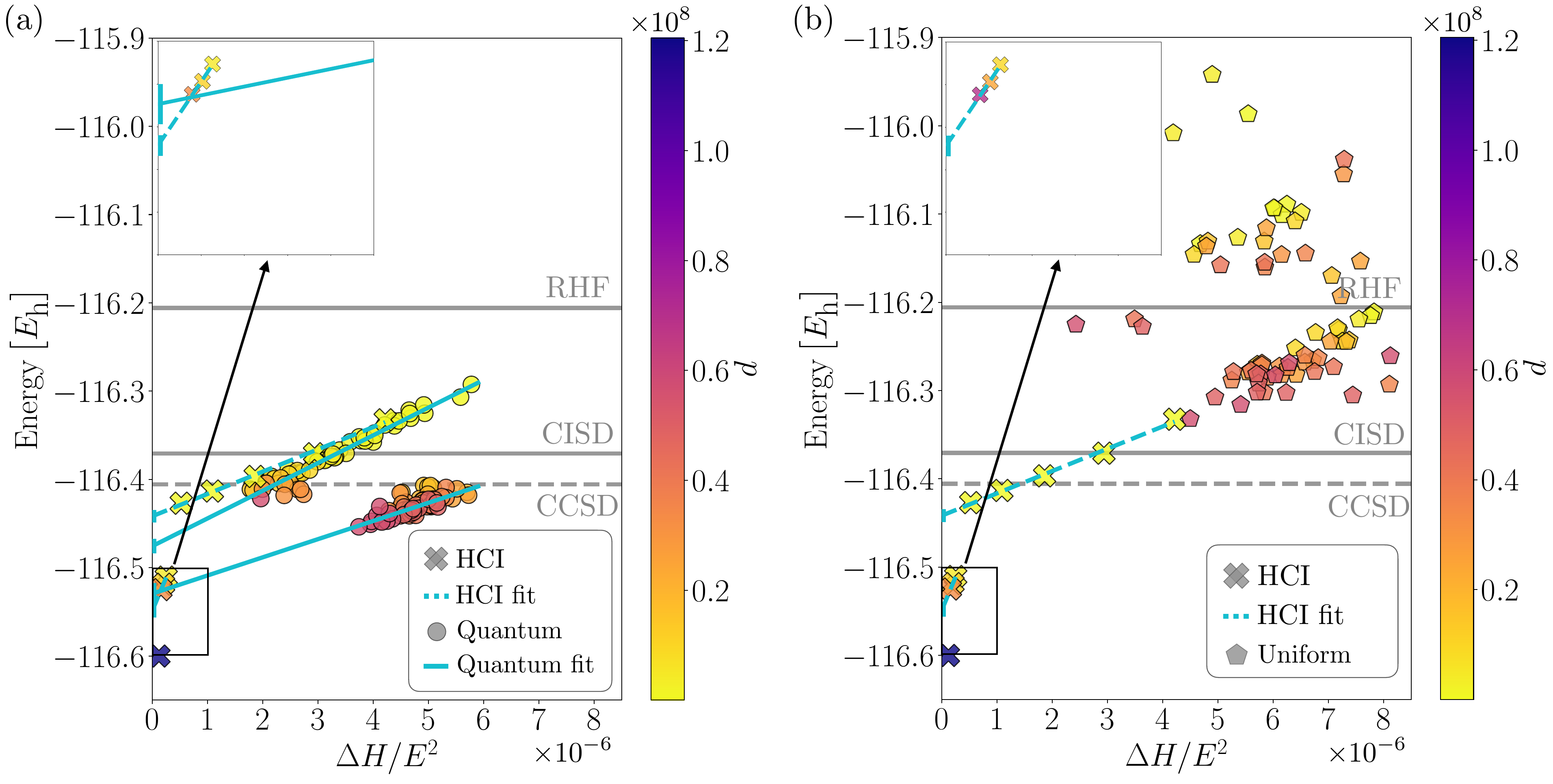}
    \caption{Energy-variance analysis in [2Fe-2S] of the approximate eigenstates obtained by running the estimator with configuration recovery on measurement outcomes from the quantum processor (panel \textbf{(a)}) and from the uniform distribution over the set of configurations that have the right particle number (panel \textbf{(b)}). Crosses show the energy-variance relation of HCI for reference, alongside the fit to a dashed line. Dots and pentagons correspond to energy-variance points obtained from the estimator applied to different batches of configurations. The number of determinants in each batch is encoded in the color bar. The inset is a zoom into the HCI extrapolation of the second Eigenstate. }
    \label{fig:Fe2S2_SCORE-vs-UniformRP}
\end{figure*}

\paragraph{Comparison of the quality of the estimator with configuration recovery, applied to quantum data and configurations drawn from the uniform distribution.--}
We first compare the performance of SQD with configuration recovery applied to measurement outcomes from the quantum processor versus configurations  drawn from the uniform distribution over configurations in the right particle sector. In both cases we consider $K = 10$ batches of electronic configurations and the estimator is applied to $2.4576 \cdot 10^6$ sampled configurations. The self-consistent recovery procedure is warm-started from the orbital occupancy vector $\vett{n}$ obtained after running ten self-consitent iterations with $d = \textrm{250K}$. Since the configuration recovery is warm-started, for $d>\textrm{250K}$ we apply two self-consistent iterations. In both scenarios, the same values of $d$ are considered. Recall that configuration recovery procedure recovers configurations that have the wrong particle number due to noise, thus the application of the estimator to configurations sampled from the uniform distribution in the right particle sector is equivalent to applying the SCI eigenstate solver to the random configurations. Fig.~\ref{fig:Fe2S2_SCORE-vs-UniformRP} shows a comparison of the energy-variance analysis obtained from the estimator on measurement outcomes from the quantum processor and on uniformly random configurations  with the right particle number. When the configurations come from measurement outcomes from a quantum processor, the energy-variance analysis shows two clusters of points following a linear relation. The two clusters correspond to two eigenstates of the Hamiltonian. The extrapolated energy is in good agreement with the energy extrapolation from HCI calculations. The energy-variance analysis on the estimator run on the random configurations  with the right particle number shows a different behavior. The energy-variance pairs are not clustered as clearly as in the previous scenario. For the largest values of $d$ considered, we observe that a first cluster begins to form, which coincides with the eigenstate of the highest energy that both HCI and SQD applied to quantum measurement outcomes reveal. However, for these values of $d$ the HPC estimator run on random samples is not capable of revealing the second eigenstate of lower energy that the estimator run on quantum data obtains.

\begin{figure*}
    \centering
    \includegraphics[width=1\linewidth]{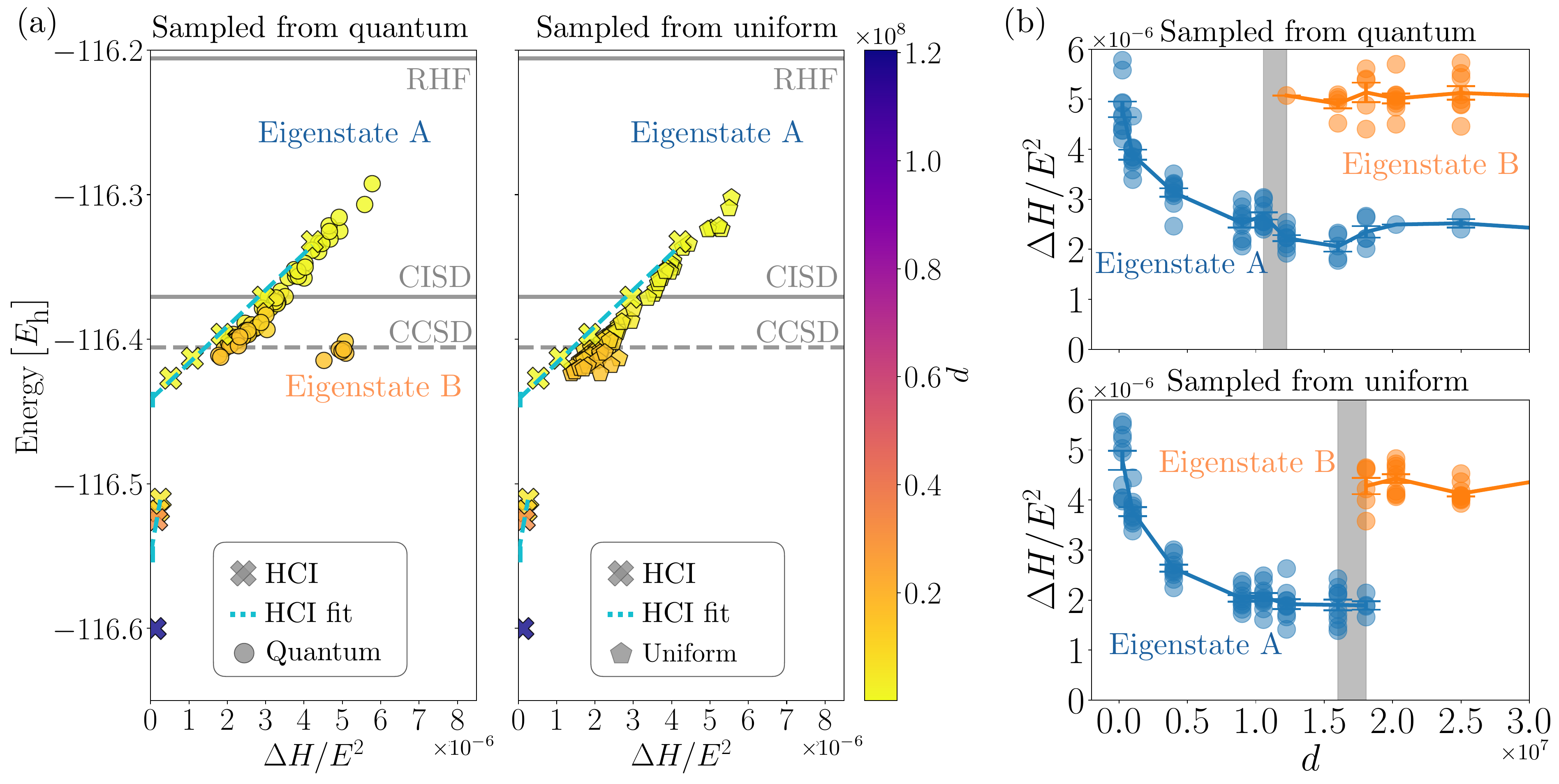}
    \caption{ Comparison of the energy-variance analysis of the low-energy spectrum of [2Fe-2S], obtained from the estimator with  configuration recovery applied to quantum data and configurations drawn from the uniform distribution. \textbf{(a)} The left panel shows energy-variance analysis of the low energy spectrum obtained by applying the estimator with configuration recovery to measurement outcomes from the quantum computer, while the right panel shows the same for samples from the uniform distribution on the Fock space. Crosses show the energy-variance relation of HCI for reference, alongside the fit to a (blue, dashed) line. Dots and pentagons correspond energy-variance points obtained from different batches of samples from the estimator. Subspace dimensions $d$ are indicated by the colors of the points (corresponding to the colorbar), up to $d = \textrm{16M}$, which is in the neighbourhood of the transition from Eigenstate A to B in the analysis applied to quantum samples. \textbf{(b)} Scatter plots of $\Delta H/E^2$ (horizontal axis on energy-variance plots) obtained from the estimator with configuration recovery on different batches of configurations, as a function of of the number of determinants $d$ in each batch. The solid lines show the mean value of $\Delta H/E^2$ as a function of $d$, and the error bars correspond to the standard error of the mean. Different groups of data points are color coded according to which eigenstate they come from. The grey region indicates the interval of $d$ where the transition from Eigenstate A to Eigenstate B occurs. Top panel corresponds to the application of the estimator with configuration recovery to configurations sampled from a quantum processor while the bottom panel corresponds to the application of the estimator with configuration recovery to configurations sampled from a uniform distribution over the Fock space.}
    \label{fig:Fe2S2_SCORE-vs-Uniform}
\end{figure*}

We also compare the performance of the HPC estimator with configuration recovery run on the data obtained from the quantum processor versus on configurations drawn from the uniform distribution over the full Fock space. In both cases, we consider $K = 10$ batches of electronic configurations and the estimator is applied to $2.4576 \cdot 10^6$ sampled configurations . The self-consistent recovery procedure is warm-started from the orbital occupancy vector $\vett{n}$ obtained after running ten configuration recovery iterations with $d = \textrm{250K}$. Since the configuration recovery procedure is warm-started, for $d>\textrm{250K}$ we apply two self-consistent operations. Fig.~\ref{fig:Fe2S2_SCORE-vs-Uniform} (a) shows the comparison of the energy-variance analyses in both scenarios for values of the subspace dimension up to $d = \textrm{16M}$. The qualitative behavior is similar, showing a cluster of points following a line that coincides with Eigenstate A of the Hamiltonian. However, $d \approx \textrm{16M}$ is enough for the estimator run on quantum samples to reveal the second cluster corresponding to Eigenstate B, while greater values of $d$ are required in the case where the samples are obtained from the uniform distribution over the full Fock space. Being able to resolve eigenstates of lower energy with smaller values of $d$ shows a more efficient description of the low-energy physics of the problem.

To quantify more precisely for which value of $d$ the Eigenstate B cluster of energy-variance pairs appears, we study the evolution of $\Delta H/E^2$ as a function of $d$. A jump in $\Delta H/E^2$ reveals the discovery of a cluster in the energy-variance plane. Fig.~\ref{fig:Fe2S2_SCORE-vs-Uniform} (b) shows the $\Delta H/E^2$ as a function of $d$ in both scenarios. While a large enough value of $d$ allows resolving Eigenstate B from the estimator with configuration recovery applied to the uniformly random samples, this transition requires the projection and diagonalization of the Hamiltonian in a subspace whose dimension is $d>\textrm{16M}$. The observation of the same transition on the configurations sampled from the quantum processor requires at most $d = \textrm{12.25M}$ configurations. The difference in these dimensions is of at least $\textrm{3.75M}$ electronic configurations.

These results indicate that the quantum processor provides a useful signal that is used by SQD to procedure more accurate ground-state representations than using the same estimator on samples from uniform distribution.



\end{document}